\documentclass[12pt]{iopart}

\pdfoutput=1
\usepackage{iopams}  
\usepackage{graphicx}
\usepackage{subcaption}
\usepackage{placeins}

\usepackage[symbol]{footmisc}

\begin{document}

\title[Time- and energy-resolved effects in the boron-10 based Multi-Grid ...]{Time- and energy-resolved effects in the boron-10 based Multi-Grid and helium-3 based thermal neutron detectors}

\author{A. Backis$^{a, b, *}$\footnote[0]{$^*$Corresponding author}, A. Khaplanov$^b$, R. Al Jebali$^{b, a}$,  R. Ammer$^b$, I. Apostolidis$^b$, J. Birch$^c$, C.-C. Lai$^{b, c}$,  P. P. Deen$^{b, d}$, M. Etxegarai$^b$, N. de Ruette$^b$, J. Freita Ramos$^b$,  D. F. F\"orster$^e$,  E. Haettner$^b$, R. Hall-Wilton$^{b, f, a}$, D. Hamilton$^a$, C. H\"oglund$^{b, g}$, P. M. Kadletz$^b$, K. Kanaki$^b$, E. Karnickis$^b$, O. Kirstein$^b$, S. Kolya$^b$, Z. Kraujalyte$^b$, A. Laloni$^b$, K. Livingston$^a$, O. L\"ohman$^h$, V. Maulerova$^{b, i}$,  N. Mauritzon$^{b, i}$, F. M\"uller$^e$, I. Lopez Higuera$^b$, T. Richter$^b$, L. Robinson$^b$, R. Roth$^{j}$, M. Shetty$^b$, J. Taylor$^b$, R. Woracek$^b$ and W. Xiong$^b$}

\address{$^a$University of Glasgow, Glasgow, United Kingdom}
\address{$^b$European Spallation Source ERIC (ESS), Lund, Sweden}
\address{$^c$Link\"oping University, Link\"oping, Sweden}
\address{$^d$Niels Bohr Institute, University of Copenhagen, Copenhagen, Denmark}
\address{$^e$Zentralinstitut f\"ur Engineering, Elektronik und Analytik, Forschungszentrum J\"ulich GmbH, J\"ulich, Germany}
\address{$^f$Università degli Studi di Milano-Bicocca, Piazza della Scienza 3, 20126 Milano, Italy}
\address{$^g$Impact Coatings AB, Westmansgatan 29G, SE-582 16 Link\"oping, Sweden}
\address{$^h$Technische Universit\"at Darmstadt, Hochschulstra{\ss}e 8, D-64289 Darmstadt}
\address{$^i$Lund University, Lund, Sweden}
\address{$^{j}$EWCON, \"Orkelljunga, Sweden}

 \ead{a.backis.1@research.gla.ac.uk}
\vspace{10pt}
\begin{indented}
\item[]October 2020
\end{indented}

\newpage

\begin{abstract}
    The boron-10 based Multi-Grid detector is being developed as an alternative to helium-3 based neutron detectors. At the European Spallation Source, the detector will be used for time-of-flight neutron spectroscopy at cold to thermal neutron energies. The objective of this work is to investigate fine time- and energy-resolved effects of the Multi-Grid detector, down to a few $\mu$eV, while comparing it to the performance of a typical helium-3 tube. Furthermore, it is to characterize differences between the detector technologies in terms of internal scattering, as well as the time reconstruction of $\sim$ $\mu$s short neutron pulses. The data were taken at the Helmholtz Zentrum Berlin, where the Multi-Grid detector and a helium-3 tube were installed at the ESS test beamline, V20. Using a Fermi-chopper, the neutron beam of the reactor was chopped into a few tens of $\mu$s wide pulses before reaching the detector, located a few tens of cm downstream. The data of the measurements show an agreement between the derived and calculated neutron detection efficiency curve. The data also provide fine details on the effect of internal scattering, and how it can be reduced. For the first time, the chopper resolution was comparable to the timing resolution of the Multi-Grid detector. This allowed a detailed study of time- and energy resolved effects, as well as a comparison with a typical helium-3 tube.
\end{abstract}

\vspace{2pc}
\noindent{\it Keywords}: Neutron detectors, Gaseous detectors, Boron-10, Multi-Grid detector, Helium-3, Time resolution 
%\maketitle

\newpage

\section{Introduction}
The European Spallation Source (ESS), is currently under construction in Lund, Sweden \cite{ESS_web, Garoby_2017, ESS_TDR}. Due to the sparsity and increased cost of helium-3 gas during the past two decades \cite{Zeitelhack}, as well as the performance limitation due to the expected high flux from the source, alternatives to the traditional helium-3 based neutron detectors are being developed. One such alternative is the boron-10 based Multi-Grid detector \cite{Magdalena2012, mg2012temp, mg2012b, mg2013, Birch2014, MGback, CNCS}, invented at the Institut Laue–Langevin (ILL) \cite{mg_patent}, and jointly developed with ESS thereafter. The Multi-Grid detector is a large area cold to epithermal neutron detector, designed for time-of-flight neutron spectroscopy \cite{copley_1993} at the upcoming CSPEC \cite{CSPEC} (cold neutron spectroscopy) and T-REX \cite{TREX} (thermal neutron spectroscopy) instruments at ESS. 

Neutron spectroscopy is a demanding technique in terms of neutron detector performance, requiring a high detection efficiency and low noise levels. It also needs a broad detector coverage, tens of square meters, and hence a low cost per unit active detector area is important. The spectroscopy information is primarily extracted from inelastic neutron scattering of a sample, measuring small changes in energy between incident and scattered neutrons. This probes sample properties such as molecular vibrations, quantum excitations and motions of atoms.  As the measured energy changes are in the order of meV, the method requires a high time resolution and good knowledge of interaction location to correctly derive the energy of the scattered neutrons \cite{Ehlers_2011, Granroth_2010, ANDERSEN2020163402}.

For neutron detection, the Multi-Grid detector employs multiple $^{10}$B$_4$C thin films coated on aluminum substrates \cite{Carina_2012, Carina_2015, Schmidt_2016}. The  $^{10}$B$_4$C deposition is carried out by the ESS Detector Coatings Workshop in Link\"oping, where the deposition is done with an industrial deposition system, using physical vapor deposition DC magnetron sputtering \cite{Schmidt_2016}. The coated substrates are stacked perpendicularly to the incident neutrons, and placed in a 3D-position sensitive multi-wire proportional chamber (MWPC), see Section 2.4 for details. Incident neutrons are absorbed in the coating, whereupon one of the conversion products is ejected into the gas volume. In the gas, charge is released and collected, and each neutron event is assigned a time-of-flight ($tof$), as well as a hit position. This information is then used to calculate the neutron flight distance ($d$).

The neutron time-of-flight and flight distance are relative to a neutron chopper. A neutron chopper is a device containing an efficient neutron absorber, such as gadolinium or boron, and is placed blocking the incident neutron beam. The chopper rotates at a high frequency, and contains an opening for neutrons to pass periodically. That is, the incident neutron beam is ``chopped" into pulses. Each time a neutron pulse is transmitted, a start time, commonly know as $T_0$, is recorded. Then, when a neutron is detected in the detector, a stop time $T$ is acquired. The time-of-flight is then calculated according to $tof = T - T_0$, and the neutron flight distance $d$ is the distance between the chopper and the detection location in the detector, i.e. the hit position.

Using the time-of-flight and flight distance, the incident neutron energy $E_n$ can be calculated on an event-by-event basis. This is done by calculating the neutron velocity $v_n$, according to $v_n = d/tof$, which gives the energy using $E_n \propto v_n^2$. However, a small fraction of neutrons undergo internal scattering before being detected. This is primarily due to the aluminum inside the detector, see Section 3.3. Consequently, these scattered neutrons are assigned an incorrect flight distance, as the additional scattered travel path cannot be accounted for. This changes the energy line shape reconstruction, adding a small distribution of neutrons with incorrectly assigned energies \cite{DIAN2018173, Dian_2019, Eszter2019thesis}. A good understanding of this effect is important, as spectroscopy instruments depend on a well understood energy line shape. This is especially true for quasi-elastic scattering analysis \cite{qens}, where subtle details in the line shape are studied at the edges of the energy peak.

For cold neutrons, the detector time and energy resolution are also critical. The full resolution of the instrument depends on the pulse broadening from the chopper system, sample scattering and the detector. Therefore, it is desired to keep the detector energy resolution finer than the resultant resolution of the remaining components. This is especially important for the slowest neutrons, as the resolution of the chopper system is highest at these energies. One of the main components of the detector resolution relates to the uncertainty of when and where the neutron conversion reaction occurs in the detector. In a helium-3 tube, this reaction can happen anywhere within the gas, while for the Multi-Grid detector, the reaction can only occur at discrete intervals, corresponding to the position of the conversion layers \cite{Dian_2019}.

This work contains three separate investigations, each characterizing an important aspect of the Multi-Grid detector. First, the neutron detection efficiency of the Multi-Grid detector is derived and compared to the theoretical prediction. Then, the magnitude of the internal scattering, as well as the effect it has on the line shape reconstruction, is investigated. This is done by comparing two different Multi-Grid prototypes, one with internal topological shielding and one without, and examining the difference in energy line shape. Finally, the energy resolution of the Multi-Grid detector and a helium-3 tube are accessed and compared.

\section{Instrumentation and experimental setup}
The measurements were conducted at the Helmholtz-Zentrum Berlin (HZB), at the BER II research reactor \cite{ber2}. At the facility, a series of measurements were done at the V20 beamline \cite{V20_instrument, V20_choppers, Maulerova_2020}. During the course of the measurements, three different detectors were used: one 10 bar helium-3 tube and two variations of the latest prototype of the Multi-Grid detector. The setup consisted of a beam-monitor, slits and a lightweight fast rotating Fermi-chopper \cite{FORSTER2018298}. The detectors were situated just after the Fermi-chopper, as illustrated in figure~\ref{fig:fig2}.

Three direct beam measurements were conducted, one for each of the three detectors, keeping the rest of the setup constant. In addition to this, two background measurements were done, one for each of the Multi-Grid prototypes, where the direct beam was blocked by the helium-3 tube. The helium-3 tube was covered with 5 mm Mirrobor shielding \cite{Mirrotron}, 80 \% B$_4$C (natural boron) content in weight, at the back. The back-shielding, in combination with the high neutron absorption efficiency of the helium-3 tube in the measured wavelength range, resulted in transmission levels below the instrument background level for almost all data points. The exception was for the data point at 1.2 \AA{}, where the transmission was at the acceptable level of $<5 \cdot 10^{-4}$.

\subsection{The ESS test beamline V20}
\begin{figure}[h!]
\centering
\includegraphics[width=1\linewidth]{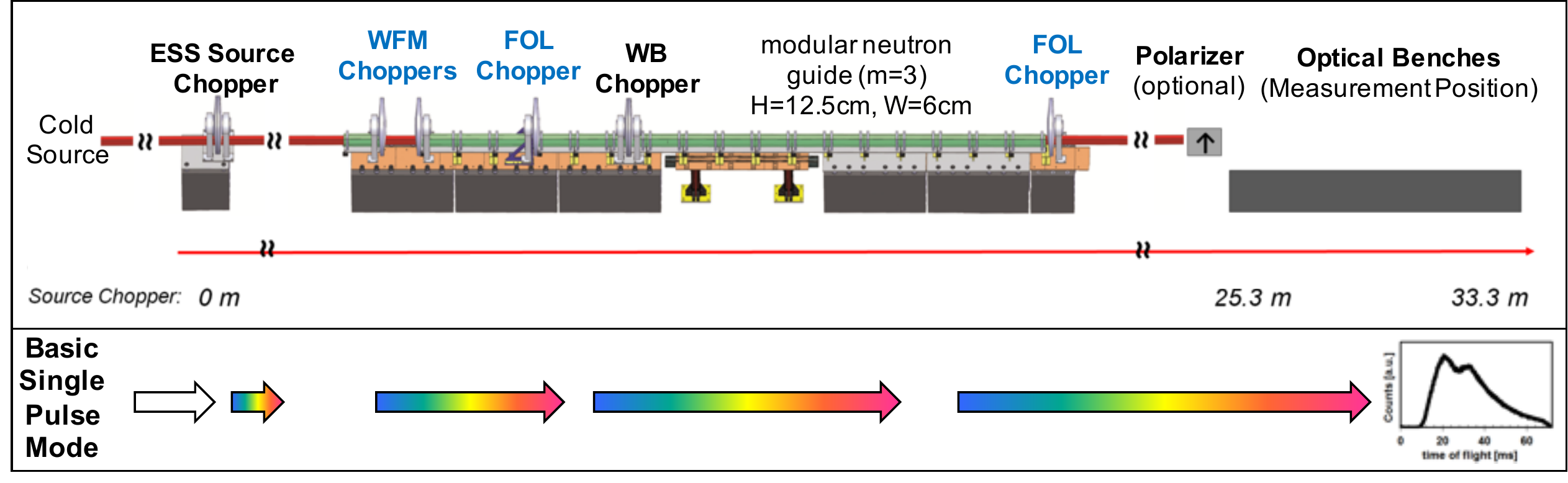}
\caption{Illustration presenting the layout of the V20 beamline, showing the main components and their locations. At 21.7 m from the moderator, the ESS source chopper is located, which, together with the WB chopper, cuts the flux into wide pulses at 14 Hz. There are also WFM and FOL choppers, above presented in blue, followed by an optional polarizer. This allows for two different modes, ``Basic Single Pulse Mode'' and ``WFM Mode''. For these measurements, single pulse mode was used, which is illustrated above. The optical benches is where the experiment specific setup is positioned.$^1$}
\label{fig:fig1}
\end{figure}

The V20 is a cold to thermal neutron beamline. It was commissioned as an ESS test beamline, with a chopper system designed to mimic the long pulses which will be obtained at ESS. This is done by a pair of double disc choppers: a source chopper and a wavelength band (WB) chopper. Together, these deliver pulses at 14 Hz, 60 ms pulse width, where the wavelengths in each pulse range from approximately 1 to 10 \AA{}. The beam line is also equipped with wavelength frame multiplication (WFM) choppers, which, together with a pair of frame overlap (FOL) choppers, cut the long pulse from the source chopper to smaller bunches. This option was, however, not used for the herein reported work, and the corresponding choppers were parked in open position.  Instead, the choppers were used in the ``Basic Single Pulse Mode'', as illustrated in figure~\ref{fig:fig1}. 

\footnotetext[0]{$^1$Reprinted from ``Nuclear Instruments and Methods in Physics Research Section A: Accelerators, Spectrometers, Detectors and Associated Equipment'', Vol. 839, R. Woracek, T. Hofmann, M. Bulat, M. Sales, K. Habicht, K. Andersen and M. Strobl, ``The test beamline of the European Spallation Source – Instrumentation development and wavelength frame multiplication'', pp 102-116, 2016, with permission from Elsevier.}

\begin{figure}[h!]
\centering
\includegraphics[width=1\linewidth]{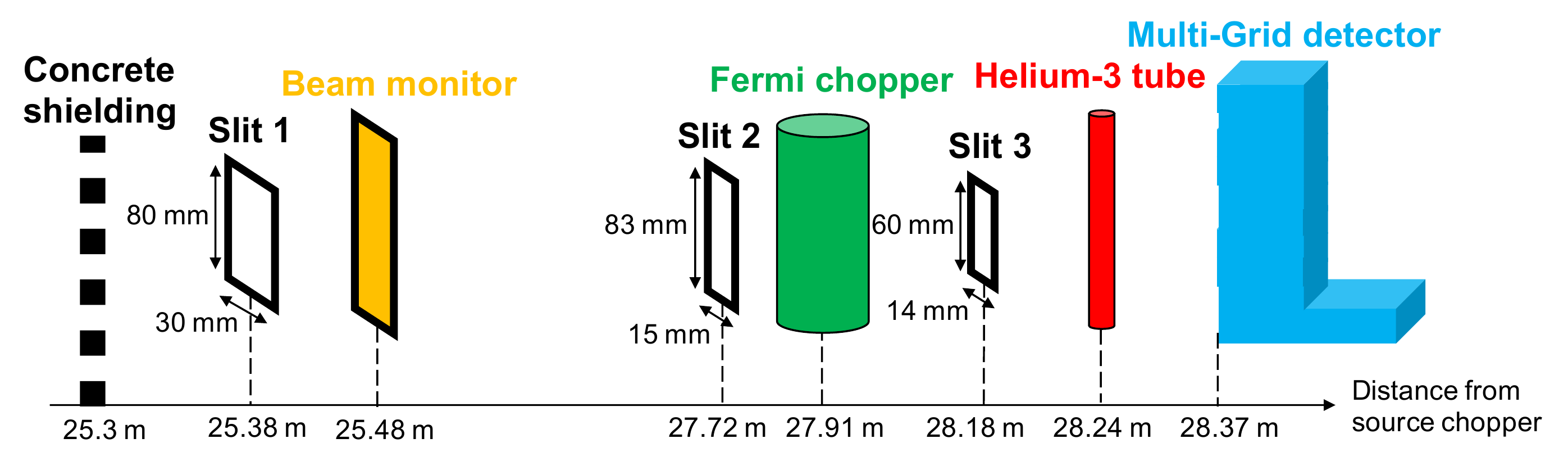}
\caption{Schematic illustration depicting the experimental setup assembled at the optical benches section at the V20 beamline. The location of the components are shown in relation to the source chopper. The values mark the center position of the components, except for the Multi-Grid detector, where it marks the front. The setup includes, downstream from the source, a beam monitor (orange), a Fermi chopper (green) and slits (black) in between defining a 14~$\times$~60~mm$^2$ rectangular beam. At the end, either the helium-3 tube (red) or Multi-Grid detector (blue) is used.}
\label{fig:fig2}
\end{figure}

The current setup was assembled at the ``Optical Benches'' position, at the last part of the beamline. This is outlined in figure~\ref{fig:fig2}, where all the components and their locations are presented. First, a slit confines the area of the incident neutron beam to a few cm$^2$. The incident flux is then recorded by a low helium-3 pressure beam monitor \cite{bme5}. After a further slit collimation, halving the beam width, a Fermi-chopper is used to cut and shape the long incident pulse to a few tens of $\mu$s short pulses. Finally, the beam is collimated to a 14~$\times$~60~mm$^2$ rectangular beam before reaching the detectors. This is either the helium-3 tube or the Multi-Grid detector. During the direct beam measurements with the Multi-Grid detector, the helium-3 tube was removed from the beam path. Then, during the background measurements, it was re-installed such that it was blocking the direct beam from reaching the Multi-Grid detector. Note that the detectors are placed in sequence, due to the limitation of the setup, and that the Multi-Grid detector is situated approximately an additional 50~\% further downstream of the Fermi-chopper than the helium-3 tube. The signal-to-background ratio at the instrument is approximately 10$^3$, and is influenced by gamma radiation levels and straying neutrons in the vicinity of V20.

\subsection{Low pressure helium-3 filled beam monitor}

\begin{figure}[h!]
        \centering
        \includegraphics[height=1.7in]{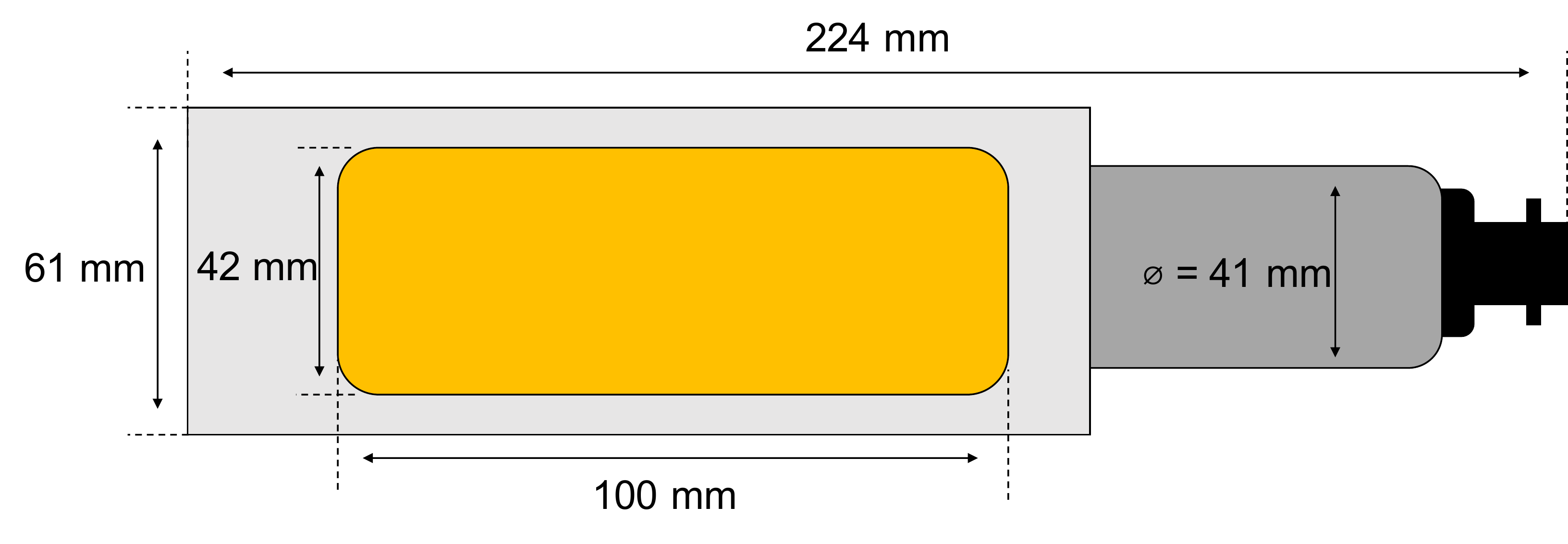}
    \caption{Low pressure helium-3 filled beam monitor. A drawing of the beam monitor, not to scale, is presented showing the details of the geometry. The left region (orange and light grey) is a box, while the middle region (dark grey) is a cylinder, and the right region (black) is a connector.}
    \label{fig:bm_drawing}
\end{figure}

The beam monitor used was a low pressure helium-3 proportional counter \cite{Maulerova_2020, Maulerova_2020_2, fatima_2016} from Eurisys Mesures, currently Mirion Technologies \cite{mirion}. The detector has an active area of 100~$\times$~42~mm$^2$, with a 40~mm active depth. The thickness of the aluminum window is 4~mm in the neutron beam path, including 2~mm inlet + 2~mm outlet. An overview of the geometry is seen in figure~\ref{fig:bm_drawing}.

The gas-filling is a 1.3 bar Ar-CH$_4$ (90:10 volume ratio) mixture, with a low pressure of helium-3. The neutron detection efficiency is  $\sim$10$^{-5}$ at 1.8~\AA{}. Thus, the incident neutron flux is only marginally reduced, corresponding to the low neutron absorption in the helium-3 gas and scattering in the 4 mm aluminum window.

\FloatBarrier

\subsection{Lightweight fast rotating Fermi-chopper}
\begin{figure}[h!]
    \centering
    \begin{subfigure}{0.5\textwidth}
        \centering
        \includegraphics[height=2.3in]{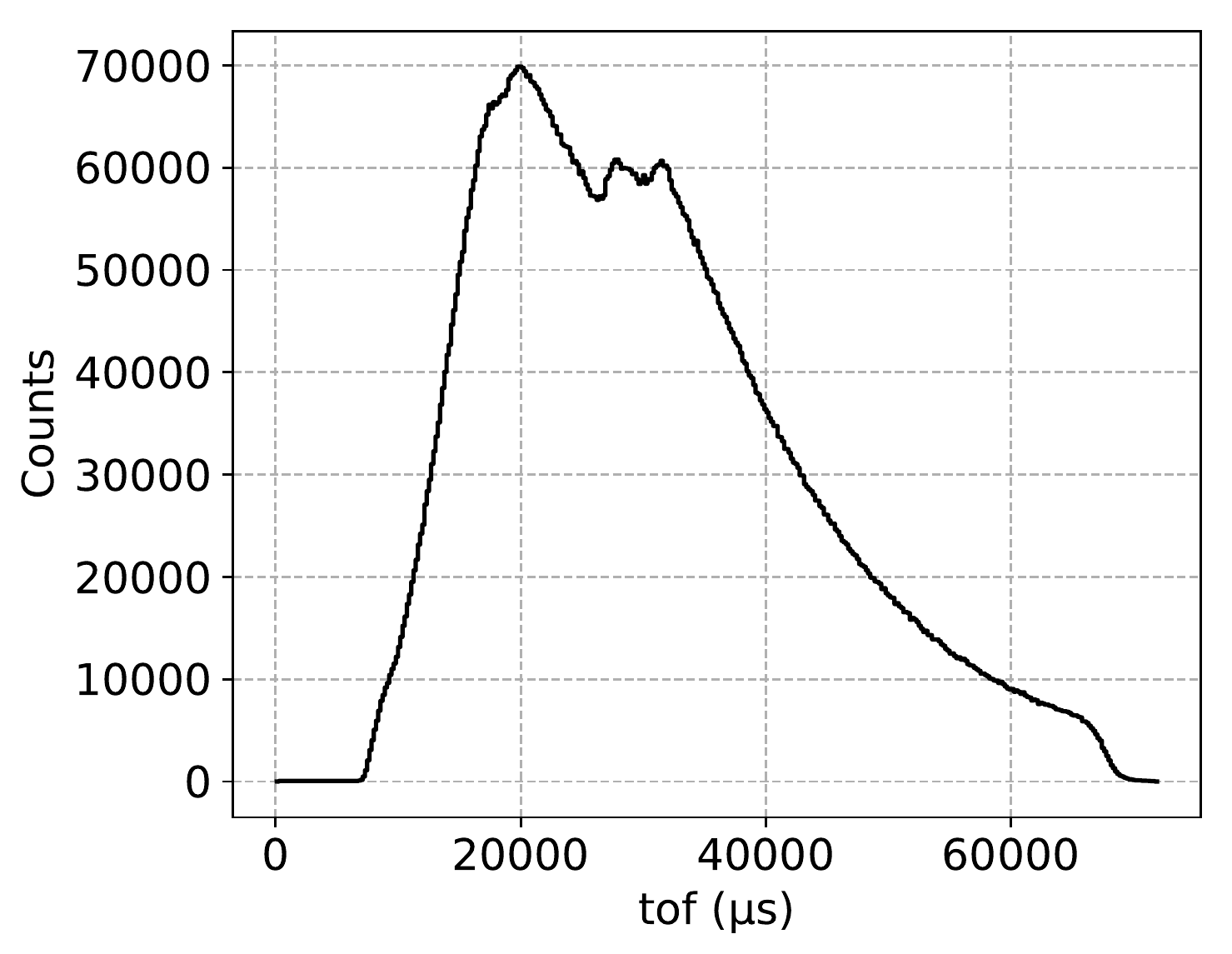}
        \caption{}
        \label{fig:fig3a}
    \end{subfigure}%
    \begin{subfigure}{0.5\textwidth}
        \centering
        \includegraphics[height=2.3in]{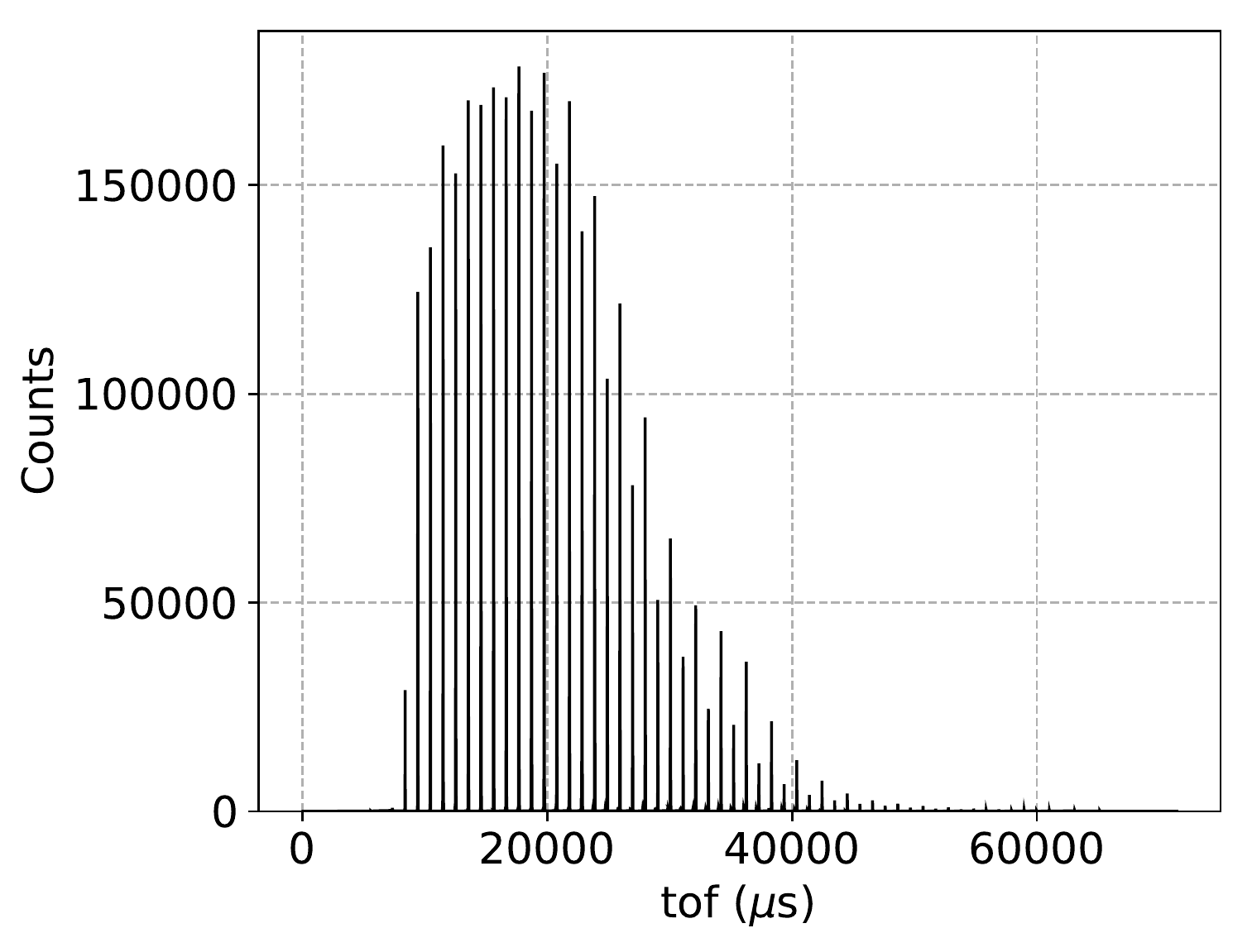}
        \caption{}
        \label{fig:fig3b}
    \end{subfigure}
    \begin{subfigure}{1\textwidth}
        \centering
        \includegraphics[height=2.65in]{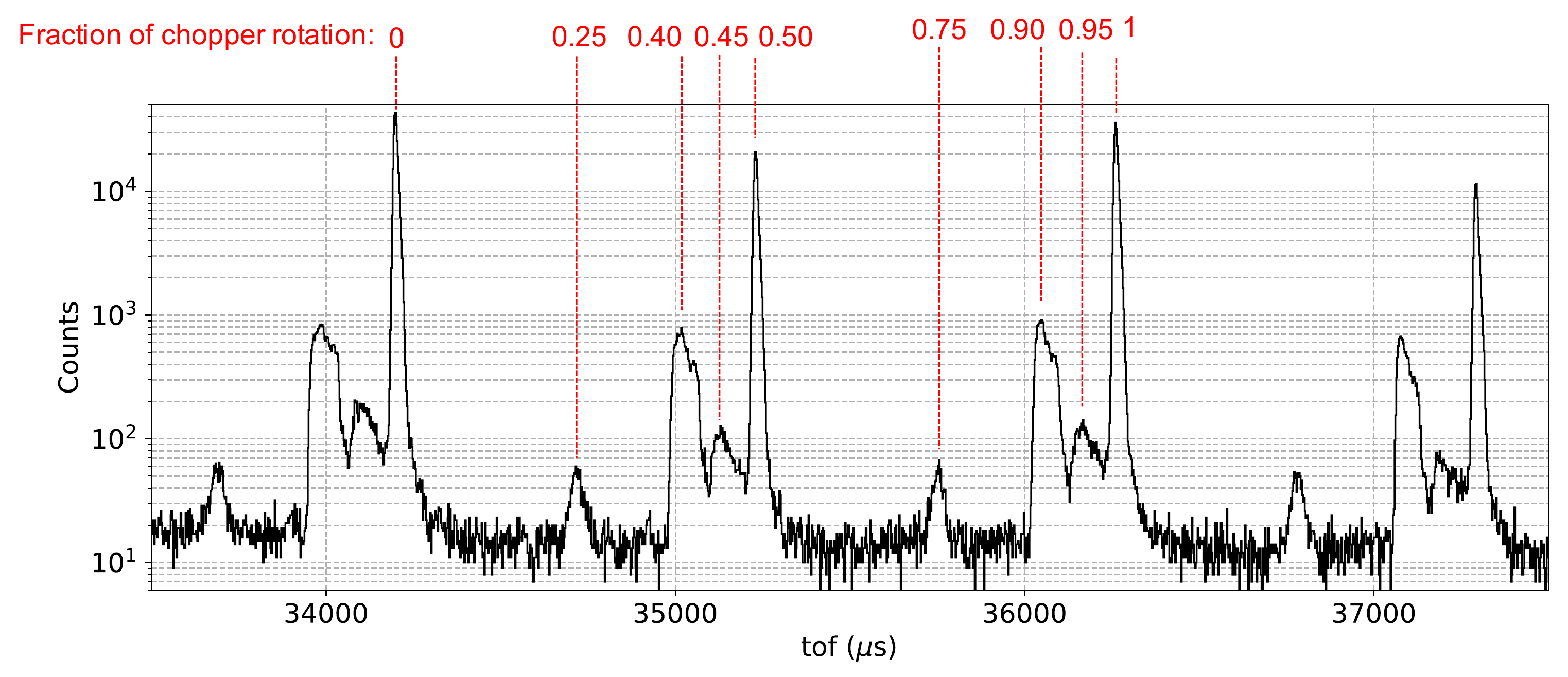}
        \caption{}
        \label{fig:fig3c}
    \end{subfigure}
    \caption{Time-of-flight spectra before and after the Fermi-chopper. The time offset is arbitrary. In (a), a spectrum collected with the beam monitor is presented, showing the wide pulse from the source chopper. In (b), a spectrum from the helium-3 tube is shown, illustrating how the chopper splits the $\sim$ 60 ms long source pulse into a series of $\sim$ 10 $\mu$s short pulses. In (c), a magnification of a portion of (b) is shown. On the log scale, it is clear that there are side peaks which are 1-2 orders of magnitude smaller than the main peaks.}
    \label{fig:fig3}
\end{figure}

For the measurements, an experimental 120 mm high, 27 mm in diameter, lightweight Fermi-chopper was used. The purpose was to produce a series of very short,  $\sim$~10~$\mu$s, and sharp pulses in time, to study the energy dependent scattering within the detector between pulses. The Fermi-chopper is synchronized to the 14 Hz source chopper frequency with a clock multiplier of 35, resulting in a 490 Hz rotation frequency. The chopper uses two rectangular chambers for neutron transmission, and a gadolinium neutron absorber on the walls. The resulting transmission of the Fermi-chopper is presented in figure~\ref{fig:fig3}. In figure~\ref{fig:fig3a}, the time-of-flight spectrum before the Fermi-chopper is shown, collected using the beam-monitor, while in figure~\ref{fig:fig3b} the time-of-flight spectrum after the Fermi-chopper is shown, collected using the helium-3 tube.

The chopper system produces a series of sharp pulses. However, as can be seen in figure~\ref{fig:fig3c}, there are additional features introduced. First, odd and even peaks are alternating in intensity. This is due to an asymmetry in the two rectangular Fermi-chambers, appearing every half rotation of the chopper blades. This is further discussed in \cite{FORSTER2018298}. Second, two distinct side peaks are identified, suppressed by 1-2 orders of magnitude. The same side peaks are present in data from the Multi-Grid detector. Thus, this feature is not a detector dependent effect. Additionally, there are 2-3 orders of magnitude suppressed peaks appearing midway between the main peaks, coinciding with every quarter rotation of the chopper blades. This was shown to correspond to a small misalignment of the incident neutron beam, allowing neutrons to pass the side of the Fermi package every quarter rotation. These peaks does not cause a problem for the current analysis, as they are easily distinguishable from the main peaks.

\FloatBarrier

\subsection{Helium-3 filled proportional counter}

\begin{figure}[h!]
        \centering
        \includegraphics[height=1.7in]{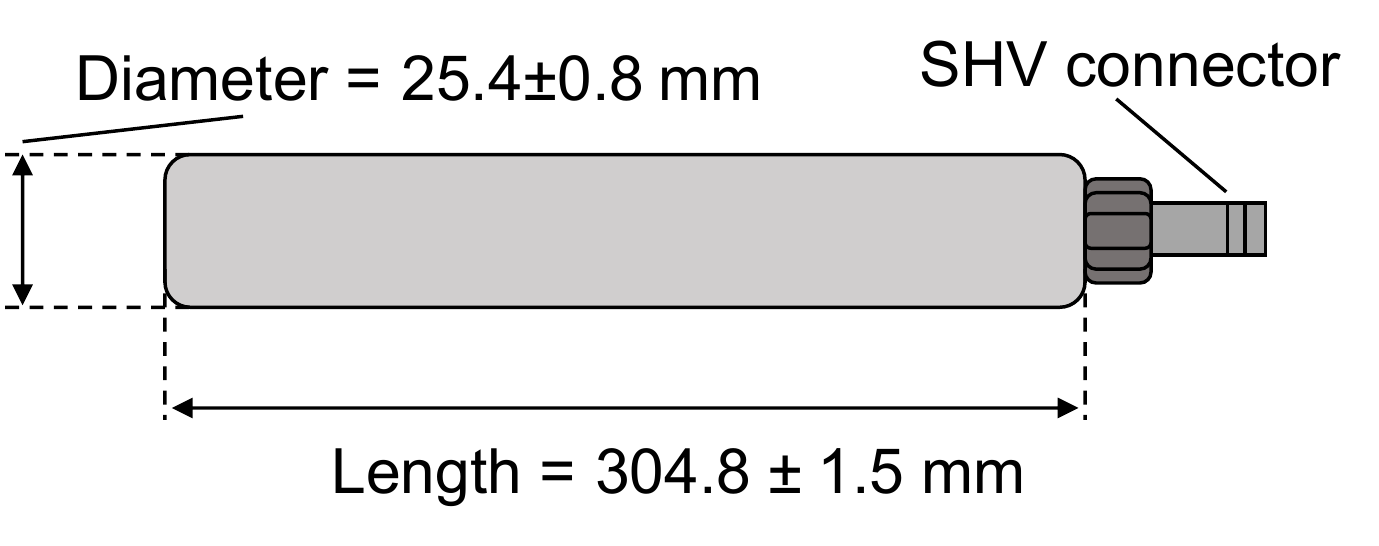}
    \caption{Helium-3 filled proportional counter. A drawing of the tube, not to scale, is presented showing the details of the geometry.}
    \label{fig:fig4}
\end{figure}

For the measurements, a helium-3 filled proportional counter from Reuter-Stokes \cite{He3tube} was used. The tube has a total gas pressure of 10 bars, split between 9.85 bar helium-3 and 0.15 bar quenching gas. The active diameter of the tube is 25.4 mm, with a length of 305 mm, and it is enclosed in 0.51 mm walls of stainless steel. The full geometry is presented in figure~\ref{fig:fig4}. During the measurements, the tube was wrapped in Mirrobor shielding \cite{Mirrotron}, leaving a small window for incident neutrons.

\FloatBarrier

The tube is a non-position sensitive detector, operated at 1350 V. Gamma events were rejected using a constant charge discriminator threshold on all events. The read-out system consists of a multi-channel analyzer, the FAST ComTeC MCA4 \cite{MCA4}, which returns time-of-flight and collected charge for each neutron event. In case of a pile-up detection during the charge integration process, the event is also assigned a ``pile-up flag". The shaping time was set to 1~$\mu$s.

Note that it was not a strict requirement for the helium-3 tube used in this measurement to have the specific characteristics described above. The main purpose of the helium-3 tube was to have a well-understood technology to ``bench-mark'' the Multi-Grid detector, while the secondary purpose was to use it as a flux normalization. The performance comparison with the Multi-Grid detector will, of course, vary if a different tube is selected. The selected tube, however, is a good choice because it has a gas pressure representative of what is commonly used for helium-3 tubes in spectroscopy instruments.

\subsection{The Multi-Grid detector}

\begin{figure}[h!]
    \centering
    \begin{subfigure}{0.5\textwidth}
        \centering
        \includegraphics[height=2.3in]{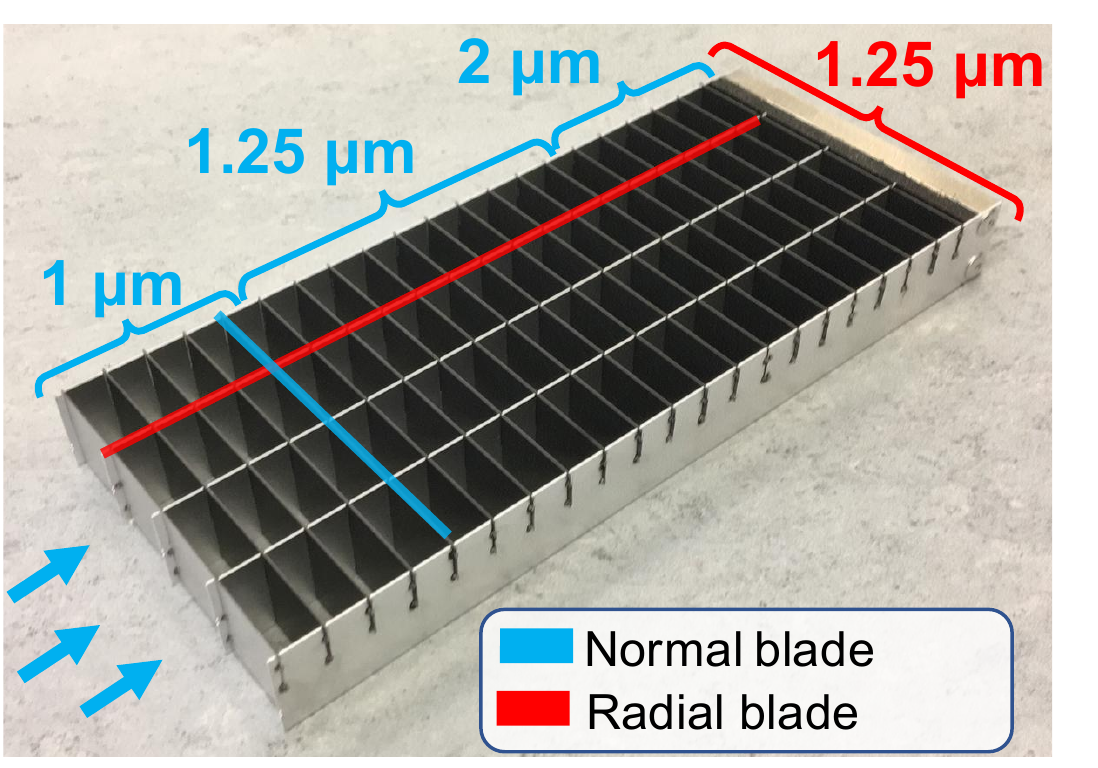}
        \caption{}
        \label{fig:fig6a}
    \end{subfigure}%
    \begin{subfigure}{0.5\textwidth}
        \centering
        \includegraphics[height=2.2in]{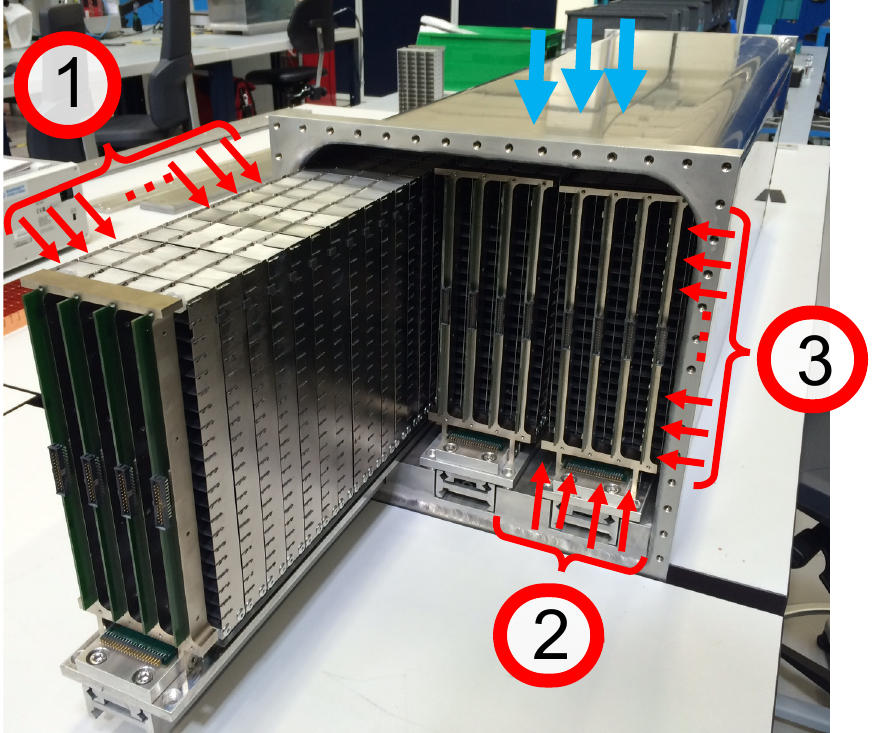}
        \caption{}
        \label{fig:fig6b}
    \end{subfigure}
    \caption{Pictures depicting the internal structure of the Multi-Grid detector. In (a), the basic building block of the detector, a grid, is shown. An example of a normal (blue) and a radial (red) blade is presented in the figure. The coating thicknesses of the different blades are shown within the blue and red brackets. The 5~mm Mirrobor sheet is seen at the back. In (b), it is shown how the grids are stacked in three rows and inserted into the vacuum tight gas vessel. Point (1) shows the 40 stacked grids, point (2) shows the 4 wire rows, and point (3) shows the 20 wire layers. Incident neutrons are presented as blue arrows in both pictures.}
    \label{fig:fig6}
\end{figure}

The Multi-Grid detector has a modular design, see for example \cite{Magdalena2012} for a rigorous description of the detector technology. It consists of a series of identical building blocks called $\textit{grids}$, as shown in figure~\ref{fig:fig6a}. In the prototype used here, each grid contains 21 layers of $^{10}$B$_4$C-coated aluminum substrates, 0.5 mm thick, which are called \textit{blades}. The chemical composition and the $^{10}$B-abundance in the $^{10}$B$_{4}$C-coatings were analyzed with Time-of-Flight Elastic Recoil Detection Analysis (ToF-ERDA), performed at the Tandem Laboratory~\cite{tandemlab}, Uppsala University. The analysis shows $\sim$79 at.\% (atomic percent) of $^{10}$B and $\sim$2 at.\% of impurity elements (mainly O) in the films, which is consistent with the values reported in previous publications \cite{Carina_2012, Carina_2015}. 

The blades are placed in a sequence and oriented orthogonal to the incident neutrons. The first blade is single-side coated and the remaining twenty double-side coated. These are the \textit{normal blades} (blue in figure~\ref{fig:fig6a}). Connecting the normal blades are five support blades, called the \textit{radial blades} (red in figure~\ref{fig:fig6a}). The naming convention refers to how the blades are oriented related to the incident neutrons. Together, these blades form a grid of cells. The normal blades have three different coating thicknesses, 1~$\mu m$, 1.25~$\mu m$ and 2~$\mu m$, where the blades with the thinner coating are at the front and the ones with thicker coating at the back. If the radial blades are coated, they are double-side coated with a 1.25 $\mu m$ coating thickness. At the back of the grid, a 5~mm thick Mirrobor sheet is placed. The purpose is to absorb the remainder of the incident neutrons, thus reducing back scattering.

The grids are stacked in three separate columns, see (1) in figure~\ref{fig:fig6b} for an example of a column, where each column contains 40 grids. Between the cells in the grids, anode wires are stretched, one wire per stack of cells, see (2) and (3) in figure~\ref{fig:fig6b}. These wires are then connected to a high voltage, while the cathode grids are put to ground potential. By placing the wires and grids in an Ar-CO$_2$ (80:20 volume ratio) filled vessel, a MWPC is established. The grids, as well as the wires, are electrically insulated from each other. This results in a position resolution defined by the cell size, where each cell is 22.5~$\times$~22.5~$\times$~10~mm$^3$. This is defined as a $\textit{voxel}$, and there are a total of 9600 voxels in the current prototype (40 height $\times$ 4$\cdot$3 width $\times$ 20 depth). The corresponding active detector surface area is approximately 0.24~$m^2$.

The induced charge released from the boron-10 neutron capture reaction is collected by the wires and grids. Each neutron event is then assigned a 3D-position based on coincidences in time between signals from wires and grids. Each event is also assigned a time-stamp, corresponding to the time-of-flight from the source chopper to the location of that event within the Multi-Grid detector. Note that each wire is adjacent to at least two $^{10}$B$_4$C-coated aluminum substrates, four if the radial blades are coated, and that incident neutrons can be converted in either one. The information of which coating the conversion took place is not recorded.

For the measurements, two separate prototypes of the Multi-Grid detector were used. These prototypes were originally designed and constructed as demonstrators for measurements at the thermal spectrometer SEQUOIA \cite{Granroth_2010} in August 2018, at the Spallation Neutron Source (SNS) \cite{sequoia_paper}. The prototypes are therefore optimized for thermal neutron energies. The two prototypes are identical in all aspects, except for the grid specifications, which are summarized in table \ref{table:table1}. The grids in Detector 1 is made with radiopure aluminum from Praxair \cite{praxair}, with less than 1 ppb (parts per billion) radioactive impurities such as Th and U, while the grids in Detector 2 is made with commercially available Al5754. The impact of radioactive impurities in the Al5754 alloy, such as Th and U, is alpha emissions which raises the time-independent background level in the detector \cite{MGback}. Furthermore, the impact of parasitic reactions due to neutron absorption in the remaining elements in the alloy, most prominently manganese, is a photon yield (cm$^{-3}$s$^{-1}$) 1 order of magnitude below that of aluminum \cite{Eszter2019thesis}. These effects, however, do not impact the current measurements, as the extra flat background can be subtracted and the photon events removed using a software cut.

For these measurements, the parameter of interest is the $^{10}$B$_4$C-coating on the radial blades, see figure~\ref{fig:fig6a}. This coating is only present in Detector 2.

\begin{table}[h!]
\centering
\caption{Summary of grid specifications relevant for internal scattering for the two Multi-Grid detectors used during the measurements.}
\label{table:table1}
\begin{tabular}{|l|c|c|l}
\cline{1-3}
\textbf{}                   & \multicolumn{1}{l|}{\textbf{Detector 1}} & \multicolumn{1}{l|}{\textbf{Detector 2}} &  \\ \cline{1-3}
\textbf{Radiopure aluminum} & Yes                                      & No                                       &  \\ \cline{1-3}
\textbf{$^{10}$B$_4$C-coated radial blades, 1.25 $\boldsymbol{\mu}$m}     & No                                       & Yes                                      &  \\ \cline{1-3}
\end{tabular}
\end{table}

The purpose of coating the radial blades is to reduce the effect of internally scattered neutrons. By adding the extra coating, scattered neutrons can be absorbed more quickly. This reduces the average extra flight distance by scattered neutrons, which in turn improves the accuracy of the energy reconstruction. In addition to this, the overall neutron detection efficiency for incident divergent neutrons, i.e. incident neutrons traveling at a path crossing the radial blades, is increased. This is due to the additional converter material introduced. As a consequence, neutrons are statistically absorbed closer to the detector entrance, reducing the travel path through aluminum within the detector. This lowers the number of interaction opportunities in aluminum, reducing the overall amount of internally scattered neutrons. This further improves the accuracy of the energy reconstruction.

The read-out electronics is the mesytec VMMR-8/16 \cite{mesytec}, which can handle 2048 channels simultaneously. In this setup, 360 channels are used, corresponding to the 80~$\cdot$~3~=~240 wires and 40~$\cdot$~3~=~120 grids. The data is transported using optical fibres. Each recorded event contains channel id, related to the wire or grid which collected the charge, as well as the time-of-flight (46~bits) and collected charge (12~bits). All events which occur within a pre-defined coincidence window are received together. A majority of the events consist of one wire and a few grids. By combining coincidences between wires and grids, neutron events are reconstructed with a $(x, y, z)$-hit location, time-of-flight and collected charge. If more than one grid fired within coincidence, the grid with the most collected charge is used for the position reconstruction.

\FloatBarrier

\section{Method and analysis}

This study concerns three properties of the Multi-Grid detector: neutron detection efficiency, internal neutron scattering, and time- and energy resolution. For the efficiency and resolution analysis, data from Multi-Grid Detector 1 are used, while for the scattering analysis data from both Detector 1 and 2 are needed. The method and analysis procedure is described below, starting with a reduction of the raw data by an event selection procedure.

\subsection{Event selection}

The Multi-Grid detector is designed for cold- to epithermal neutron detection. However, as the detector also has a certain low, but not negligible, gamma sensitivity \cite{Khaplanov_2013}, it is necessary to remove gamma events before proceeding with the analysis. This is done by studying the pulse height spectra (PHS), which shows the distribution of charge collected from events in the detector.

\begin{figure}[h!]
    \centering
    \begin{subfigure}{0.5\textwidth}
        \centering
        \includegraphics[height=2.3in]{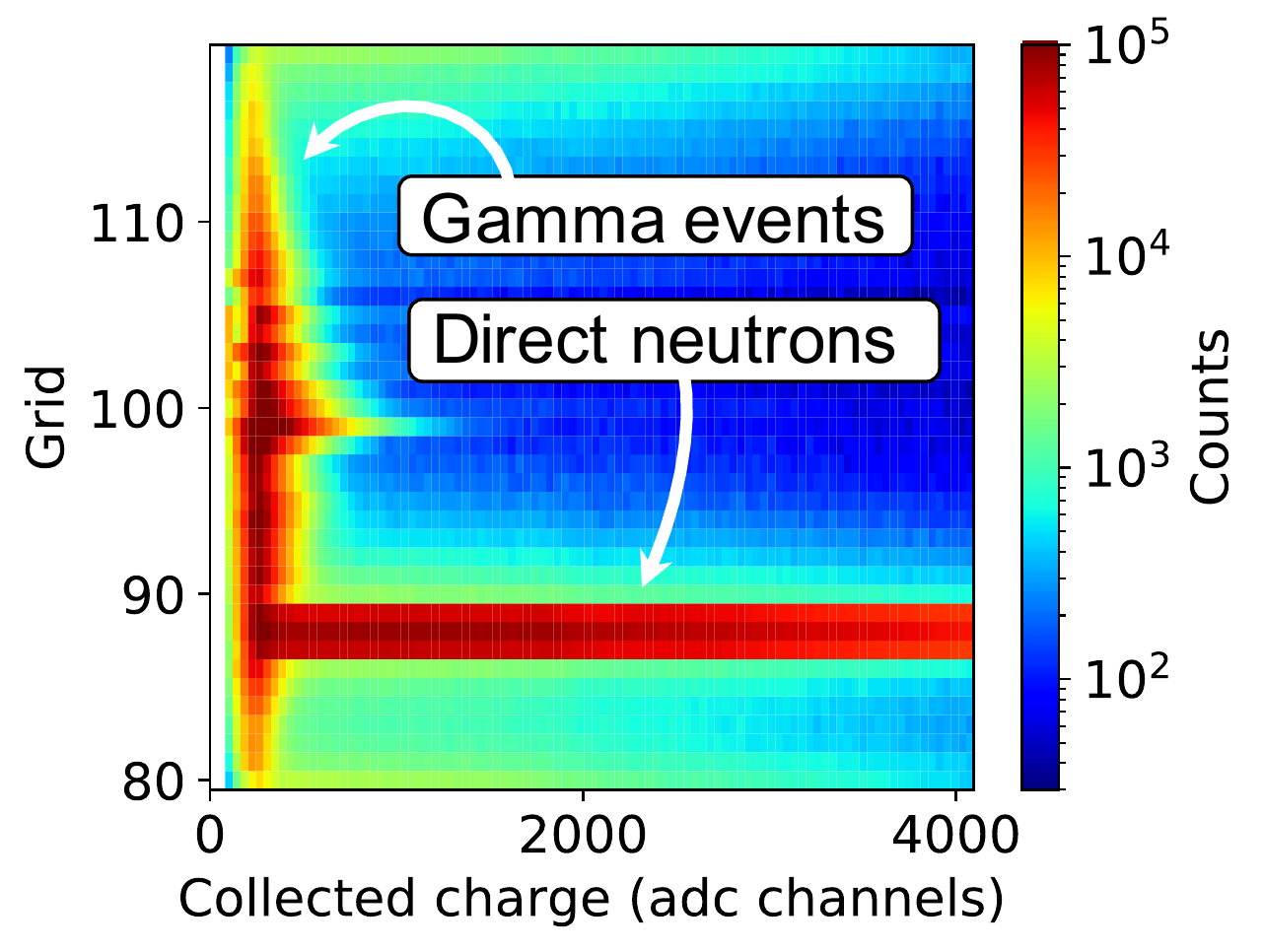}
        \caption{}
        \label{fig:grid_vs_adc_small}
    \end{subfigure}%
    \begin{subfigure}{0.5\textwidth}
        \centering
        \includegraphics[height=1.6in]{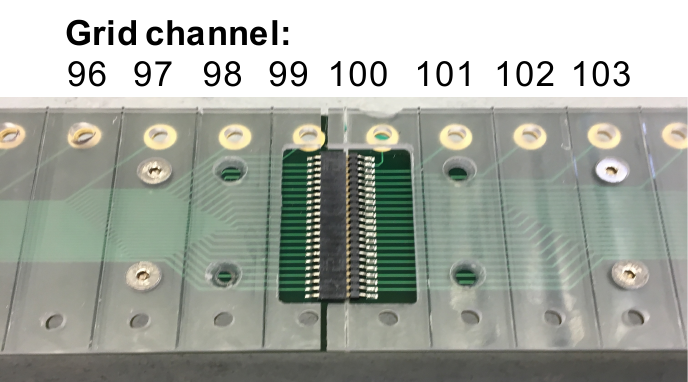}
        \caption{}
        \label{fig:pcb_connector}
    \end{subfigure}
    \caption{Event selection on data from the Multi-Grid detector. In (a), the PHS (x-axis) are shown for all individual grids (y-axis). The numbering refers to the electronic channels related to the grids, where Grid 80 is at the bottom and Grid 119 at the top. In the figure, data from Detector 1 is used as an example. In (b), the exposed PCB connectors are shown. The center grid, Grid 99, is placed directly above the connectors.}
    \label{fig:temp}
\end{figure}

Gamma events have a clear signature in PHS. This is presented in figure~\ref{fig:grid_vs_adc_small}, where the PHS (x-axis) is plotted for each grid (y-axis). The y-axis is expressed in electronic channels, where Grid 80 is at the bottom and Grid 119 at the top (channel 0 to 79 are reserved for wires). The x-axis is expressed in 12 bit analog-to-digital converter (ADC) channels, ranging from 0 to 4095. The gamma events are concentrated in a distribution at the low ADC values, 100 to 500 ADC channels, while neutrons span the full range. Notably, neutrons from the direct beam are seen hitting only three grids at the lower part of the detector, as shown by the three horizontal red stripes.

There is an elevated noise level in the middle grid, Grid 99, where the events cover a larger range in the PHS. This middle grid sits directly on top of a junction between the two unshielded PCB connectors connected halfway up the detector, see figure~\ref{fig:pcb_connector}. The increased rate could, therefore, be due to crosstalk between the exposed connectors and the grid. Alternatively, a local physical offset would put the wire closer to the center grid, which would increase the gain there in a discrete step. As the nominal pitch between grids and wires is only 5 mm, a deviation on the fraction of mm would be sufficient to cause a noticeable effect.

For the purpose of the following analysis, the exact distribution of the gamma spectrum is not critical. Here, a constant ADC software threshold at 600 ADC channels is applied, such that the gamma events are rejected. However, for the scattering analysis, a more aggressive ADC cut is required to clean the data further. This is because the rate of scattered events is, in the middle region of the detector at parts of the lambda spectrum, comparable to the elevated noise rate. Fortunately, the noise events are concentrated at the low ADC values, while the scattered neutron events span the full ADC range. Therefore, a cut at 1200 ADC channel is used instead for the scattering analysis, which removes the remainder of the noise present in the middle grid, while keeping a majority of the events from scattered neutrons.

In addition to the ADC-cuts, a multiplicity cut is performed. The multiplicity of an event denotes the number of wires and grids which fired within the coincidence window. As the conversion products have a finite range, there is a limit on how large multiplicities proper neutron events can have. Therefore, a study of the multiplicity can be used as an additional filter to remove gamma events and false coincidences. For this study, events with wire multiplicity 1 and grid multiplicity 1 to 5 is kept. The maximum grid multiplicity is larger than that for wires, as the released charge spread more easily to adjacent grids than adjacent wires.

\subsection{Efficiency}

Neutron detection efficiency is defined as the fraction of incident neutrons which are detected, i.e. the number of detected neutrons divided by the total number of incident neutrons. This is described in equation (\ref{eq:efficiency_def}),

\begin{eqnarray}
\label{eq:efficiency_def}
\epsilon = \frac{S^{detected}}{S^{incident}},
\end{eqnarray}
where $\epsilon$ is the neutron detection efficiency, $S^{detected}$ is the sum of detected neutrons and $S^{incident}$ is the sum of incident neutrons. For the Multi-Grid detector, the quantity $S^{detected}_{MG}$ is accessed from the number of counts in the data, while $S^{incident}_{MG} $ is not measured directly. Instead, it is estimated from a separate measurement with the helium-3 tube. Using these two measurements, one with the Multi-Grid detector and one with the helium-3 tube, the neutron detection efficiency for the Multi-Grid detector is derived.

The incident neutrons on the Multi-Grid detector $S^{incident}_{MG}$, i.e. integrated absolute flux, is calculated according to equation (\ref{eq:incident_flux_est}),

\begin{eqnarray}
\label{eq:incident_flux_est} 
S^{incident}_{MG} = S^{detected}_{He-3} \cdot \frac{1}{\epsilon_{He-3}} \cdot \frac{BM_{MG}}{BM_{He-3}},
\end{eqnarray}
where $S^{detected}_{He-3}$ is the detected neutrons in the helium-3 tube and $\epsilon_{He-3}$ is the calculated efficiency of the helium-3 tube. $BM_{MG}$ is the integrated beam monitors counts over the full wavelength spectrum from the measurement with the Multi-Grid detector, and $BM_{He-3}$ the corresponding number from the separate run with the helium-3 tube. The inverse of the helium-3 efficiency is used to scale $S^{detected}_{He-3}$ to approximate the total number of neutrons incident on the detector, $S^{incident}_{He-3}$. The fraction of the beam monitor counts is then used to scale this value to make it comparable to the integrated flux on the Multi-Grid detector. That is, it accounts for the difference in flux during the two separate measurements. 

Inserting equation (\ref{eq:incident_flux_est}) into (\ref{eq:efficiency_def}), a formula for deriving the Multi-Grid efficiency as a function of the incident neutron wavelength $\lambda_n$ is thus written according to equation (\ref{eq:efficiency_calc}),

\begin{equation}
\label{eq:efficiency_calc} 
\epsilon_{MG}(\lambda_n) = S^{detected}_{MG}(\lambda_n)\cdot \left ( S^{detected}_{He-3}(\lambda_n) \cdot \frac{1}{\epsilon_{He-3}(\lambda_n)} \cdot \frac{BM_{MG}}{BM_{He-3}} \right )^{-1},
\end{equation}
where $\epsilon_{MG}(\lambda_n)$ is the derived Multi-Grid efficiency.

\begin{figure}[h!]
    \centering
    \begin{subfigure}{0.5\textwidth}
        \centering
        \includegraphics[height=2.3in]{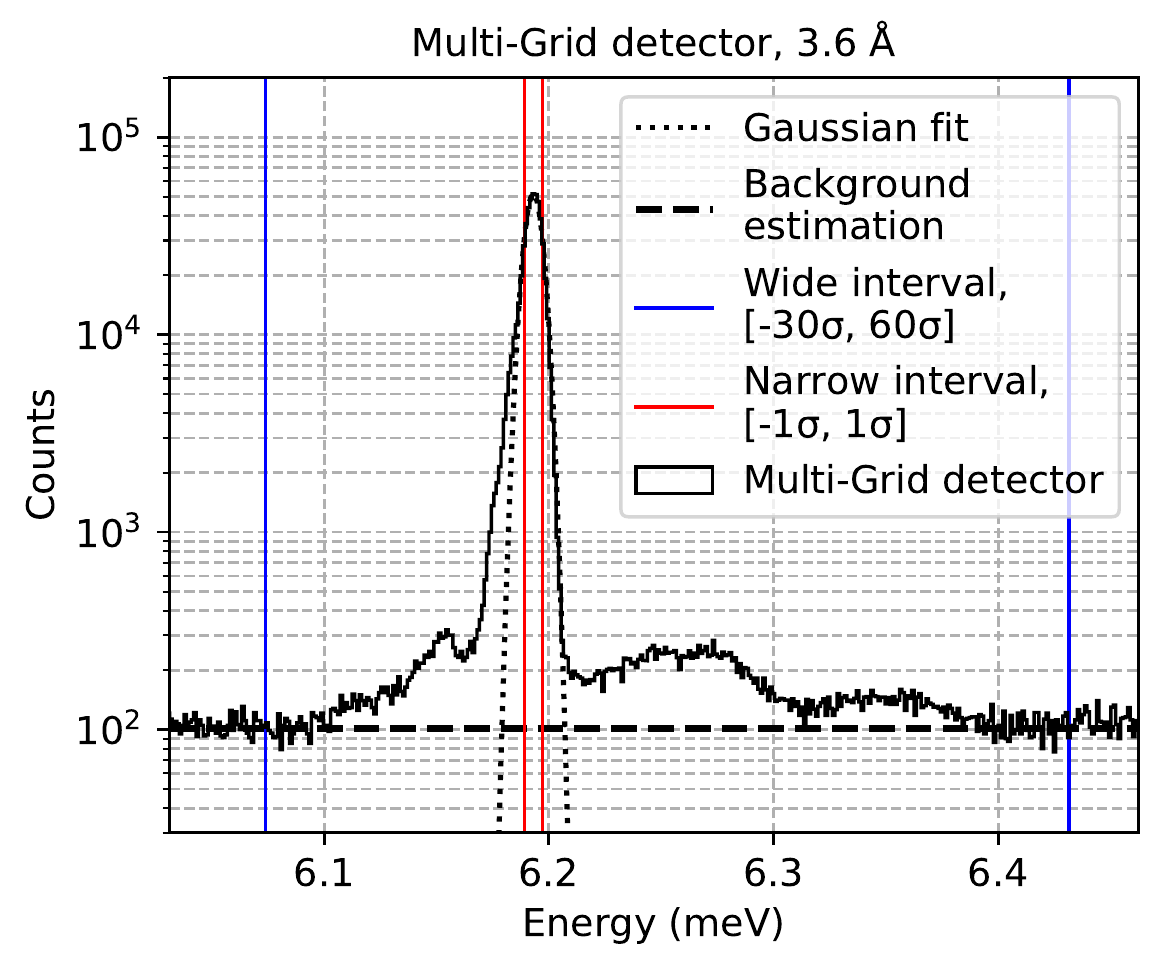}
        \caption{}
        \label{fig:fig10a}
    \end{subfigure}%
    \begin{subfigure}{0.5\textwidth}
        \centering
        \includegraphics[height=2.3in]{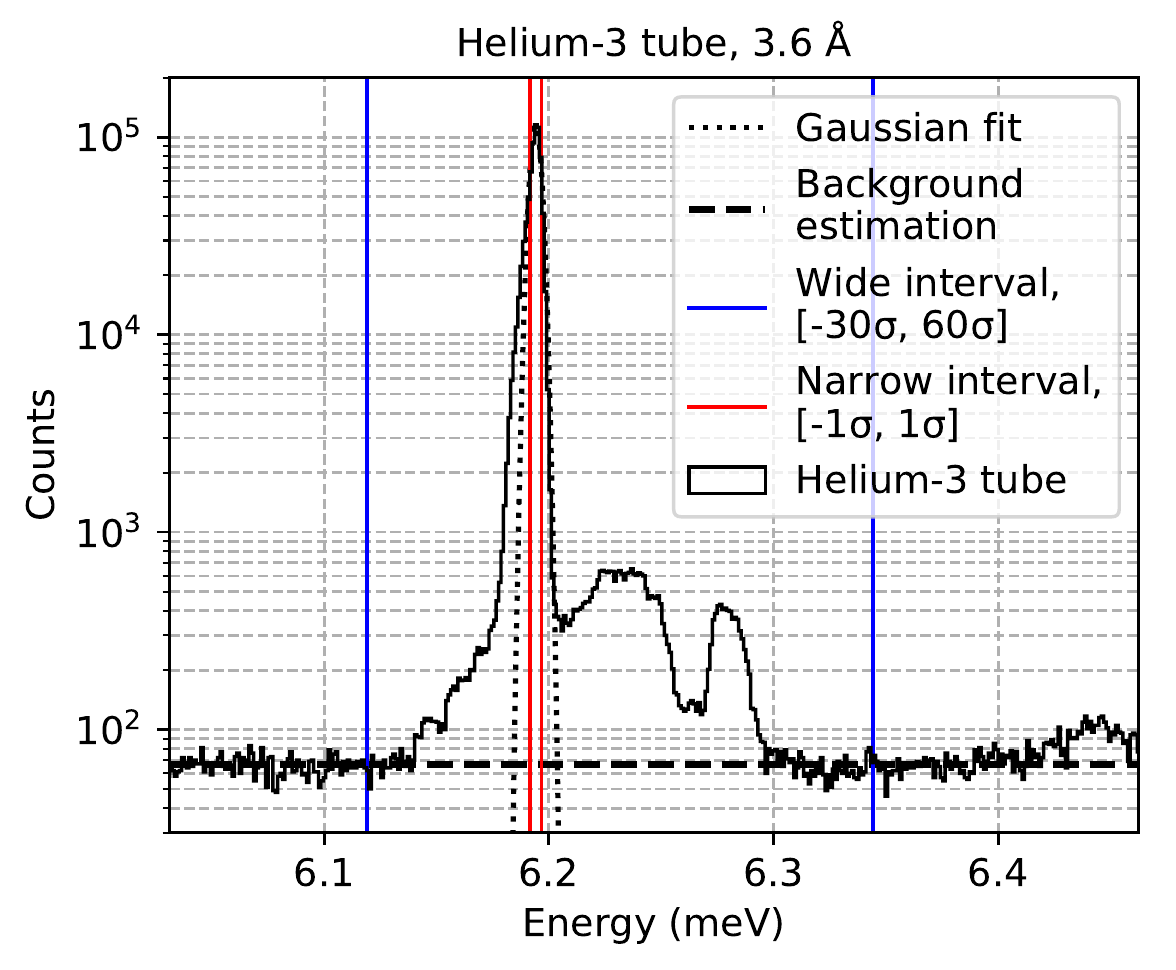}
        \caption{}
        \label{fig:fig10b}
    \end{subfigure}
    \begin{subfigure}{\textwidth}
        \centering
        \includegraphics[height=2.2in]{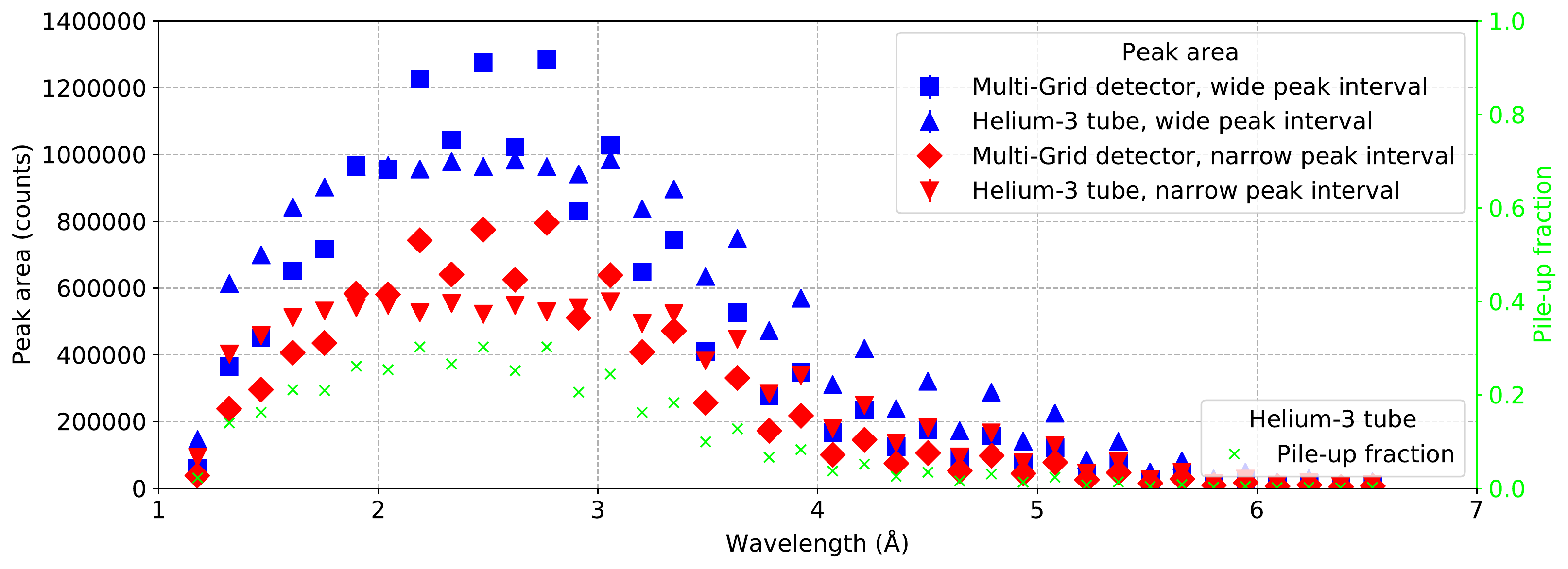}
        \caption{}
        \label{fig:fig10c}
    \end{subfigure}
    \caption{Peak areas used for efficiency calculation. In (a) and (b), an example of peaks at 3.6 \AA{} are presented for the Multi-Grid detector and helium-3 tube. In the plots, the Gaussian fit (dotted black), the background estimation (dashed black) and the narrow- (red) and wide (blue) integration intervals are also presented. In (c), integrated peak area is shown as a function of neutron wavelength. The peak area for the Multi-Grid detector (squares and diamonds) and helium-3 tube (triangles) is presented for the narrow peak interval (red) and wide peak interval (blue). The fraction of pile-up events encountered in the helium-3 read-out system (green) is plotted on the separate right-hand y-axis.}
    \label{fig:integration}
\end{figure}

$S^{detected}_{MG}(\lambda_n)$ and $S^{detected}_{He-3}(\lambda_n)$ are calculated by integrating counts above background for each peak from the Fermi-chopper in the wavelength spectra. This is presented in figure~\ref{fig:integration}.

As the experimental setup does not produce well defined peaks, recall the parasitic peaks shown in figure~\ref{fig:fig3c}, there is an uncertainty on the correct way to integrate the peak area. This uncertainty is accounted for by introducing two intervals: one very narrow, $\pm \sigma$, encompassing only the peak center, and one wide, encompassing the full peak, as well as any parasitic peaks. By using the narrow interval, effects from the parasitic peaks can be rejected. However, this also means that any differences in peak shape between the Multi-Grid detector and helium-3 tube are not accounted for. By using the two intervals, an estimate of this systematic uncertainty on the efficiency is obtained.

The peak areas are presented in figure~\ref{fig:fig10c} as a function of wavelength, where the areas from the narrow- (red) and wide (blue) are presented together. The highest fluence rate in the spectra, at around 2.5 \AA{}, is approximately 2~$\cdot$~$10^{6}$~s$^{-1}$~cm$^{-2}$. The fluence rate in the peak was estimated using data from the beam monitor, located a few meters upstream from the helium-3 tube.

$\epsilon_{He-3}(\lambda_n)$ is calculated to account for the absorption in the gas and  the scattering in the stainless steel tube. If a neutron is absorbed, it is considered detected. The neutron absorption probability is determined using the helium-3 absorption cross-sections, gas pressure, and neutron travel distance in the tube. The travel distance depends on where the neutron hits the tube along the tube diameter, i.e. the neutron has a longer travel distance if it hits the center of the tube than at the edge, and the calculation accounts for this by using the average perpendicular tube depth as neutron travel distance. The acquired absorption probability is then scaled by the fraction of the incident neutron flux lost due to scattering in the steel tube. This fraction is estimated as the scattering probability scaled by the heuristic factor 0.5, as not all neutrons which scatter are lost. Corrections for factors such as wall effect and dead zones in the gas are not taken into account in this approximation.

\begin{figure}[h!]
    \centering
    \begin{subfigure}{0.5\textwidth}
        \centering
        \includegraphics[height=2.1in]{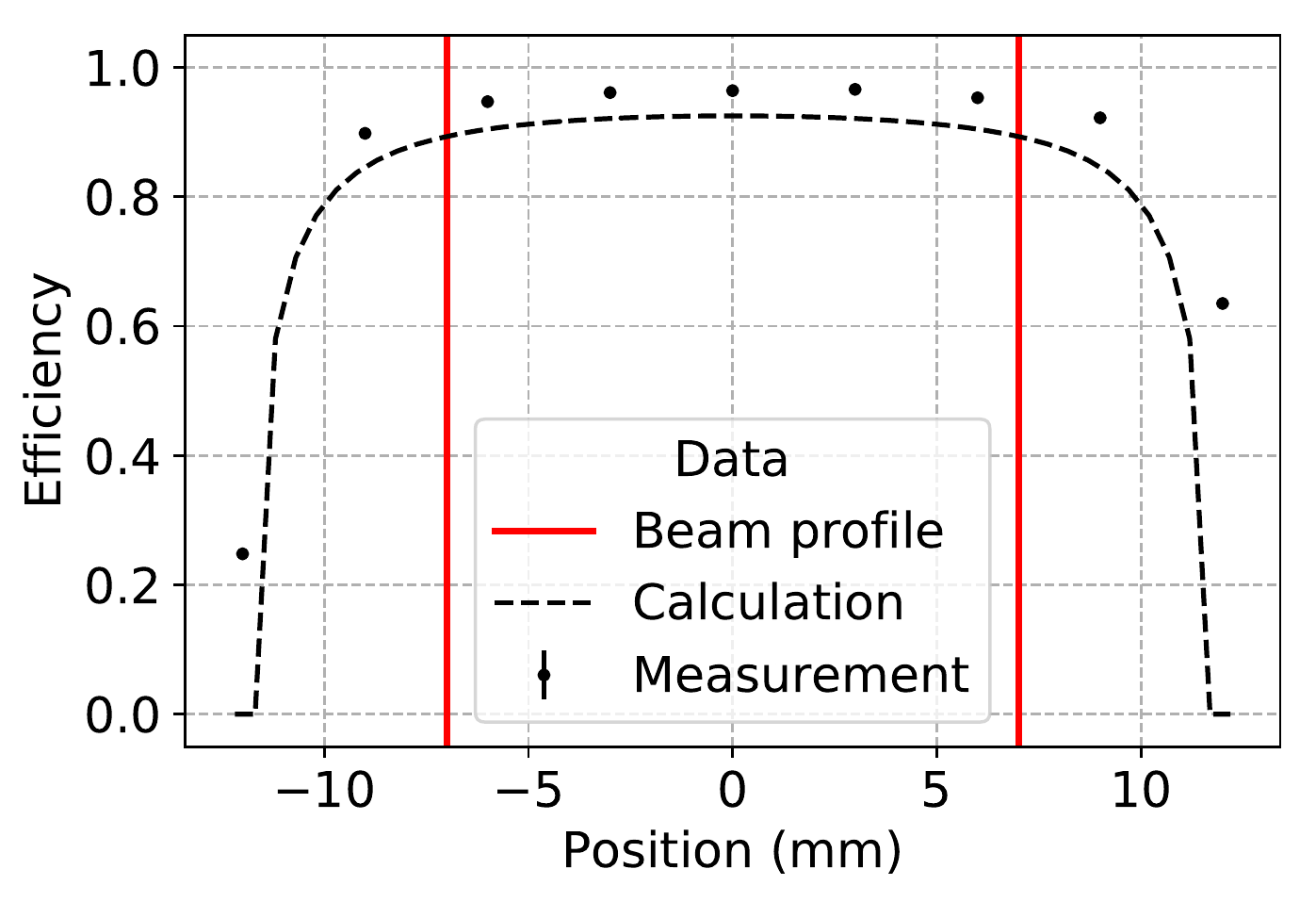}
        \caption{}
        \label{fig:fig5a}
    \end{subfigure}%
    \begin{subfigure}{0.5\textwidth}
        \centering
        \includegraphics[height=2.1in]{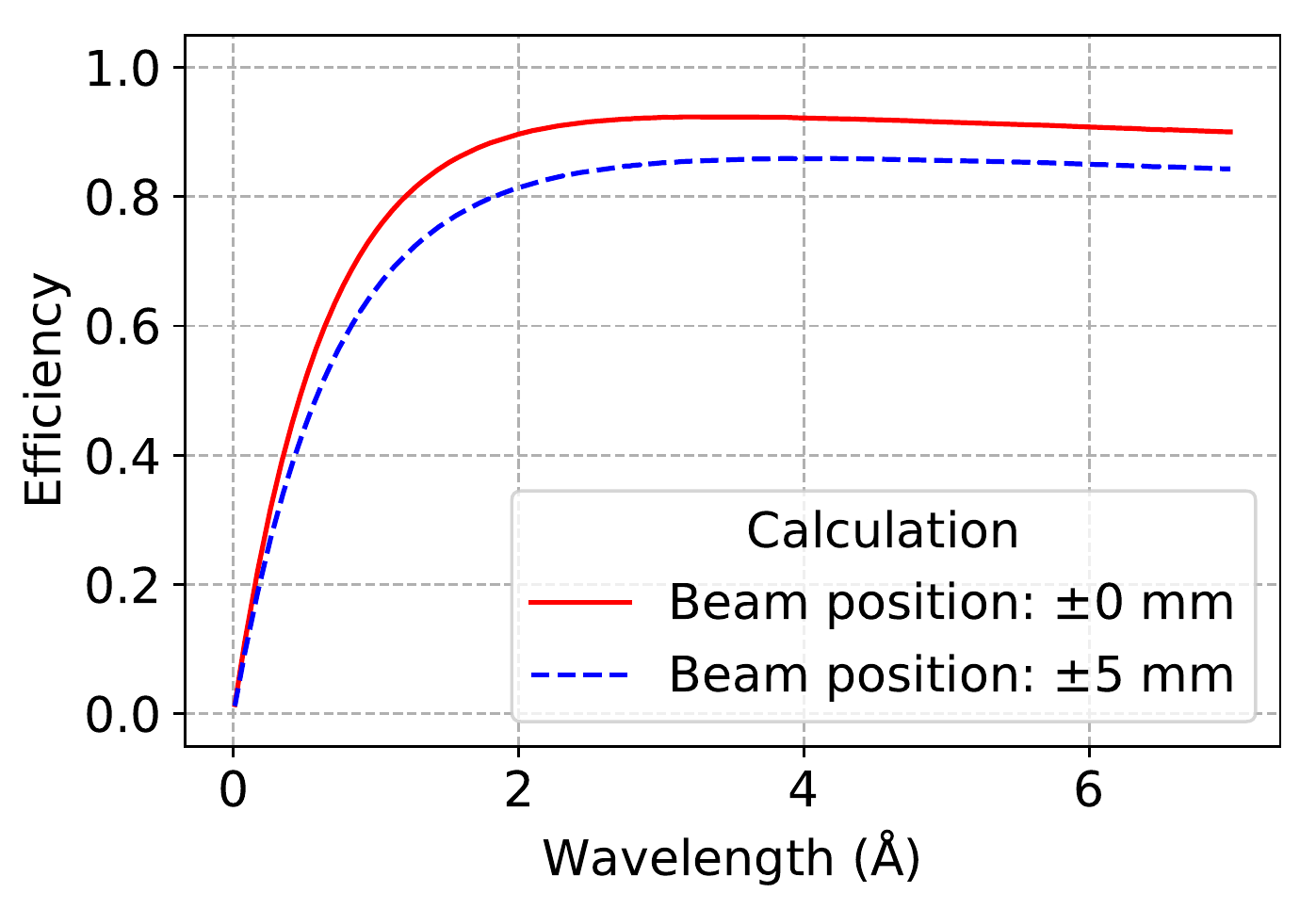}
        \caption{}
        \label{fig:fig5b}
    \end{subfigure}
    \caption{Neutron detection efficiency of the helium-3 tube. In (a), the efficiency at 2.5 \AA{} is shown as a function of displacement from the tube center. The neutron beam width (red lines) is shown together with the calculation (black line) and data (black points), which was gathered during a previous measurement at ILL. In (b), the calculated efficiency is presented as a function of wavelength. Each value is averaged over the tube width hit by the beam, showing it centered (red) and at $\pm$5 mm offset (blue).}
    \label{fig:fig5}
\end{figure}

In figure~\ref{fig:fig5a}, the calculated efficiency is plotted as a function of position along the tube diameter, together with measurement data from a previous measurement \cite{Pfeiffer_2016, frapi_unpublished}. The data agree to within a few percentages, confirming the validity of the calculation. The offset is accounted for in later calculations as a systematic uncertainty. In figure~\ref{fig:fig5b}, the average efficiency across the tube diameter is shown as a function of neutron wavelength. Two curves are presented, one with the beam centered on the tube and one with a 5~mm offset. This is to account for the systematic uncertainty of beam alignment, which is taken into consideration in later calculations.

\begin{figure}[h!]
    \centering
    \begin{subfigure}{0.5\textwidth}
        \centering
        \includegraphics[height=2.1in]{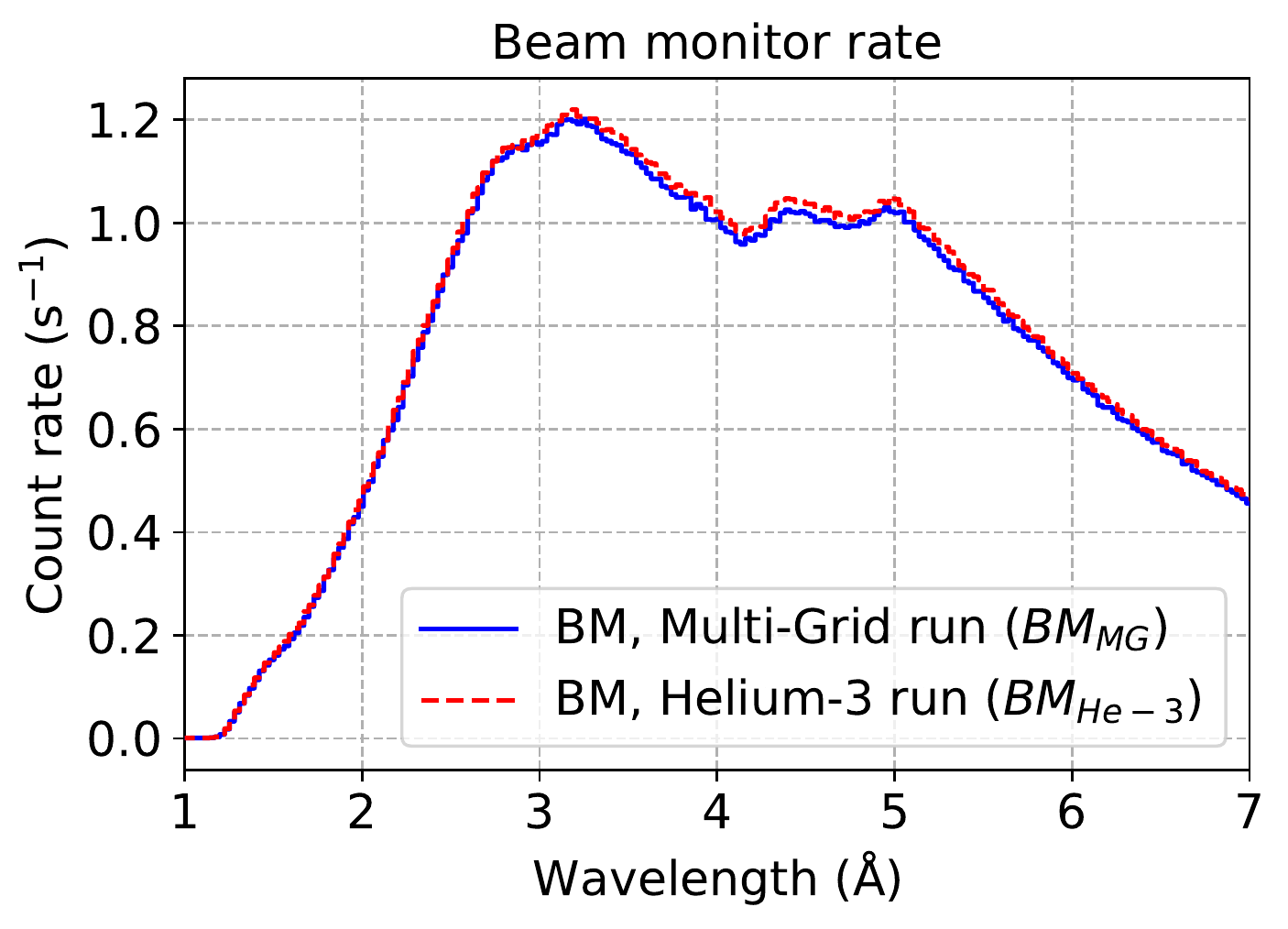}
        \caption{}
        \label{fig:bm_counts}
    \end{subfigure}%
    \begin{subfigure}{0.5\textwidth}
        \centering
        \includegraphics[height=2.1in]{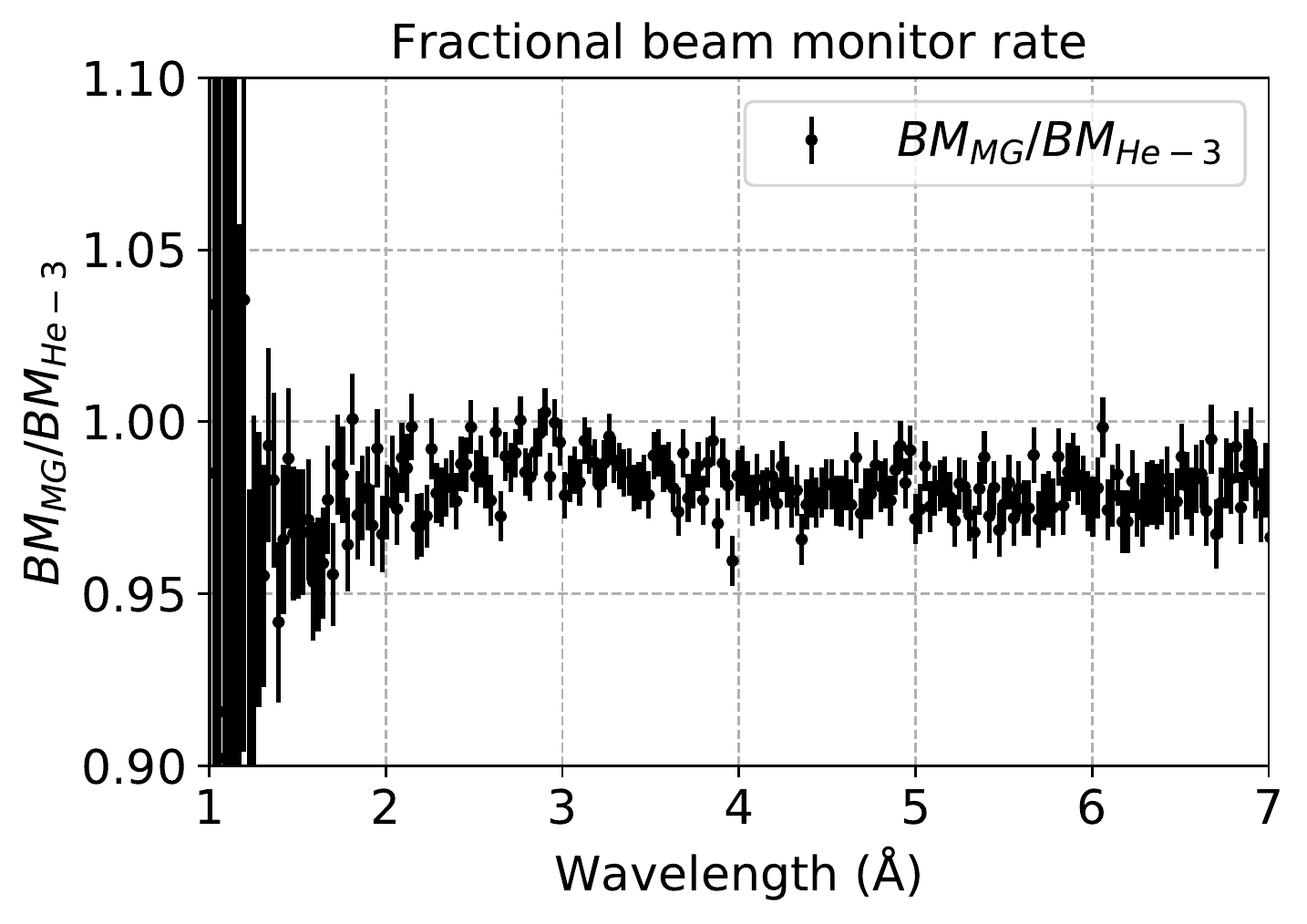}
        \caption{}
        \label{fig:bm_fraction}
    \end{subfigure}
    \caption{Beam monitor data corresponding to the two different measurements, one with the Multi-Grid detector (blue) and one with the helium-3 tube (red). In (a), the histogrammed beam monitor counts, normalized by measurement duration, are presented as a function of neutron wavelength. In (b), the fractional rates, Multi-Grid detector over helium-3 tube, are shown.}
    \label{fig:bm}
\end{figure}

In figure~\ref{fig:bm}, the beam monitor data are plotted, corresponding to the separate measurements with the Multi-Grid detector and helium-3 tube, $BM_{MG}$ and $BM_{He-3}$, respectively. In figure~\ref{fig:bm_counts}, the rates are shown, while in figure~\ref{fig:bm_fraction} the fractional rate, $BM_{MG}/BM_{He-3}$, is presented. It is seen that incident flux during the two measurements were similar, as the fraction is only a few percentages below 1. It is also seen that the fraction is wavelength independent within a few percentages. The high uncertainty around 1~\AA{} is due to poor statistics.

\begin{figure}[h!]
\centering
\includegraphics[width=1\linewidth]{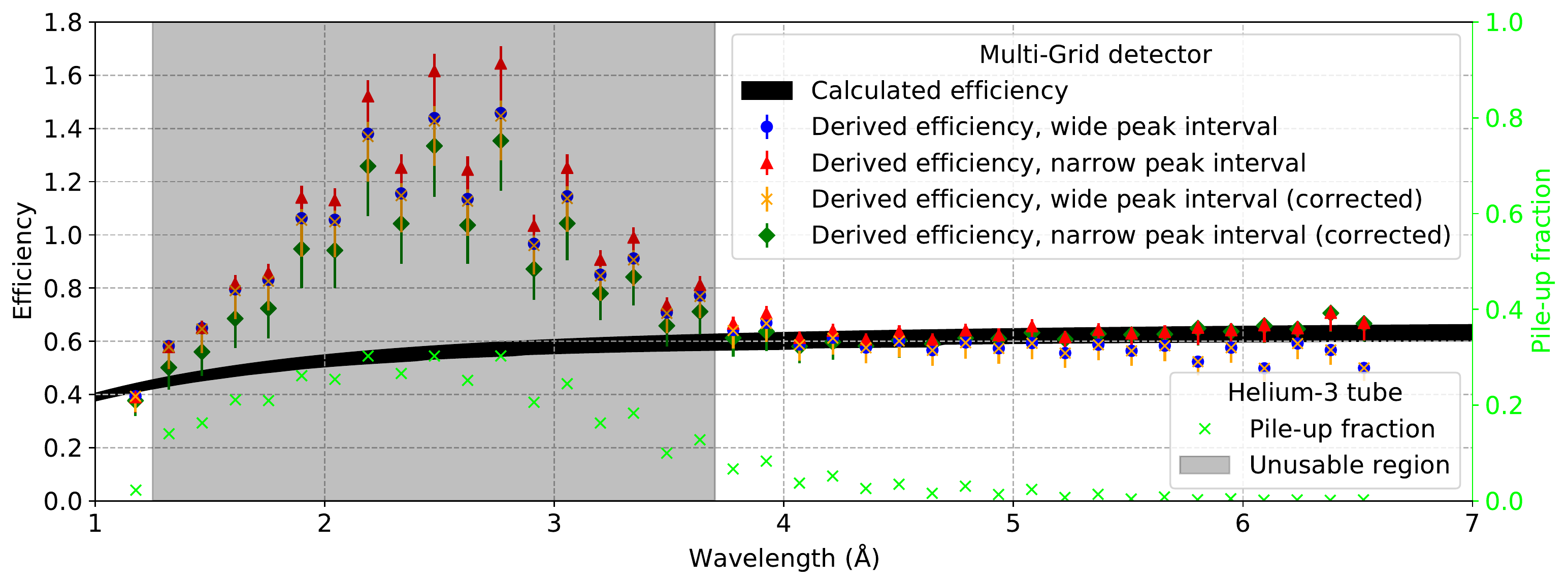}
\caption{Derived Multi-Grid efficiency (blue circles and red triangles) plotted against the calculated efficiency (black band). The width of the black band shows the uncertainty range of the efficiency calculation. In plot, the derived efficiency where the helium-3 data has been corrected to account for the saturation in the helium-3 tube (orange crosses and green diamonds) are also shown. The fraction of pile-up events encountered in the helium-3 read-out system (green crosses) is plotted on the separate right-hand y-axis. The grey region covers the unusable portion of the data, caused by saturation in the helium-3 detector system.}
\label{fig:derived_mg_eff}
\end{figure}

The derived Multi-Grid efficiency is presented in figure~\ref{fig:derived_mg_eff}. In the plot, the derived efficiency for the two integration intervals (red and blue) are compared to the analytical prediction (black). The error bars show systematic uncertainties. The calculation for the analytical prediction includes the attenuation of neutrons in aluminum, the absorption probability in the $^{10}$B$_4$C-coating, as well as the escape probability of the conversion products from the coating. The incident neutrons are assumed to hit the front of the detector at a perpendicular angle. All calculations are based on derivations presented in \cite{frapi_thesis, frapi_2012, basaez2018decal}. The width of the curve indicates the systematic uncertainty on the calculation, based on the uncertainty on the input parameters, as well as flux loss due to scattering in the aluminum window.

Referring to figure~\ref{fig:derived_mg_eff}, it is seen that for long wavelengths, 4~to~6~\AA{}, the derived- and calculated efficiency agree well within the uncertainties. For wavelengths shorter than 4~\AA{} (highlighted grey area), a strong deviation from the calculation is seen. This is due to the saturation of the helium-3 detector system. The highest fluence rate in the spectra is high enough, $>$~$10^{6}$~s$^{-1}$, to cause multiple hits within the 1~$\mu$s shaping time of the read-out system. This is seen from the fraction of pile-up events in the tube (green), which follows the observed deviation. An additional reason is that the data transfer speed limit per channel, $\approx$~10$^6$~events~s$^{-1}$, is similar to the peak flux, which might cause loss of data. This could have been prevented by using a lower incident flux. However, as the scattering analysis requires the best possible signal-to-noise ratio, a high neutron flux was essential.

An attempt was made to account for the saturation in the helium-3 tube using information of the incident flux and shaping time. This is presented for the wide peak interval (orange crosses) and the narrow peak interval (green diamonds). The procedure allows for a few more data points to be within uncertainties. However, the correction is not strong enough for the majority of data points within the saturated region. This could be because the incident flux is sufficiently intense to introduce additional effects in the helium-3 tube, such as space charge effects, which further decreases the efficiency. As these additional effects are not accounted for in the correction, a deviation is still seen.

The saturation process is also the cause of the strong staggering effect between 2 and 3~\AA{}. In figure~\ref{fig:fig10c}, it is seen how the Multi-Grid detector (squares and diamonds) follow the intensity from Fermi-chopper, i.e. every other pulse is more intense, even where the flux is at the highest level. This is not the case for the helium-3 tube (triangles), which is flat between 2 and 3~\AA{}. As the oscillations between adjacent data points in the two detectors no longer match in this region, the fraction of the peak areas is not flat, as it is above 4~\AA{}. The saturation effect seen in the helium-3 detector is absent in the Multi-Grid detector.

\FloatBarrier

\subsection{Internal neutron scattering}

The main source of internal neutron scattering in the Multi-Grid detector is caused by the presence of aluminum \cite{DIAN2018173}. The other elements in the neutron beam path, boron and carbon in $^{10}$B$_4$C, have a negligible effect. This is because $^{10}$B$_4$C, although having a 2-3 times higher scattering cross-section than aluminum in the measured energy range \cite{neutron_data_booklet}, have over two orders of magnitude thinner total thickness compared to aluminum (60 $\mu$m and 14.5 mm, respectively, for neutrons incident perpendicular on the detector surface).

\begin{figure}[h!]
    \centering
    \begin{subfigure}{0.5\textwidth}
        \centering
        \includegraphics[height=2.3in]{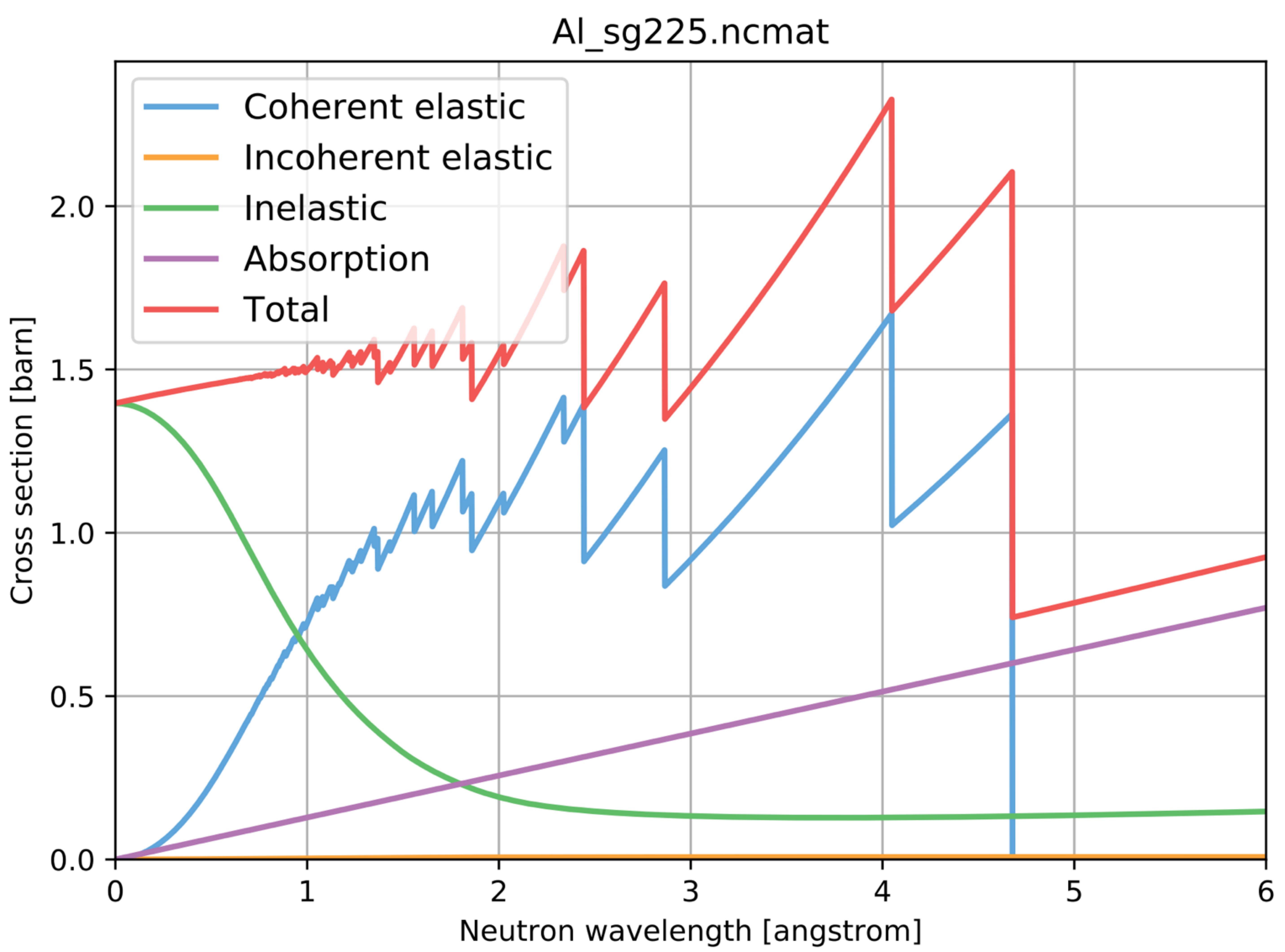}
        \caption{}
        \label{fig:scatter_cross_vs_lambda}
    \end{subfigure}%
    \begin{subfigure}{0.5\textwidth}
        \centering
        \includegraphics[height=2.0in]{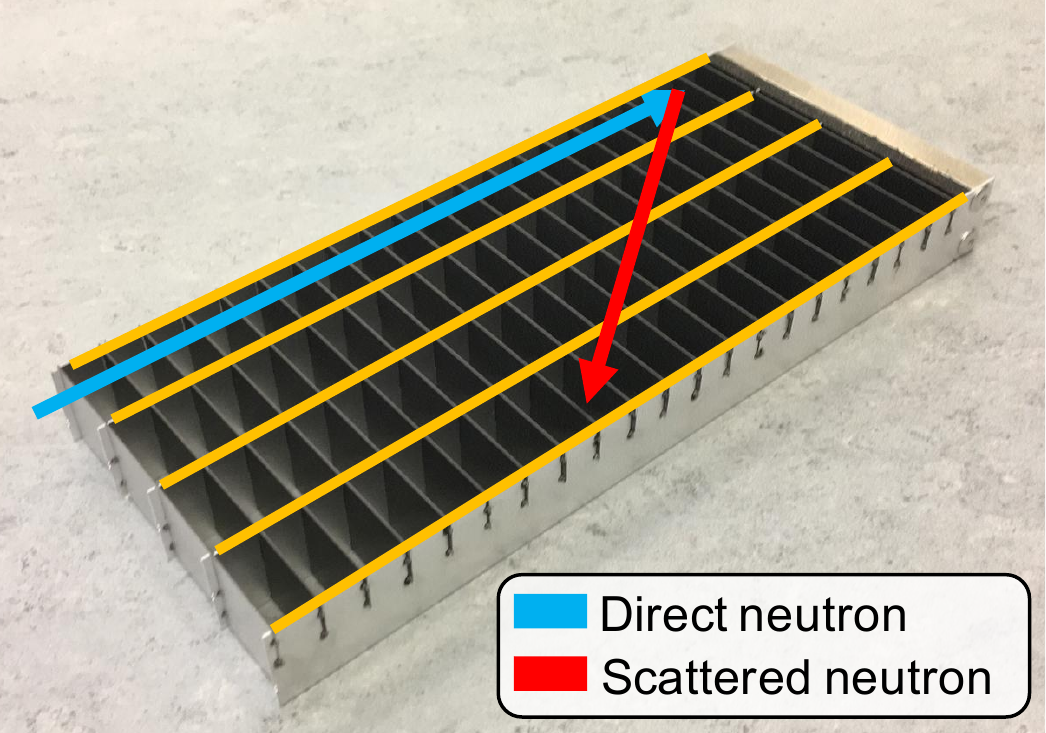}
        \caption{}
        \label{fig:grid_scattering}
    \end{subfigure}
    \caption{Neutron scattering of aluminum. In (a), cross-sections for different neutron interactions with aluminum is presented. These include elastic scattering (blue and orange), inelastic scattering (green), absorption (purple), as well as the total cross-section (red). The figure is generated using NCrystal \cite{ncrystal}. In (b), an illustration of internal scattering is shown, where an incident neutron (blue) is scattered (red). The radial blades are highlighted in orange, which can be either coated or non coated with $^{10}$B$_4$C.}
    \label{fig:al_scatter}
\end{figure}

Aluminum has a periodic crystal structure, like any other crystalline solid, which allows for coherent scattering of atoms within the same crystal lattice, as well as incoherent scattering from the individual atomic nuclei. The neutron interaction cross-sections in aluminum are presented in figure~\ref{fig:scatter_cross_vs_lambda}, generated using NCrystal \cite{ncrystal}. It is seen that  for wavelengths between 1 and approximately 4.7 \AA{}, coherent elastic scattering is dominant. The cut-off wavelength at 4.7 \AA{}, where the coherent elastic scattering drops to zero, indicates the maximum wavelength where diffraction occurs in the crystal structure, i.e. the maximum wavelength where the Bragg condition is fulfilled for the aluminum crystal lattice. For wavelengths longer than this, no coherent scattering occurs. An example of internal neutron scattering is presented in figure~\ref{fig:grid_scattering}.

The impact of the scattered neutrons shows as an added time-dependent background in the detector. This is caused by an incorrect energy reconstruction of the scattered neutrons. The neutron energy, $E_n$, is determined according to equation (\ref{eq:energy_1}),

\begin{eqnarray}
\label{eq:energy_1}
E_n  = \frac{m_n}{2} \cdot  \left ( \frac{d}{tof} \right ) ^2,
\end{eqnarray}
where $m_n$ is the neutron mass, $d$ is the source-to-detection distance, and $tof$ is the corresponding time-of-flight. However, as the distance $d$ is based on the detection voxel, this distance will be incorrect if the neutron is scattered before being detected. Moreover, as the neutron might acquire an additional flight time between scattering and detection, or a shortened flight time if the neutron gained energy through inelastic scattering in aluminum (green cross-sections in figure~\ref{fig:scatter_cross_vs_lambda}), this further distorts the energy reconstruction. Therefore, a scattered neutron has an energy reconstruction dependent on equation (\ref{eq:energy_2}),

\begin{eqnarray}
\label{eq:energy_2}
E_n' = \frac{m_n}{2} \cdot  \left ( \frac{d \pm \delta d}{tof \pm \delta T} \right ) ^2,
\end{eqnarray}
where $E_n'$ is the reconstructed energy for an internally scattered neutron, $\pm \delta T$ is the change in the flight time between scattering and detection, and $\pm \delta d$ is the change in assumed distance due to the incorrect voxel detection. Due to the large mass difference between aluminum nuclei and neutrons, most scattering can be considered elastic. Consequently, $\delta T$ will predominantly be positive, corresponding to the extra flight time between scattering and detection. Therefore, internally scattered neutrons will predominately be reconstructed with an energy $E_n' < E_n$. The exception to this is when a neutron gains energy through inelastic scattering and is scattered forward with a sufficiently small deviation, so that $\delta d$ is small. In this case, the neutron will be detected earlier than if it would not have been scattered, hence $\delta T<0$ and $E_n' > E_n$.

To minimize the impact of scattered neutrons on the energy line shape, the additional flight distance between scattering and detection should be kept as short as possible. This reduces $\delta d$ and $\delta T$ in equation (\ref{eq:energy_2}), closing in to the ideal case in equation (\ref{eq:energy_1}). To facilitate this, $^{10}$B$_4$C-coating on the radial blades in the grids can be introduced (figure~\ref{fig:grid_scattering}, orange lines).

\begin{figure}[h!]
    \centering
    \begin{subfigure}{0.5\textwidth}
        \centering
        \includegraphics[height=2.2in]{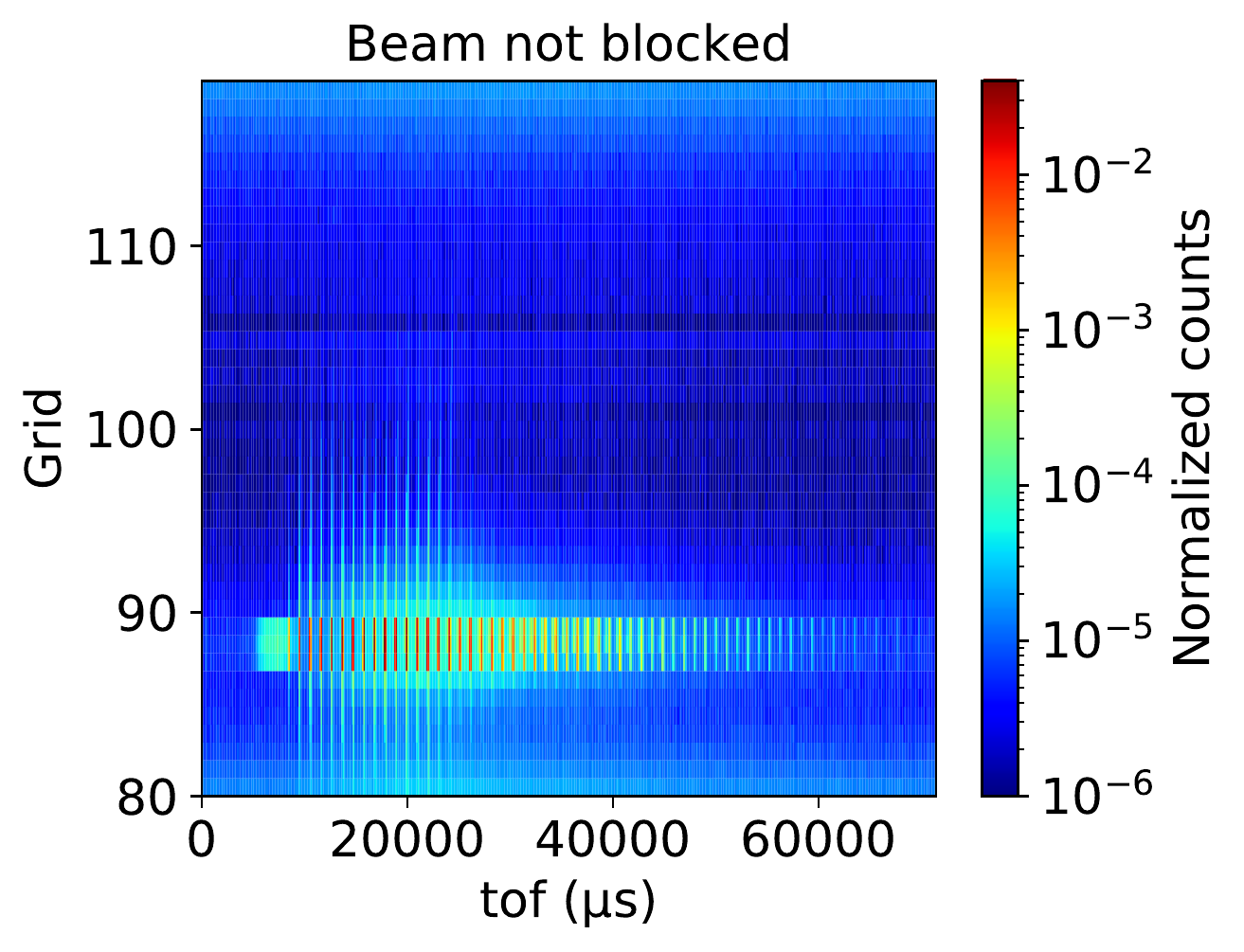}
        \caption{}
        \label{fig:grid_vs_tof_not_blocked}
    \end{subfigure}%
    \begin{subfigure}{0.5\textwidth}
        \centering
        \includegraphics[height=2.2in]{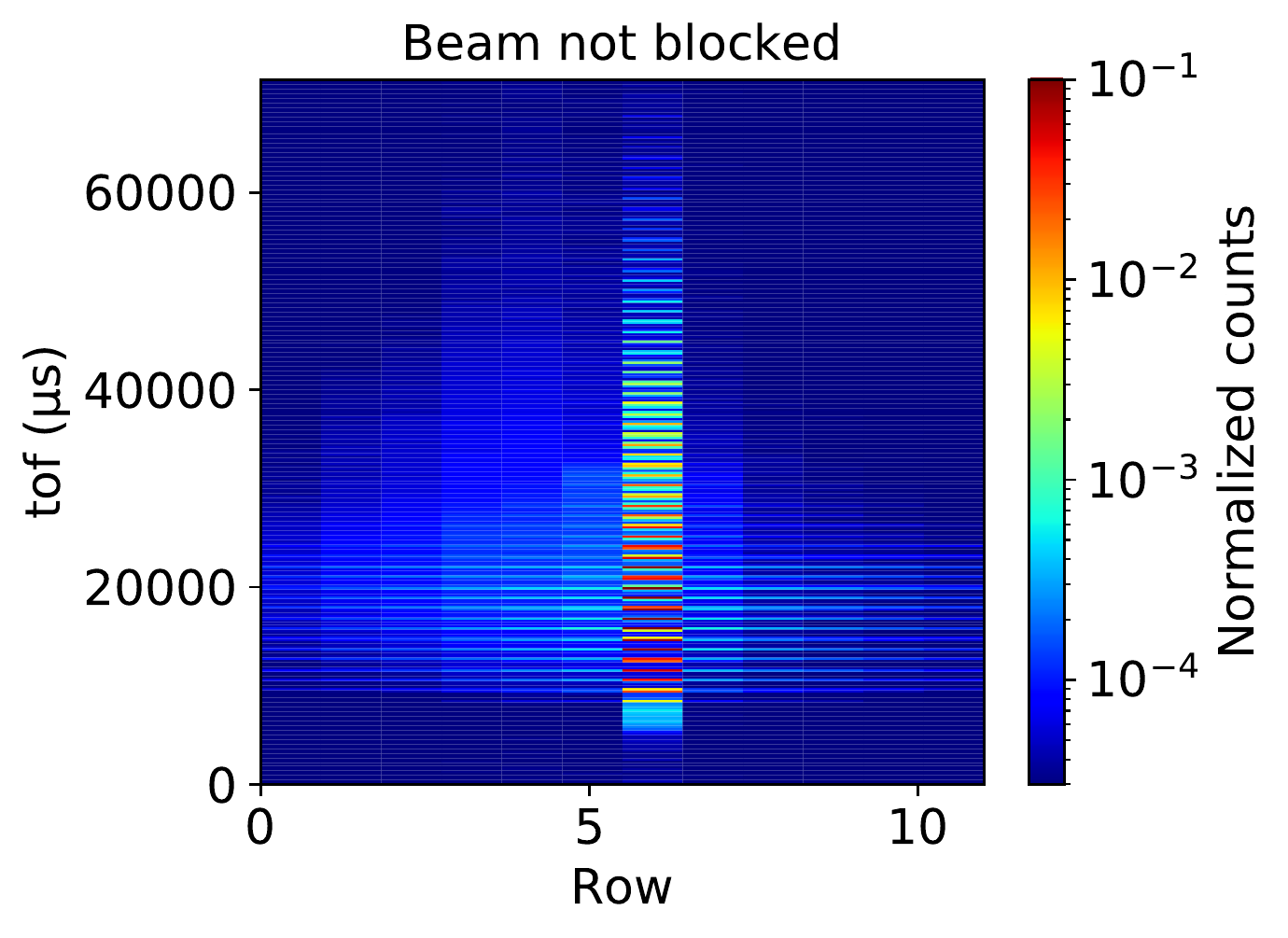}
        \caption{}
        \label{fig:row_vs_tof_not_blocked}
    \end{subfigure}
    \begin{subfigure}{0.5\textwidth}
        \centering
        \includegraphics[height=2.2in]{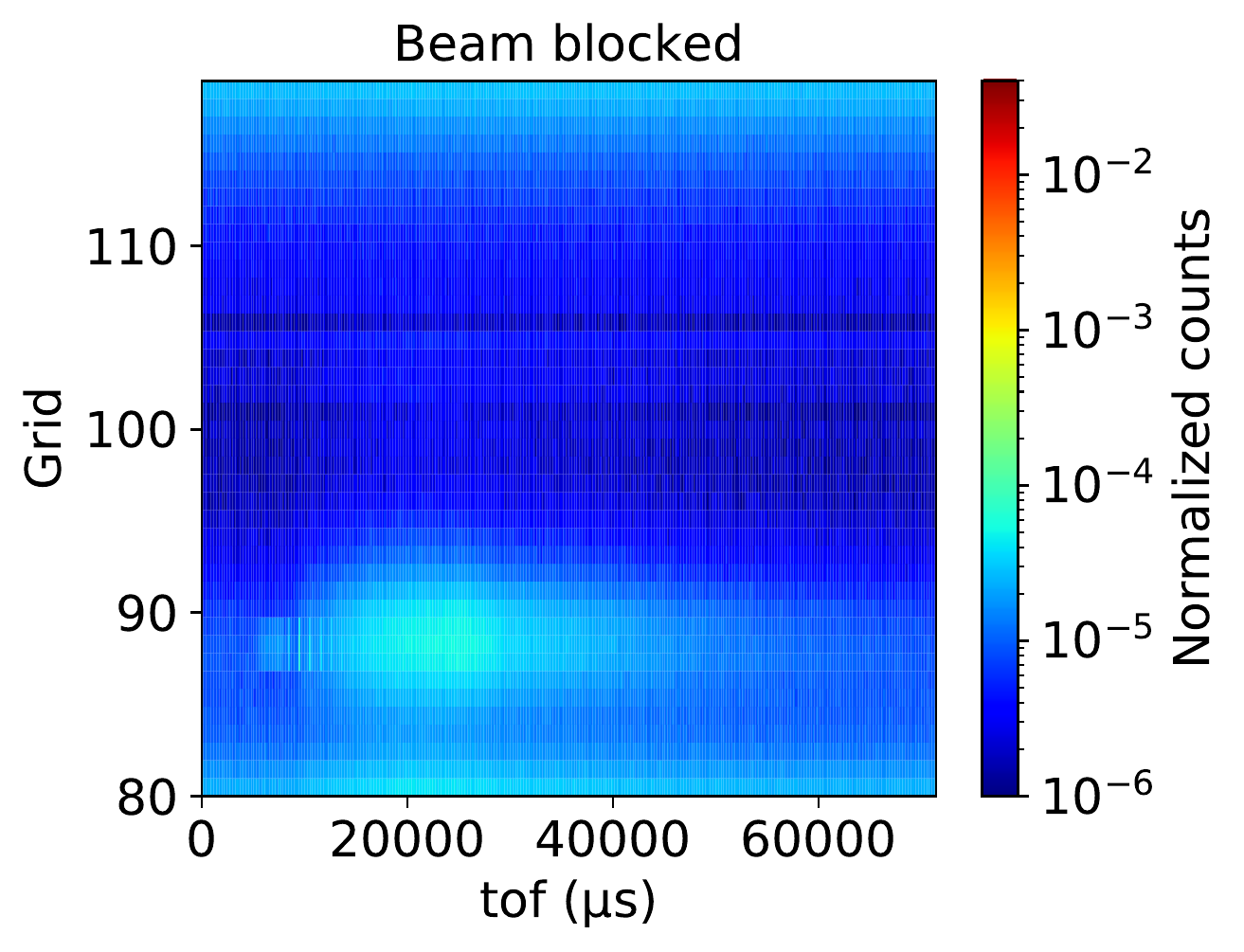}
        \caption{}
        \label{fig:grid_vs_tof_blocked}
    \end{subfigure}%
    \begin{subfigure}{0.5\textwidth}
        \centering
        \includegraphics[height=2.2in]{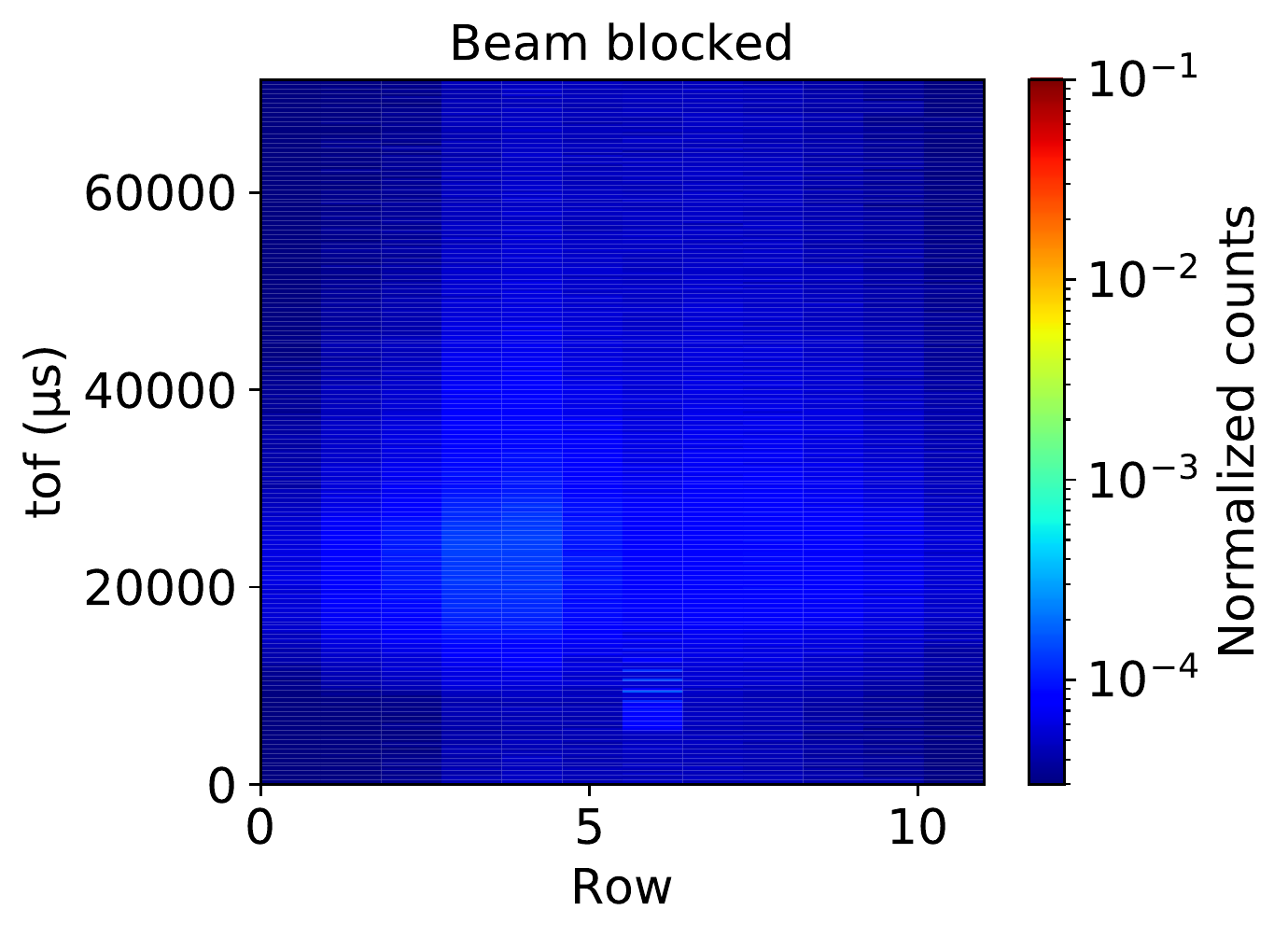}
        \caption{}
        \label{fig:row_vs_tof_blocked}
    \end{subfigure}
    \caption{Histograms comparing data from the Multi-Grid detector when the beam was not blocked, (a) and (b), and blocked, (c) and (d). The two left plots show counts in grids vs. time-of-flight, while the two right plots show counts in rows vs. time-of-flight. The counts have been normalized to the accumulated beam monitor counts from the corresponding run. Note that Detector 1 is used in this example, but  the same conclusions are drawn if Detector 2 is studied.}
    \label{fig:row_and_grid_vs_tof}
\end{figure}

To investigate the effect of the radial coating, data from Detector 1 (non-coated radial blades) and Detector 2 (coated radial blades) are compared. As the neutron beam is highly collimated, recorded events from scattered neutrons (figure~\ref{fig:grid_vs_tof_not_blocked} and \ref{fig:row_vs_tof_not_blocked}, long green stripes) can be separated from those from the direct beam. This is achieved by performing a geometrical cut, removing all events in the direct beam (grid 87-89, row 6) and keeping the scattered neutrons (everything outside grid 87-89, row 6). The volume outside the direct beam region is called the \textit{beam periphery}. Note that there are scattered neutrons in the beam region as well, and that these are rejected in this approach. This is not an issue, however, as an absolute measure of scattering is not intended, only a comparison between Detector 1 and Detector 2.

To verify that the events seen at the beam periphery are indeed internally scattered neutrons, and not from a beam halo or a similar effect, the background data is used. This is presented in figure~\ref{fig:grid_vs_tof_blocked} and \ref{fig:row_vs_tof_blocked}, which shows data recorded when the direct beam was blocked with the helium-3 tube. The helium-3 tube, 25 mm in diameter, is wide enough to stop the direct beam. However, it does not have a sufficient solid angle coverage to make a prominent blocking of a potential halo effect. That the long green lines in figure~\ref{fig:grid_vs_tof_not_blocked} and \ref{fig:row_vs_tof_not_blocked} is only seen when the helium-3 tube is removed from the direct beam, demonstrates that the effect must originate after the helium-3 tube. As the only component after the helium-3 tube is the Multi-Grid detector, this shows that the lines originate from scattering in the detector.

\begin{figure}[h!]
    \centering
    \begin{subfigure}{0.5\textwidth}
        \centering
        \includegraphics[height=2.5in]{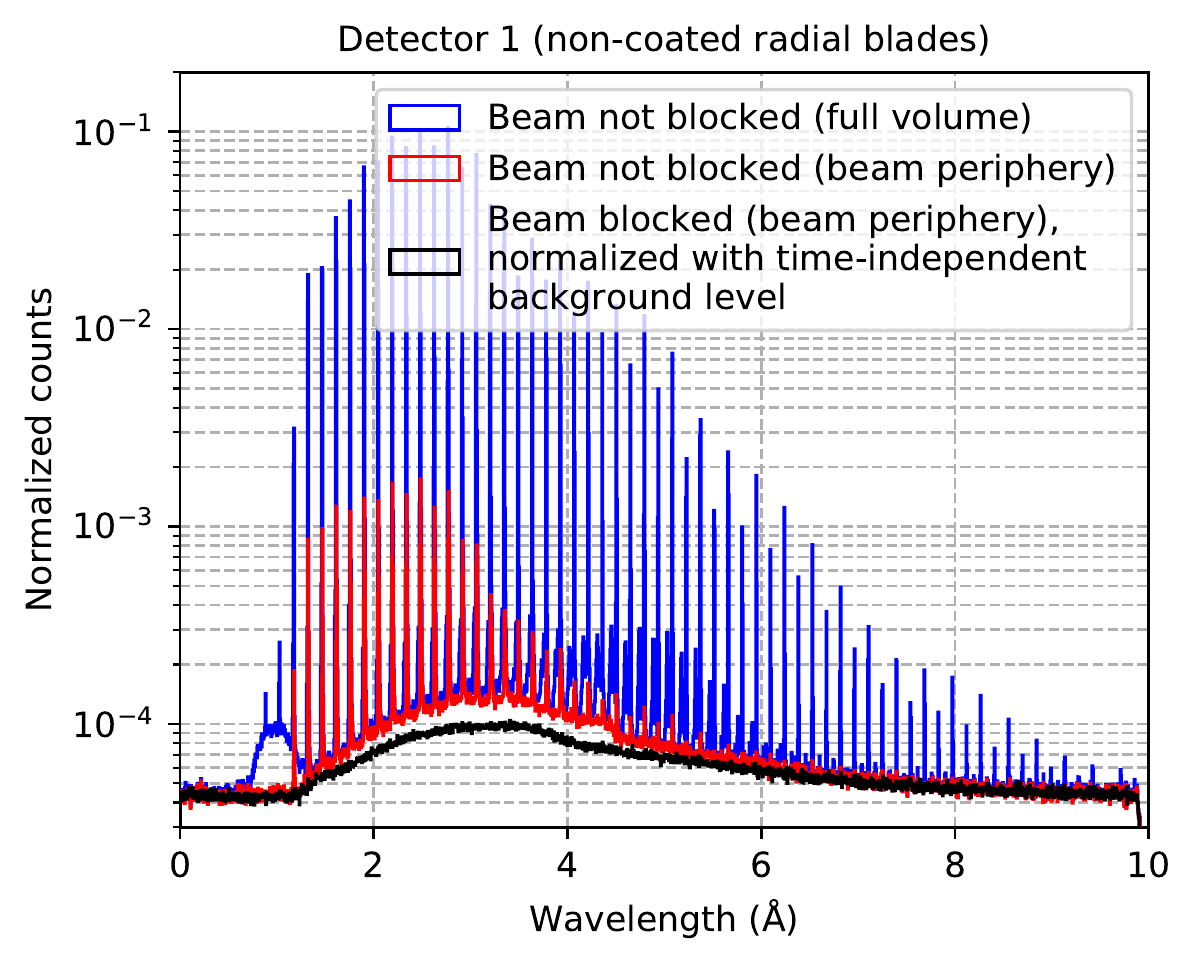}
        \caption{}
        \label{fig:nr_blocked_and_not_blocked}
    \end{subfigure}%
    \begin{subfigure}{0.5\textwidth}
        \centering
        \includegraphics[height=2.5in]{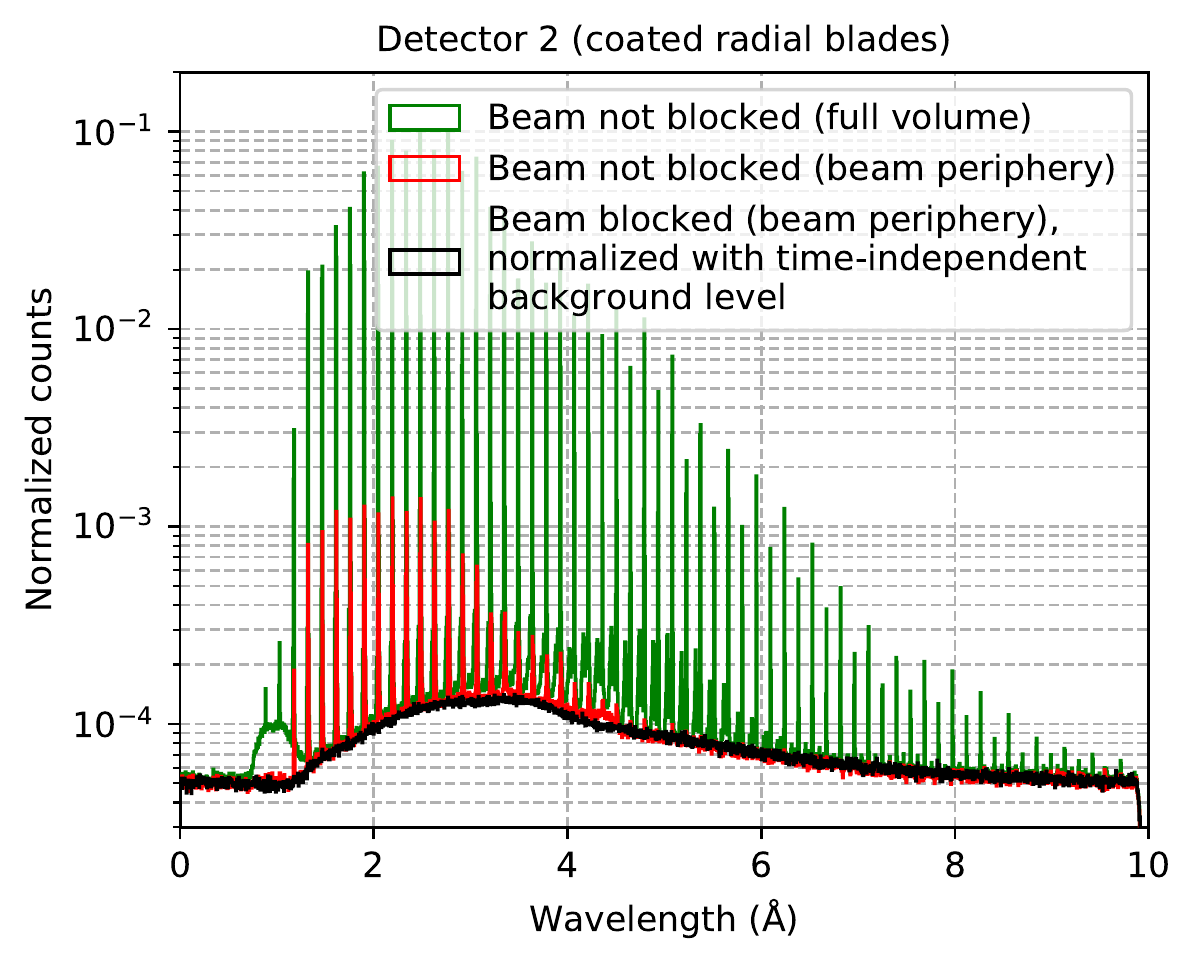}
        \caption{}
        \label{fig:r_blocked_and_not_blocked}
    \end{subfigure}
    \caption{Histograms over wavelength for the two Multi-Grid detectors. In (a), Detector 1 (non-coated radial blades) is shown, while in (b), Detector 2 (coated radial blades) is presented. Using data from when the beam was not blocked, two separate histograms are shown in each plot: data from the full volume (blue and green) and beam periphery (red). The background data (black) is also based on the beam periphery region. The plots are normalized by accumulated beam monitor counts from the individual runs. The background data has an additional normalization based on the time-independent background level during the separate measurements.}
    \label{fig:blocked_and_not_blocked}
\end{figure}

Studying data with two different cuts, full data versus no direct beam, the wavelength distribution is acquired for Detector 1 and 2. This is presented in figure~\ref{fig:blocked_and_not_blocked}. The blue and green histograms are from the full data, while the red histograms are from data where the direct beam is cut. Overlaying the curves allows for a visualization of the magnitude of scattered neutrons and how it depends on wavelength. The black histograms are from the corresponding background measurements, which has an additional normalization based on the time-independent background level in the facility. This rate is seen in the flat region between 0 and 0.5 \AA{}, and depends on the overall activity in the vicinity of V20, which cannot be account for by the beam monitor data alone. It is noted that the background data has no time correlation with the Fermi-chopper, i.e. an absence of sharp peaks.

\begin{figure}[h!]
    \centering
    \begin{subfigure}{0.5\textwidth}
        \centering
        \includegraphics[height=2.5in]{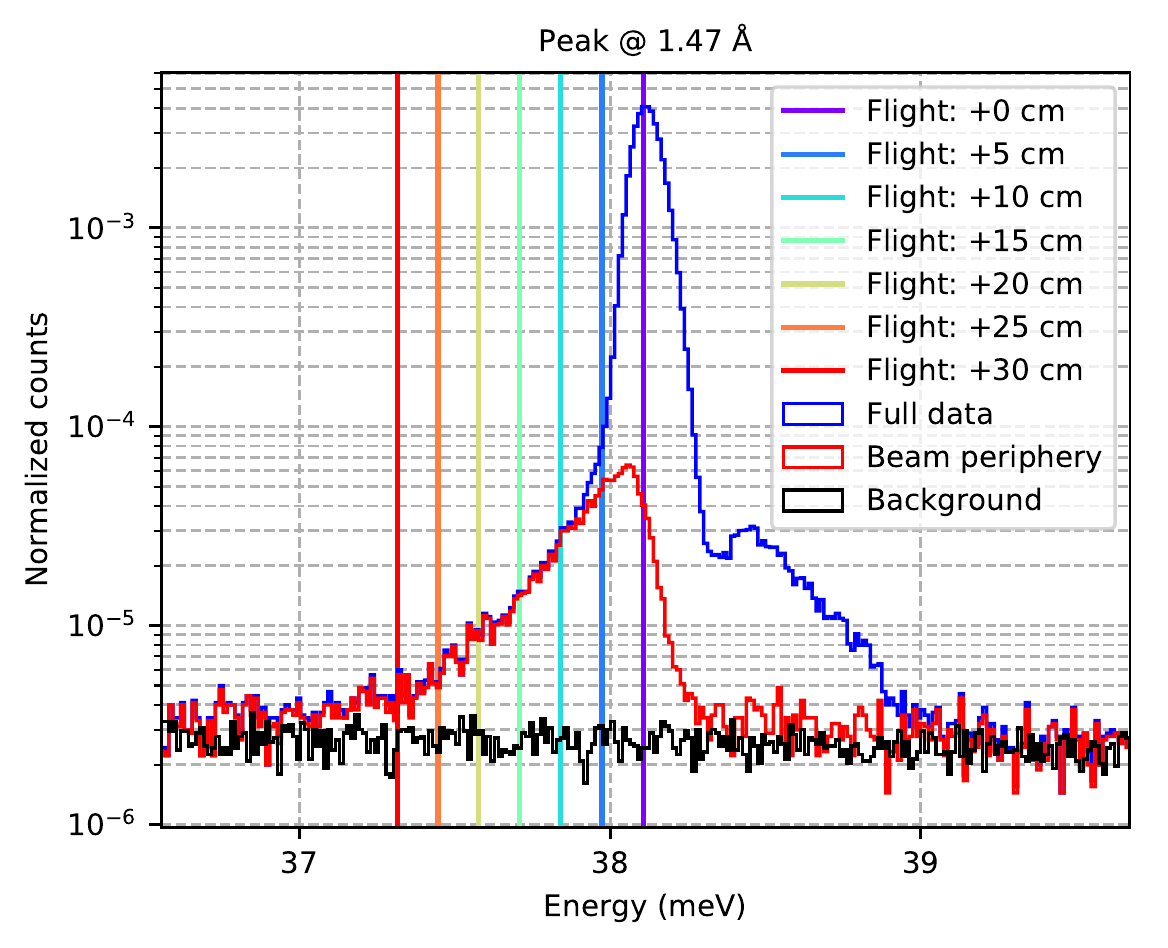}
        \caption{}
        \label{fig:mg_distance}
    \end{subfigure}%
    \begin{subfigure}{0.5\textwidth}
        \centering
        \includegraphics[height=2.5in]{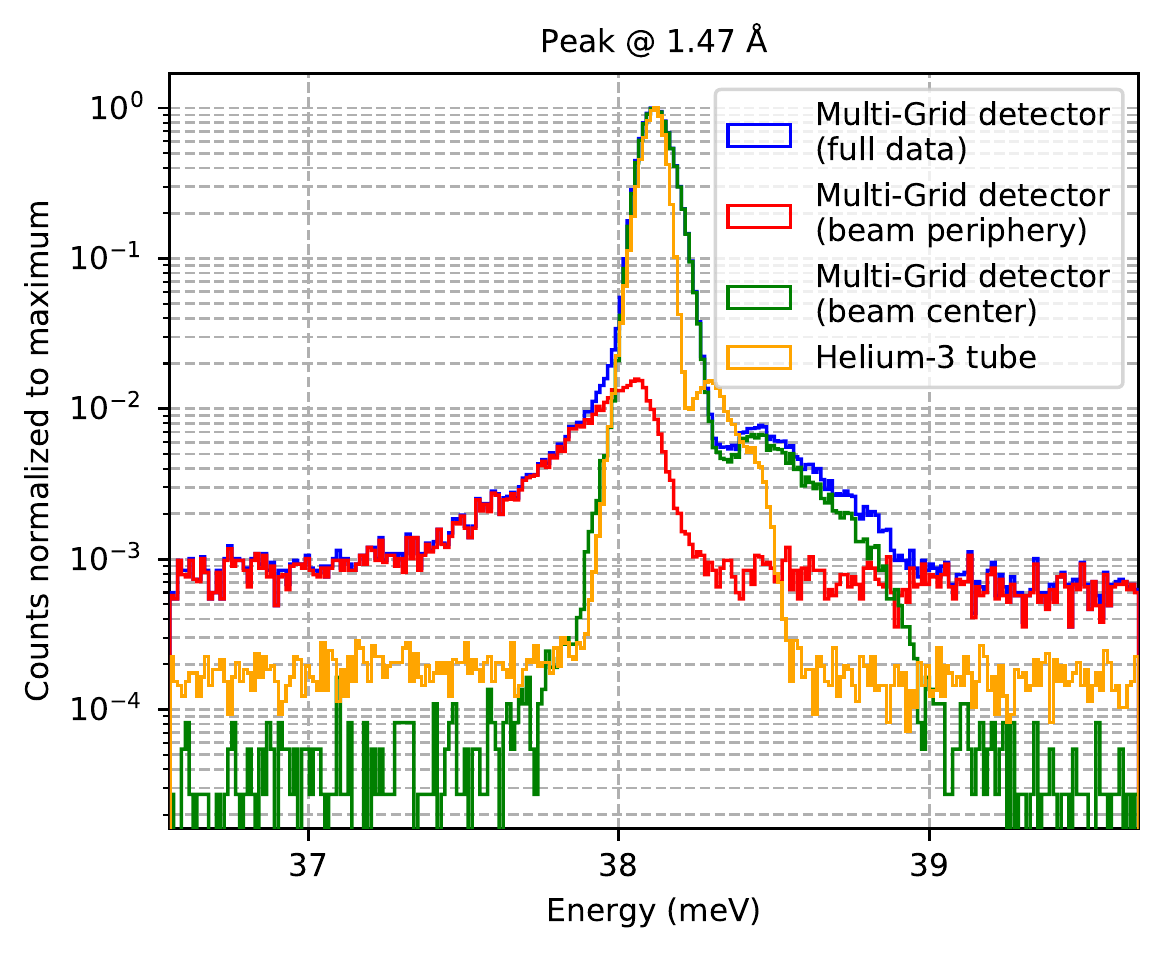}
        \caption{}
        \label{fig:mg_full_comp}
    \end{subfigure}
    \caption{Effect on peak shape by internally scattered neutrons at 1.47 \AA{} in the Multi-Grid detector (Detector 1). Histograms from the full data in the Multi-Grid detector (blue), the beam periphery (red) and beam center (green) is seen, together with the background data (black) and helium-3 data (orange). In (a), the approximated location of scattered neutrons are shown as vertical lines. Each line corresponds to the energy reconstruction for scattered neutrons with a specific extra travel distance within the detector, as specified in the legend. The counts in the histograms are normalized to the accumulated beam monitors counts. In (b), the Multi-Grid detector is compared with the helium-3 tube. The helium-3 data has been artificially shifted along the energy-axis, such that it is aligned with the peak center of the Multi-Grid detector. This is to facilitate peak comparison. The counts are normalized to peak maximum.}
    \label{fig:dist_and_full_comp}
\end{figure}

To study the energy line shape in detail, a peak at approximately 1.47~\AA{} from Detector 1 is used as an example, see figure~\ref{fig:dist_and_full_comp}. In figure~\ref{fig:mg_distance}, histograms over energy are presented. Again, the blue histogram is from the full volume, the red histogram is from the beam periphery, and the black histogram is from the background measurement. Overlaying the curves allows for a clear visualization of how the scattered neutrons affect the shape of the peak. It is observed that the majority of scattered neutrons are reconstructed with a lower energy than the peak mean, resulting in a ``shoulder" on the left side of the peak.

From the long green lines in figure~\ref{fig:grid_vs_tof_not_blocked}, it is seen that the maximum distance the scattered neutron travel within the detector is approximately 15 grids, corresponding to about 30~cm. To estimate where these scattered neutrons are reconstructed in the energy spectra, equation (\ref{eq:energy_2}) is used. This is done by using the peak mean as an approximation of the incident neutron energy, while assuming elastic scattering and that $\delta T$ is the dominant contribution to the distortion of the energy reconstruction. The colored vertical lines in figure~\ref{fig:mg_distance} shows the result of this analysis, which overlaps well with the location of neutrons detected in the beam periphery (red).

A comparison in the energy line shape with data from the helium-3 tube is presented in figure~\ref{fig:mg_full_comp}. Four histograms are shown, three from the Multi-Grid detector and one from the helium-3 tube. The separate histograms from the Multi-Grid detector corresponds to the full data, beam periphery, and beam center. On the left side of the peak, it is seen that the helium-3 data overlap well with the data from the beam center of the Multi-Grid detector (compare orange and green). This further validates that the peak shoulder in the full data (blue) is due to scattered neutrons. Note that there is additional broadening in the data from the Multi-Grid detector, as it was further away from the Fermi-chopper than the helium-3 tube. This also affects the parasitic peak on the right side of the peak center, which is present in both data sets but broader for the Multi-Grid detector.

To quantitatively compare Detector 1 and 2 in terms of internal scattering, a figure-of-merit, $fom$, is introduced. This is defined as the number of counts above background at a specified interval away from the peak center, divided by the peak area. The background is estimated to be locally flat, i.e. constant in energy, over the peak width. It is calculated on a peak-by-peak basis, based on the rate at the side of the peak. The reason the background measurements is not used for background subtraction is because of non-negligible systematic offsets. This is seen in figure~\ref{fig:nr_blocked_and_not_blocked} and \ref{fig:r_blocked_and_not_blocked}, where the background (black) does not follow the ``bump" at 3 \AA{} equally well.

The definition of the chosen $fom$ is presented in equation (\ref{eq:fom_1}),

\begin{eqnarray}
\label{eq:fom_1}
fom = \frac{1}{peakarea} \cdot \sum_{E_1}^{E_2}(counts - background),
\end{eqnarray}
where $peakarea$ is the peak area within $\pm$5$\sigma$ (using data from the full volume), $E_1$ and $E_2$ are the energy interval limits, $counts$ is the number of counts in the beam periphery, and $background$ is the flat background estimation. That is, the $fom$ captures the background subtracted counts at the edge of the peak, as a fraction of the peak area. Consequently, a small $fom$ is desirable, as this implies a low amount of scattered neutrons in the specified energy range. In words, the $fom$ can be approximately stated as:

\begin{eqnarray*}
\label{eq:fom_2}
fom = \frac{shoulder \, \, area}{peak \, \, area}.
\end{eqnarray*}

\begin{figure}[h!]
    \centering
    \begin{subfigure}{0.5\textwidth}
        \centering
        \includegraphics[height=2.5in]{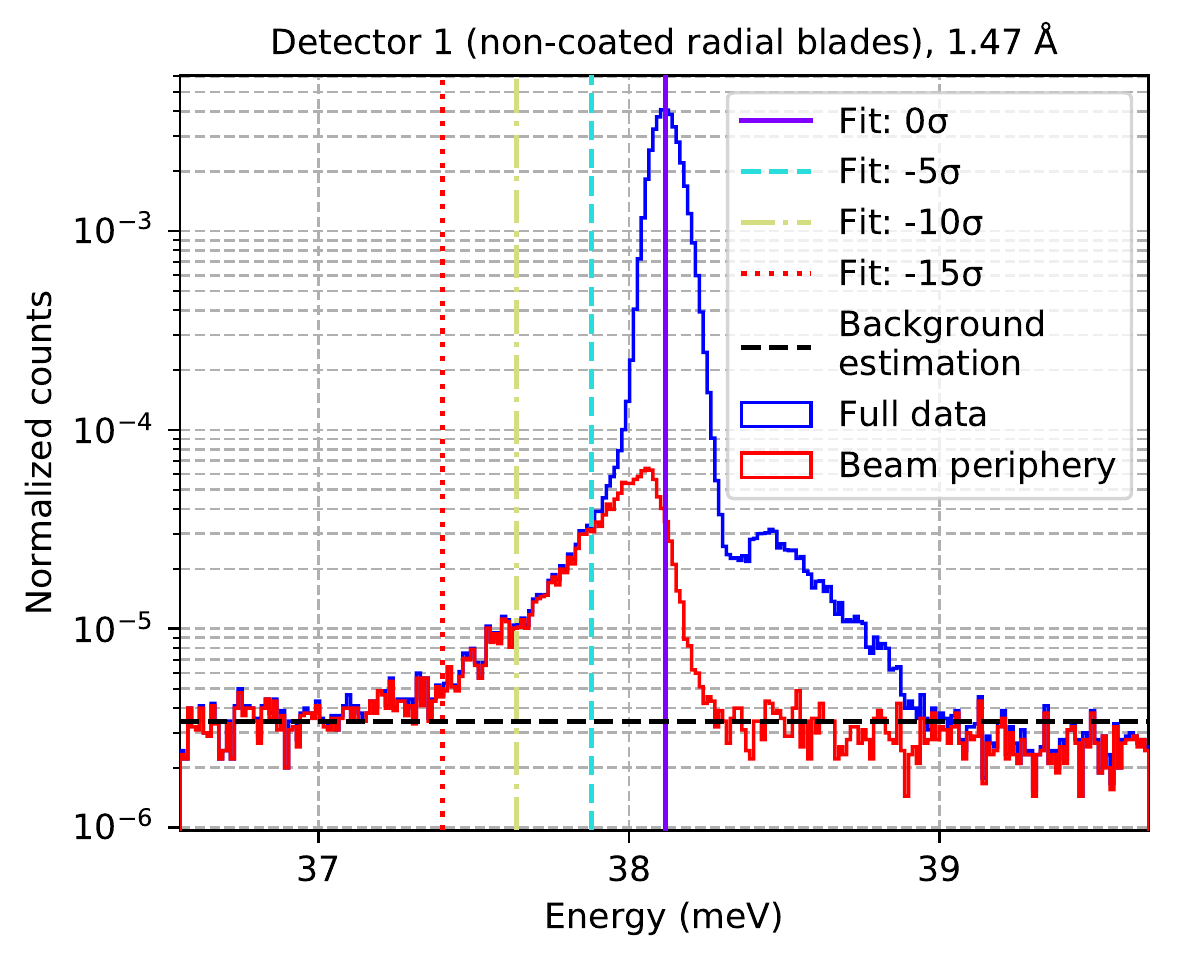}
        \caption{}
        \label{fig:mg_nr_scattering}
    \end{subfigure}%
    \begin{subfigure}{0.5\textwidth}
        \centering
        \includegraphics[height=2.5in]{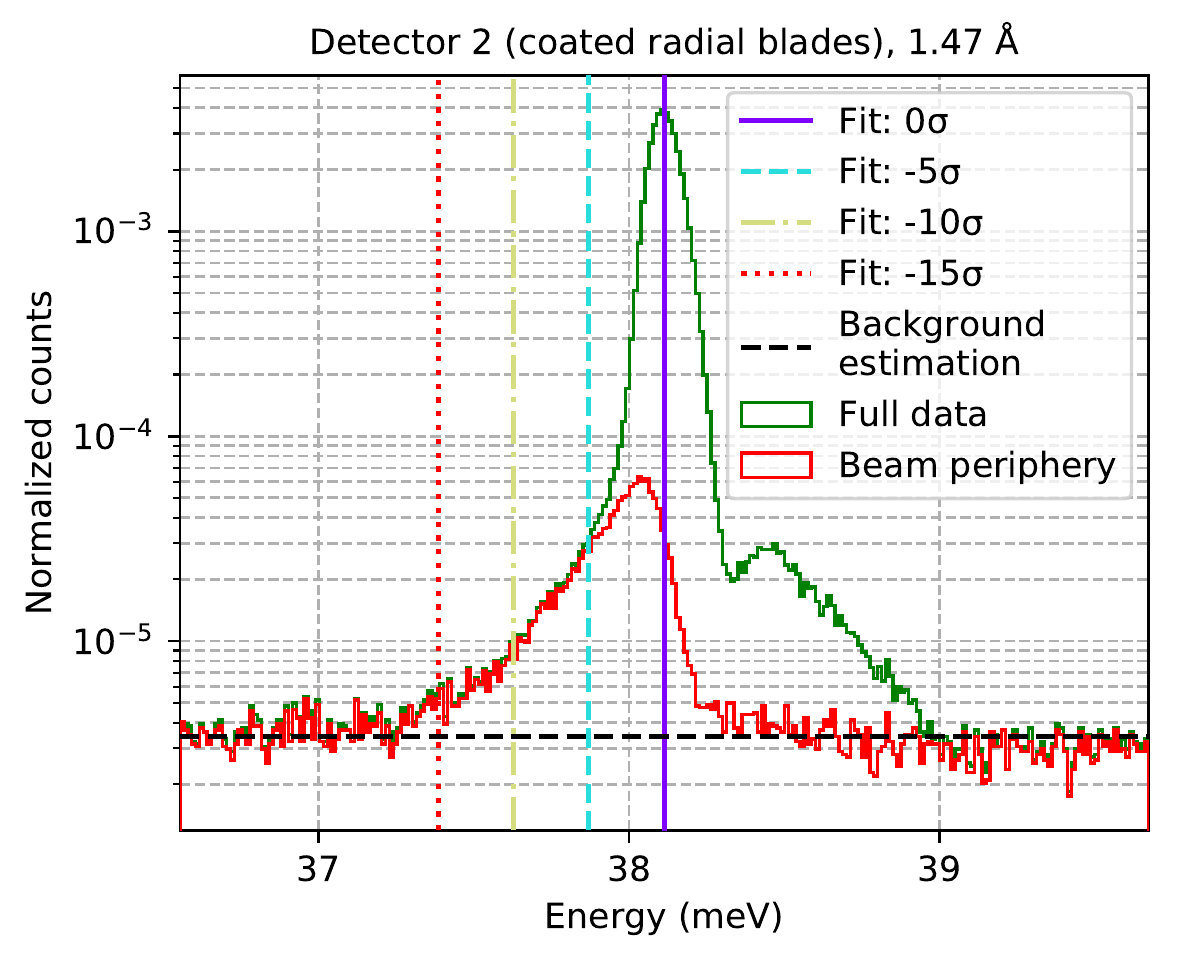}
        \caption{}
        \label{fig:mg_r_scattering}
    \end{subfigure}
    \caption{Effect on peak shape by internally scattered neutrons at 1.47 \AA{}. Histograms from the full data in Detector 1 (blue) and Detector 2 (green) are compared to data where the direct beam is removed (red). The integration limits are presented as vertical lines, including 0$\sigma$ (solid),  -5$\sigma$ (dashed), -10$\sigma$ (dash-dotted), and -15$\sigma$ (dotted). The parameter $\sigma$ is the standard-deviation around the peak center, calculated using a Gaussian fit. The background estimation is presented as a horizontal black line. The plots are normalized by accumulated beam monitor counts from the individual runs.}
    \label{fig:mg_scattering}
\end{figure}

The limits $E_1$ and $E_2$ should be selected on a peak-by-peak basis, such that the same peak region is scanned for all peaks across the wavelength spectra. Unfortunately, this is not a trivial task, as the peak shape changes with wavelength. This also complicates the comparison in $fom$ between different instruments, such as studies done at SEQUOIA and CNCS \cite{CNCS, sequoia_paper}, as the peak shape is heavily dependent on the resolution of the chopper system. Here, the peak shoulder is split into approximately three equally sized regions, where the region limits are based on the standard deviation $\sigma$ around the peak center. Each interval is $5 \sigma$ wide, see vertical lines in figure~\ref{fig:mg_scattering} for an example at 1.47~\AA{}.

\begin{figure}[h!]
    \centering
    \begin{subfigure}{\textwidth}
        \centering
        \includegraphics[height=2.4in]{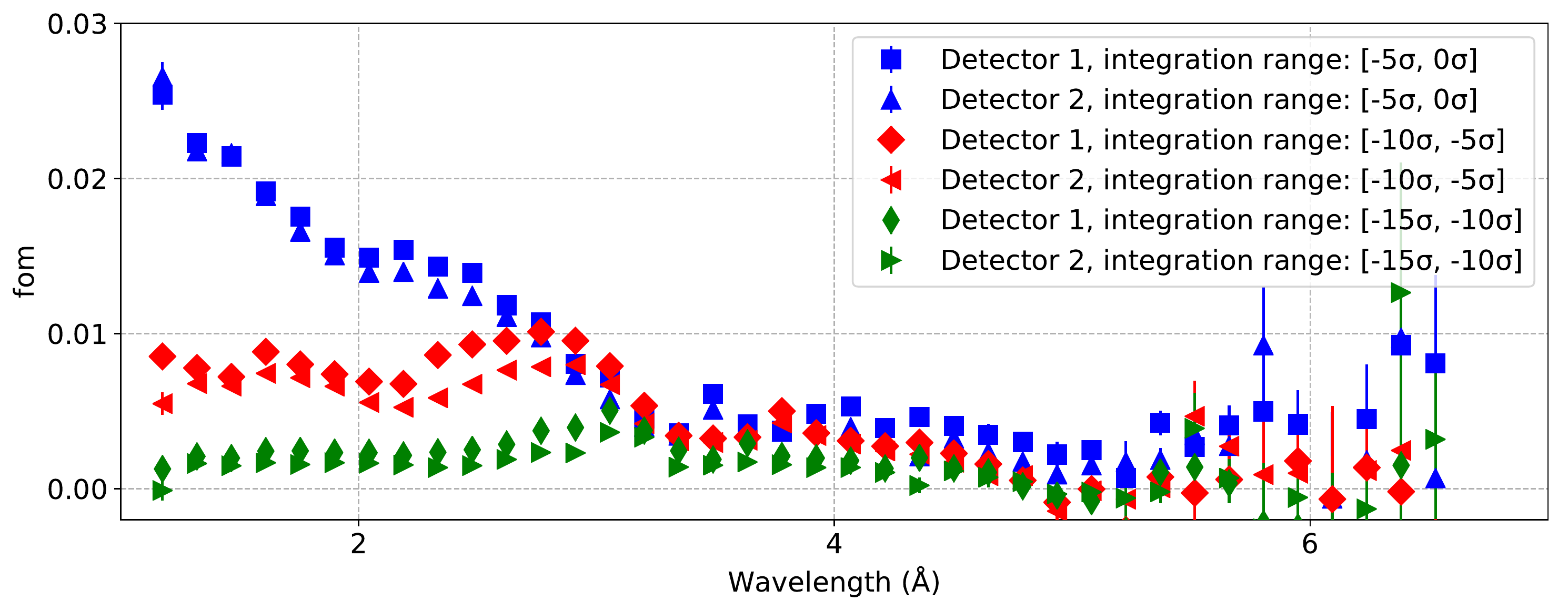}
        \caption{}
        \label{fig:fom_1}
    \end{subfigure}
    \begin{subfigure}{\textwidth}
        \centering
        \includegraphics[height=2.4in]{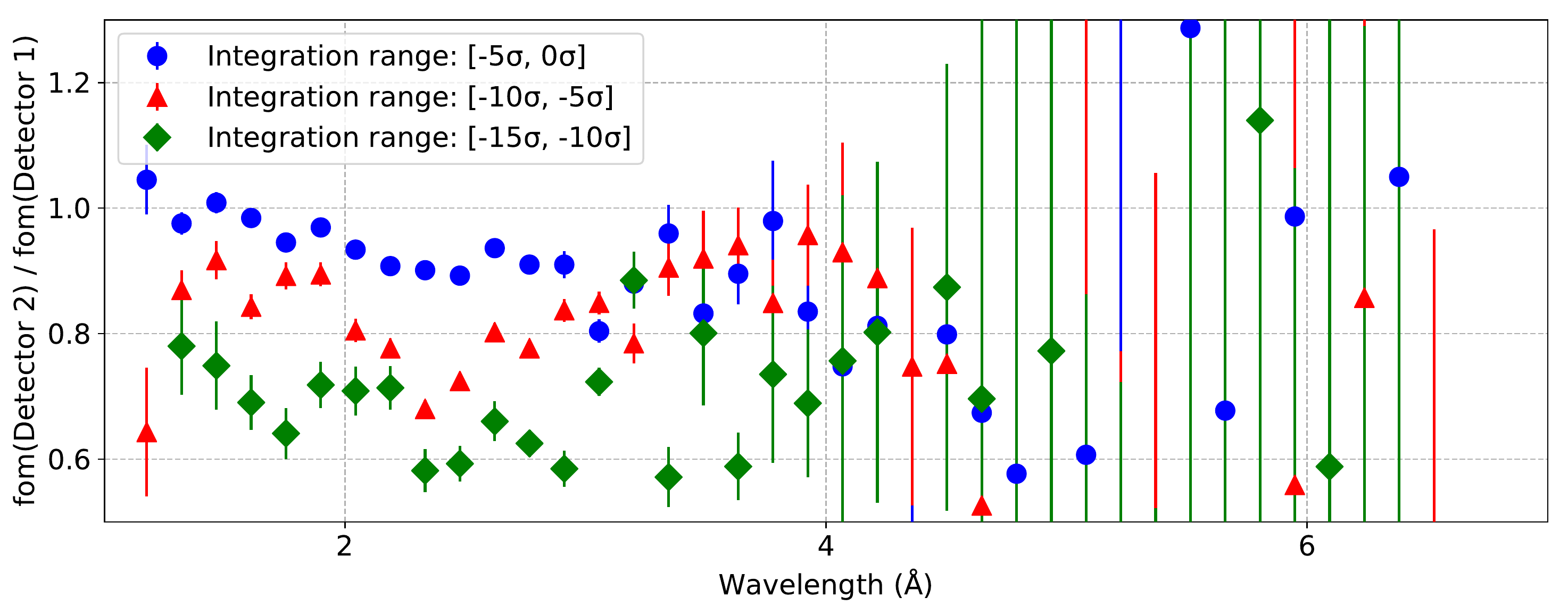}
        \caption{}
        \label{fig:fom_fractional}
    \end{subfigure}
    \caption{Comparison of the $fom$, as described in text, as a function of neutron wavelength, comparing Detector 1 (non-coated radial blades) and Detector 2 (coated radial blades). In (a), Detector 1 (squares and diamonds) and Detector 2 (triangles) are compared using three different integration ranges: [-5$\sigma$, 0$\sigma$] (blue), [-10$\sigma$, -5$\sigma$] (red), and [-15$\sigma$, -10$\sigma$] (green). In (b), the fractional $fom$, $fom$(Detector 2)/$fom$(Detector 1), is plotted. The three data sets (circles, triangles and diamonds) correspond to the three different integration intervals.}
    \label{fig:fom}
\end{figure}

The $fom$ for Detector 1 and 2 are plotted in figure~\ref{fig:fom_1}, where each data point corresponds to a peak. The $fom$ is shown for the three different integration ranges: [-5$\sigma$, 0$\sigma$], [-10$\sigma$, -5$\sigma$], and [-15$\sigma$, -10$\sigma$]. These relative intervals vary in absolute range in energy from peak to peak. The absolute energy interval, $\Delta E$, varies from approximately $\Delta E = 5 \cdot 10^{-1}$ meV at 1 \AA{}, to $\Delta E = 5 \cdot 10^{-3}$ meV at 6 \AA{}. It is observed that for wavelengths above the aluminum cut-off wavelength at 4.7 \AA{}, $fom$ is in most cases indistinguishable from zero. Furthermore, it is seen that for most wavelengths, Detector~1 (non-coated radial blades) has a larger $fom$ than Detector~2 (coated radial blades). This is visualized more clearly in figure~\ref{fig:fom_fractional}, where the fractional $fom$ is shown.  Above 4.7 \AA{}, the uncertainties are large due to low statistics above the aluminum cut-off wavelength. It is noted that the effect resultant from coating the radial blades increase with distance from the peak center, and that the scattered neutrons are attenuated more strongly further away.

\newpage

The average improvement of $fom$ by coating the radial blades are presented for the three shoulder regions in table~\ref{table:table2}.$^2$\footnote[0]{$^2$If, for background subtraction, the background measurement is used instead of the background estimation, the improvements are slightly higher. Starting from the interval closest to the peak center and going outwards, the results are: 13~$\pm$~9~\%, 22~$\pm$~7~\%, and 39~$\pm$~10~\%. However, these values are believed to be an overestimation of the effect of coating the radial blades, due to a systematic offset in the background measurement with Detector 1, as seen in figure~\ref{fig:blocked_and_not_blocked}.} The results are based on data below 4~\AA{}, where the statistical uncertainties are acceptable. Note that these numbers are from a measurement with a collimated 14 $\times$ 60 mm$^2$ rectangular beam on the Multi-Grid detector, which is not generally the mode during normal operation. The reason a thin beam was nevertheless used was an experimental limitation, as the incident beam had to fit the helium-3 tube as well, which is much more narrow than the Multi-Grid detector. The direct beam of the current measurement hits the detector close to 90$^\circ$, within a single row, in between radial blades. However, during normal operation, neutrons will in many cases hit a much larger portion of the detector surface area (with the important exception of Bragg diffraction from crystals, which can be a few cm$^2$ on the detector surface), and typically at angles with a small divergence around 90$^\circ$. This will influence the effect of coating the radial blades, as the absorption of neutrons depends on two factors: incident angle on the radial blades, and the number of radial blades crossed. From measurements at SEQUOIA \cite{sequoia_paper}, preliminary results show a greater advantage of using the coated radial blades during normal operation.

\begin{table}[h!]
\centering
\caption{Average improvement of $fom$ between 1 and 4 \AA{} by coating the radial blades. The results are presented separately for the three shoulder regions.}
\label{table:table2}
\begin{tabular}{|c|c|}
\hline
\multicolumn{1}{|c|}{\textbf{Shoulder region}} & \textbf{Average improvement}   \\ \hline
[-5$\sigma$, 0$\sigma$]                    & 7 $\pm$ 6 \%                \\ \hline
[-10$\sigma$, -5$\sigma$]                  & 16 $\pm$ 8 \%                  \\ \hline
[-15$\sigma$, -10$\sigma$]                    & 31 $\pm$ 8 \%                  \\ \hline
\end{tabular}
\end{table}

All the presented percentage errors should be interpreted as variations around the mean, not statistical uncertainties. It is not predicted that the radial blades will have the same effect across all wavelengths. Thus, a mean with a low variance is not expected. This is because the effect of the radial blades depends on the number of radial blades crossed, which depends on the scattering angle of the neutron. As the scattering angle is wavelength dependent \cite{ncrystal}, the effect vary as a function of wavelength, as was shown in figure~\ref{fig:fom_fractional}. The stated standard deviations around the mean have the intention to capture the magnitude of this variation over the measured wavelengths.

\FloatBarrier

\subsection{Time- and energy resolution}

The observed energy resolution depends on two main components: the resolution of the detector and the resolution of the chopper system. The observed peak shape is a convolution of the chopper pulse shape and the detector response shape, according to equation (\ref{eq:resolution}),

\begin{eqnarray}
\label{eq:resolution}
f_{observed} = f_{chopper} \otimes f_{detector},
\end{eqnarray}
where $f_{observed}$ is the observed energy distribution, $f_{chopper}$ is the energy distribution from the chopper system, and $f_{detector}$ is the detector energy response function.

The resolution of the chopper system, $f_{chopper}$, depends on all choppers in the setup, which in this case includes most importantly the source chopper and the Fermi-chopper. The initial pulse produced by the source chopper, which triggers the $T_0$ signal, is allowed to travel 28 meters before reaching the Fermi-chopper. The long travel length allows the neutrons with different energies to separate from each other in time, due to their different velocities. This pulse broadening is desired, as it increases the energy resolution when using the $tof$ from the $T_0$-signal to calculate neutron energy. 

The purpose of the Fermi-chopper is to cut the wide incident pulse, with neutrons well separated in energy due to their long travel path, into sharp pulses. These pulses, in contrast to the source pulse, should be kept as sharp as possible in time. This is so that any broadening due to the full detector component $f_{detector}$ is more easily observed. That is, by keeping the distance from the Fermi-chopper short, some information about the incident neutron energy is lost due to less separation in energy. However, because of the small separation in energy, any broadening due to the detector component will have a larger relative effect. Thus, keeping a small $\Delta T$, and consequently $\Delta E$, on the final pulse is important to study detector related broadening.

The pulse width in time at the detector surface, $\Delta T_{chopper}$, is determined according to equation (\ref{eq:t_chopper}),

\begin{eqnarray}
\label{eq:t_chopper}
\Delta T_{chopper} = t_{open} + d(v_{s}^{-1} - v_{f}^{-1}),
\end{eqnarray}
where $t_{open}$ is the Fermi-chopper window opening time, $d$ the Fermi-chopper-to-detector distance, and $v_{s}$ and $v_{f}$ are the velocities of the slowest and fastest neutrons in the pulse,  respectively. Therefore, to reduce the broadening, it is important to have a short opening time and to keep the Fermi-chopper as close as possible to the detector. As the width of $f_{chopper}$ grows as a function of $\Delta T_{chopper}$, the same conditions mentioned above are required to keep a sharp pulse in energy.

\begin{figure}[h!]
    \centering
    \begin{subfigure}{0.5\textwidth}
        \centering
        \includegraphics[height=2.4in]{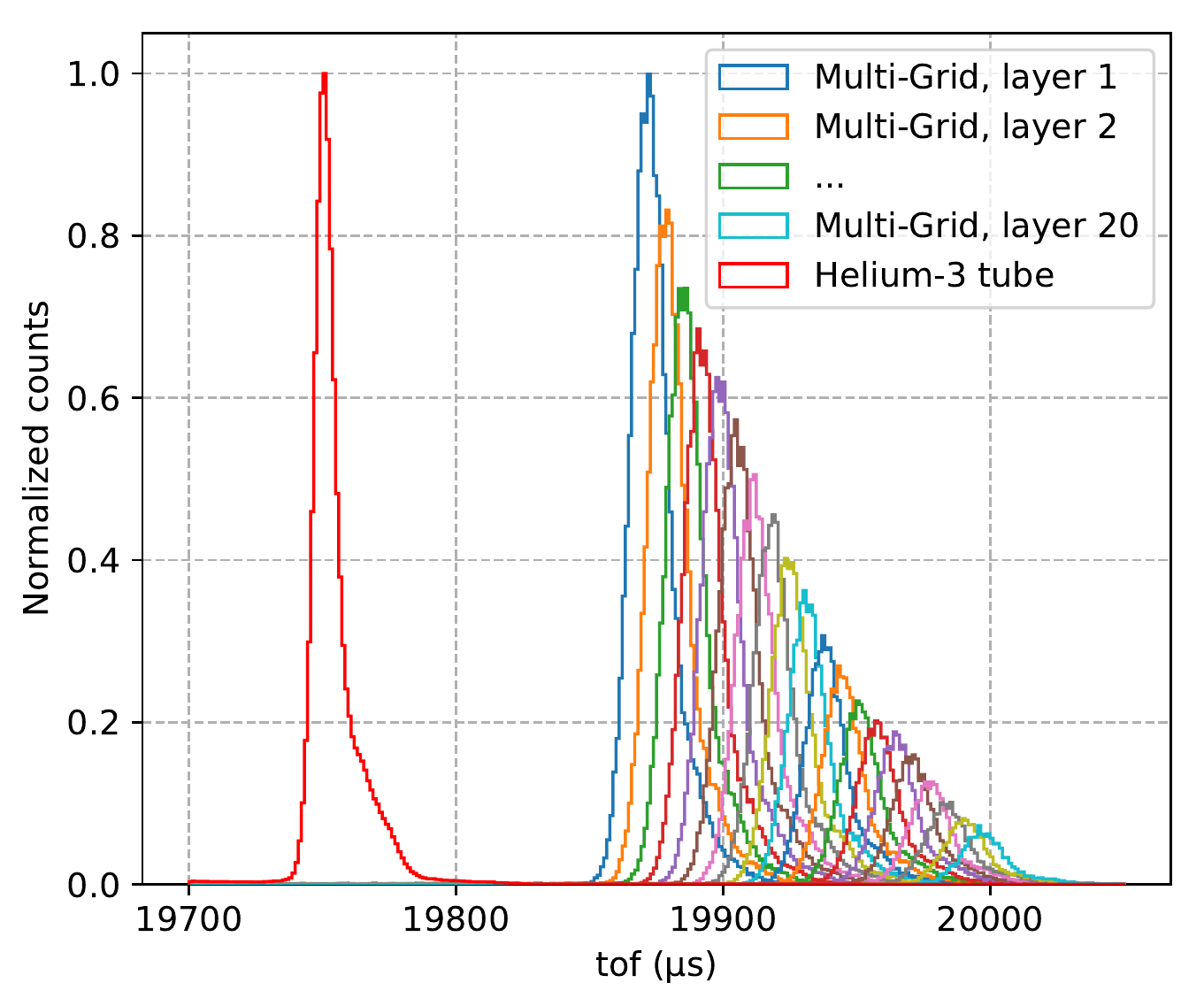}
        \caption{}
        \label{fig:tof_single_peak}
    \end{subfigure}%
    \begin{subfigure}{0.5\textwidth}
        \centering
        \includegraphics[height=2.4in]{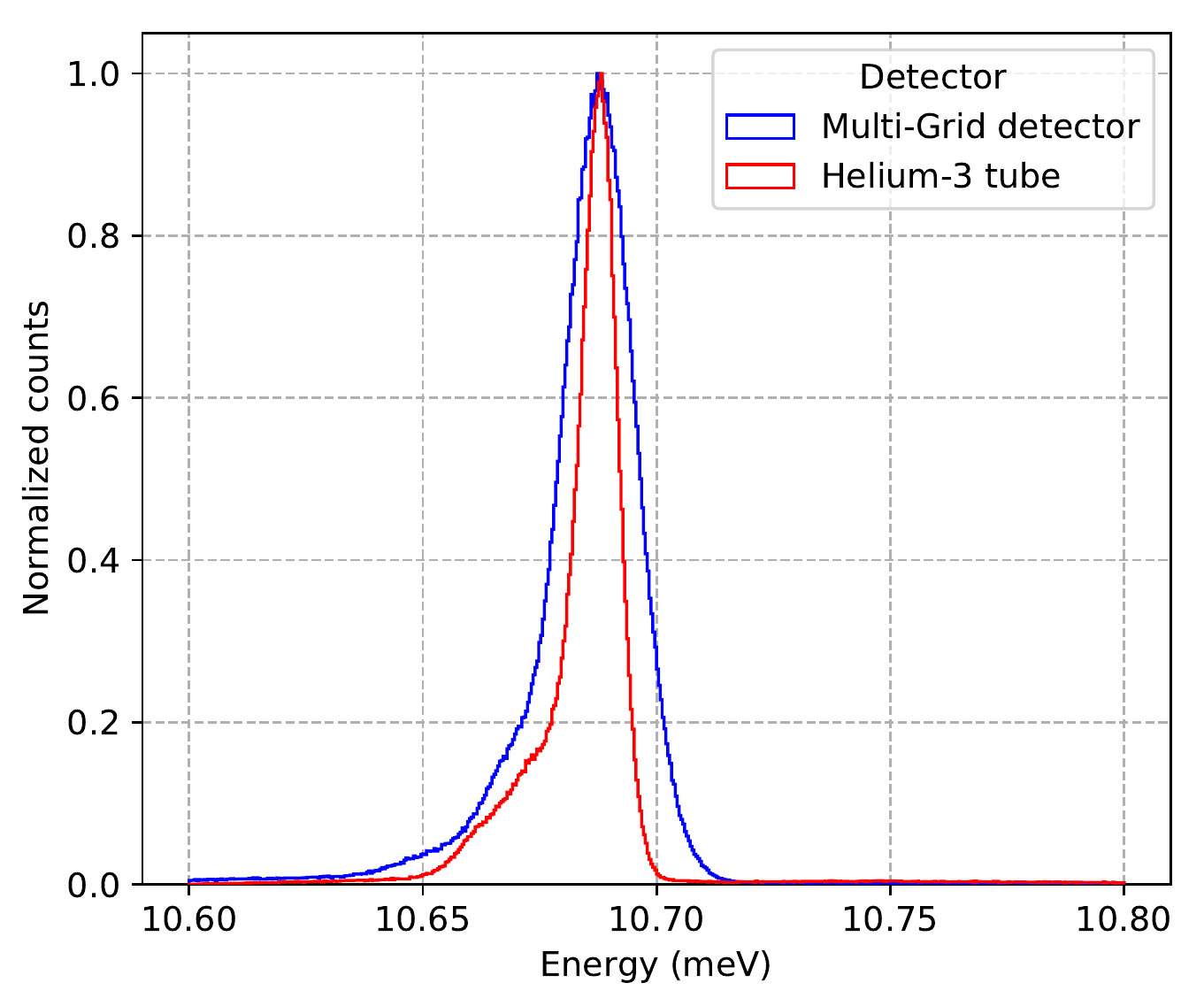}
        \caption{}
        \label{fig:energy_single_peak}
    \end{subfigure}
    \begin{subfigure}{0.5\textwidth}
        \centering
        \includegraphics[height=2.4in]{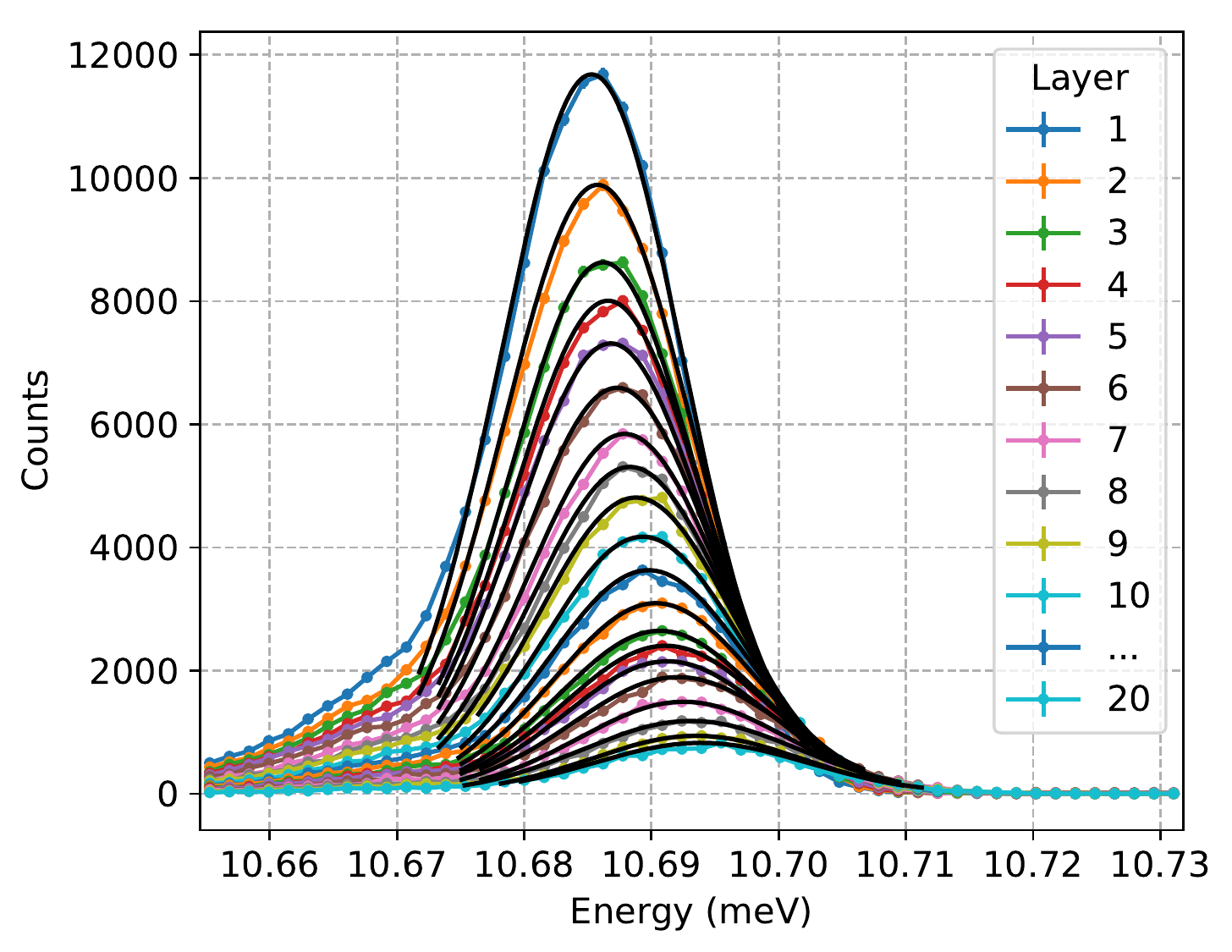}
        \caption{}
        \label{fig:energy_layers}
    \end{subfigure}%
    \begin{subfigure}{0.5\textwidth}
        \centering
        \includegraphics[height=2.4in]{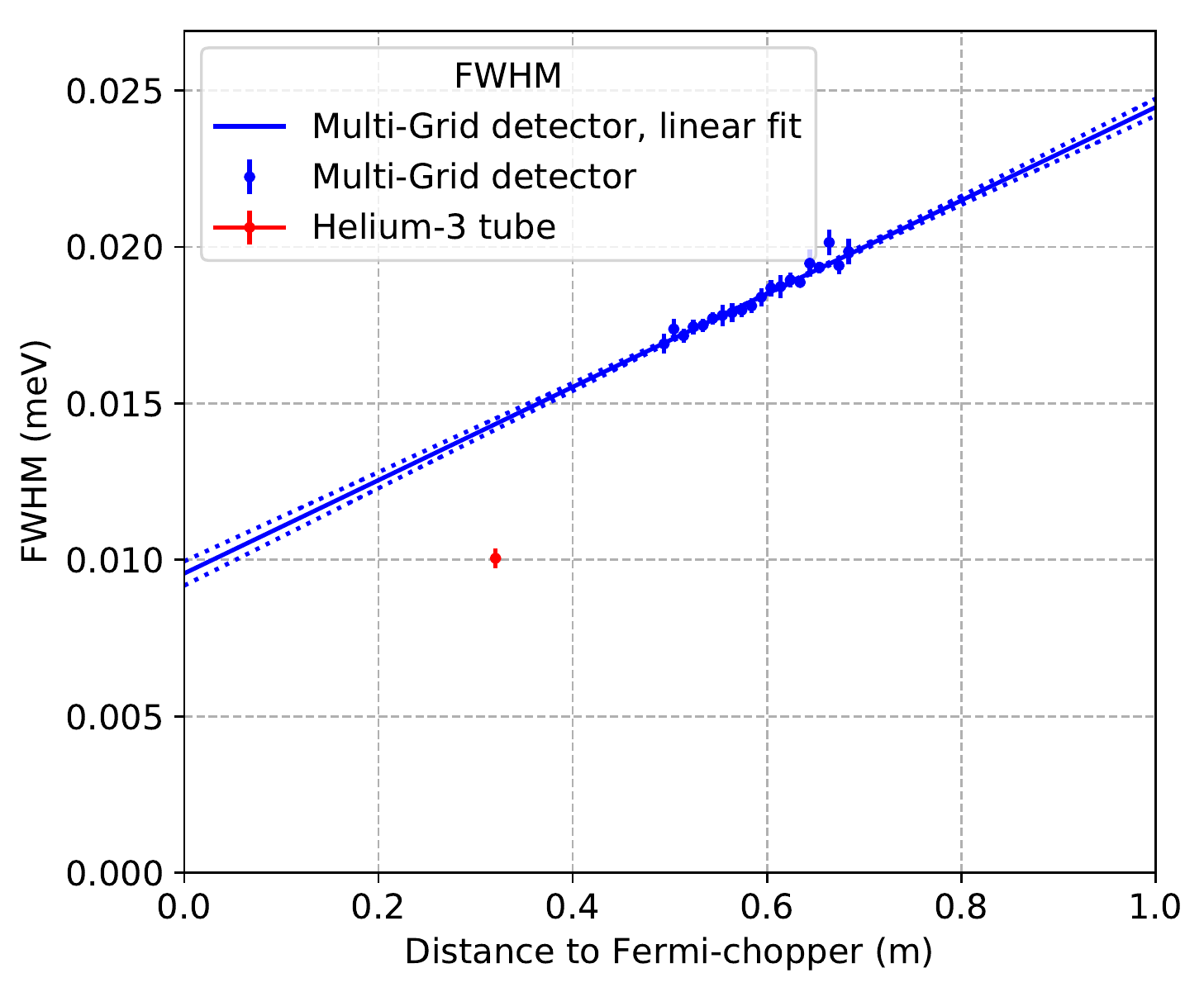}
        \caption{}
        \label{fig:inter_fitted_2}
    \end{subfigure}
    \caption{Correction for pulse broadening with distance, using 2.8 \AA{} neutrons as an example. In (a), the $tof$ from the twenty individual layers in the Multi-Grid detector, from front to back,  is presented together with the helium-3 tube (red). In (b), the corresponding energy spectra for the Multi-Grid detector (blue) and helium-3 tube (red) are shown. The histograms in (a) and (b) are normalized to peak maximum, to facilitate peak comparison. In (c), the widening in energy is presented for each layer. The distribution from each layer is fitted with a Gaussian (black) and the FWHM is extracted from the fit parameters. In (d), the FWHM from each Multi-Grid layer (blue) is shown together with the FWHM of the helium-3 tube (red), as a function of distance from the Fermi-chopper. A linear fit (solid blue line) from the Multi-Grid points is also presented together with the fit uncertainties (two dotted blue lines), demonstrating how the chopper pulse widens with distance.}
    \label{fig:single_peaks}
\end{figure}

As the helium-3 tube and the Multi-Grid detector are at different distances from the Fermi-chopper, recall figure~\ref{fig:fig2}, the width in energy spectra from Multi-Grid data are corrected to account for this offset. In figure~\ref{fig:tof_single_peak}, the $tof$ spectra are shown individually for each of the twenty wire layers in the Multi-Grid detector, together with the $tof$ from the helium-3 tube. In figure~\ref{fig:energy_single_peak}, the energy spectra corresponding to these $tof$-values are presented. In the figure, data from all twenty layers from the Multi-Grid detector are seen to be reconstructed at the same position, overlapping with the helium-3 data. However, there is an additional broadening on the Multi-Grid detector, caused by the additional distance from the Fermi-chopper.

To account for this, the peak width in energy is calculated using a Gaussian fit for each layer, see figure~\ref{fig:energy_layers}.$^3$\footnote[0]{$^3$It is observed that the peak center is at a lower energy for the front layer, ca 10.685 meV, compared to the back layer, ca 10.695 meV, with a gradual shift for inbetween layers. This is because the $^{10}$B$_4$C-coating has a higher absorption cross-section at lower energies. Consequently, the low energetic neutrons will be absorbed closer to the detector entrance than the high energetic ones. That is, the peak center will gradually shift to higher energies towards the back layers, as only the higher energetic neutrons are likely to ``survive" the full depth of the detector.} This is to follow how the peak width, defined as the full width at half maximum (FWHM), depends on the distance from the Fermi-chopper. In figure~\ref{fig:inter_fitted_2}, the FWHM for each of the layers in the Multi-Grid detector, as well as the FWHM of the helium-3 tube, are plotted as a function of the distance from the Fermi-chopper. A linear fit to the data points from the Multi-Grid detector is also shown, together with the fit uncertainties. The interpolation takes into account the widening of the pulse with distance, and allows the Multi-Grid detector to be compared to the helium-3 tube. That is, the width the Multi-Grid is interpolated to the value it would have had, at the same distance from the Fermi-chopper as the helium-3 tube. Here the procedure is shown for a peak at 2.8 \AA{} as an example, but the same analysis is done for the peaks at the remaining wavelengths.

\begin{figure}[h!]
    \centering
    \begin{subfigure}{0.5\textwidth}
        \centering
        \includegraphics[height=2.5in]{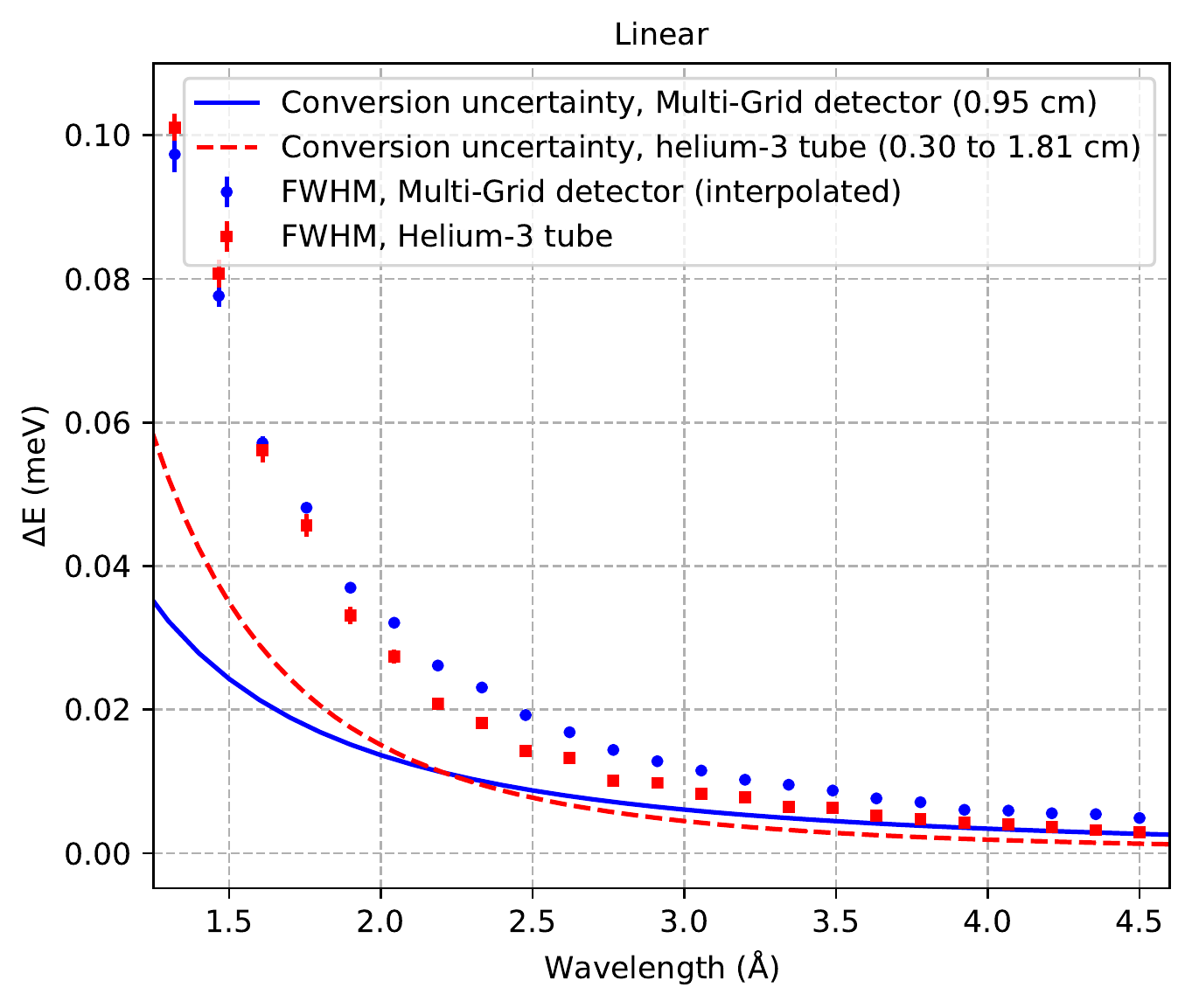}
        \caption{}
        \label{fig:interpolated_lin}
    \end{subfigure}%
    \begin{subfigure}{0.5\textwidth}
        \centering
        \includegraphics[height=2.5in]{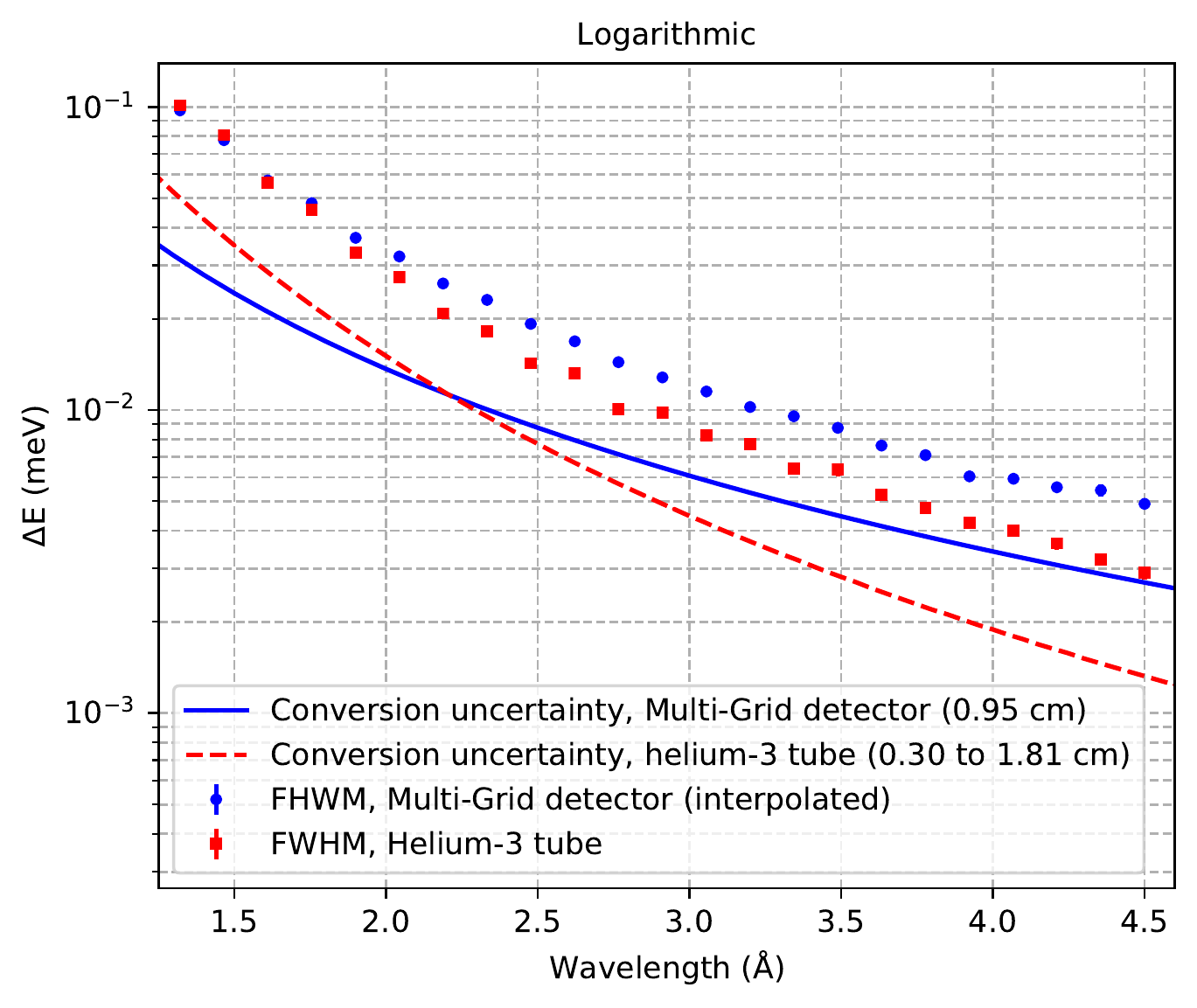}
        \caption{}
        \label{fig:interpolated_log}
    \end{subfigure}
    \caption{Comparison between the FWHM of the Multi-Grid detector (blue dots) and the helium-3 tube (red squares), as a function of neutron wavelength. The Multi-Grid values are acquired from a linear interpolation to the position of the helium-3 tube, as described in the text. The solid blue lines show the conversion location uncertainty in the Multi-Grid detector, which results from the 0.95~cm distance between adjacent converter layers. The dashed red lines present the conversion location uncertainty in the helium-3 tube, corresponding to the range covered by 95\% of the detected neutrons. In (a), the values are plotted on a linear scale, while in (b) the scale is logarithmic.}
    \label{fig:interpolated_fwhm}
\end{figure}

Using the interpolated pulse width of the Multi-Grid detector, the detector energy resolution can be compared to that of the helium-3 tube. This is shown in figure~\ref{fig:interpolated_fwhm}, where the FWHM of the two detectors are presented as a function of neutron wavelength. It can be seen that for the four data points corresponding to the shortest wavelengths measured, the Multi-Grid detector and the helium-3 tube has an equal energy resolution within error bars. For the remaining wavelengths, conversely, the helium-3 tube has a finer energy resolution. However, the trend seems to indicate that for wavelengths shorter than those measured here, the Multi-Grid might have an advantage. The difference between the detectors is more prominent at longer wavelengths, where the chopper resolution is finest.

The observed difference relates to the variation in time resolution between the two detectors technologies. The crucial difference between the detectors is the neutron absorption time distribution, which depends on the conversion location uncertainty, see blue and red lines in figure~\ref{fig:interpolated_fwhm}. These curves show how the uncertainty in conversion location translates to a corresponding uncertainty in energy. This uncertainty can vary more in the helium-3 tube than that of the Multi-Grid detector, as the absorption reaction can take place anywhere in the gas along the 2.5 cm depth of the helium-3 tube. For the Multi-Grid detector, however, the absorption only takes place at discrete intervals of 0.95~cm (1~cm~pitch~$-$~2~$\cdot$~0.025~cm aluminum thickness), where the solid conversion layers are located.

The two calculated curves are always beneath the data points as they do not account for the remaining time uncertainties in the measurement. This is illustrated in equation (\ref{eq:fwhm_calc}),

\begin{eqnarray}
\label{eq:fwhm_calc}
\textnormal{FWHM} = f(\Delta E_{chopper}, \Delta E_{conversion}, \Delta E_{rest}),
\end{eqnarray}
where $\Delta E_{chopper}$ is the energy width from the chopper,  $\Delta E_{conversion}$ is the energy width from conversion location uncertainty, and $\Delta E_{rest}$ is the width from the remainder of effects, such as broadening from the electronics, and charge collection times in the MWPC. The function $f$ calculates the FWHM, and it grows with the three parameters mentioned above. This is because the observed distribution is a convolution of the three distributions from which these widths corresponds to, as seen in equation (\ref{eq:resolution}).

\begin{figure}[h!]
    \centering
    \begin{subfigure}{0.5\textwidth}
        \centering
        \includegraphics[height=2.5in]{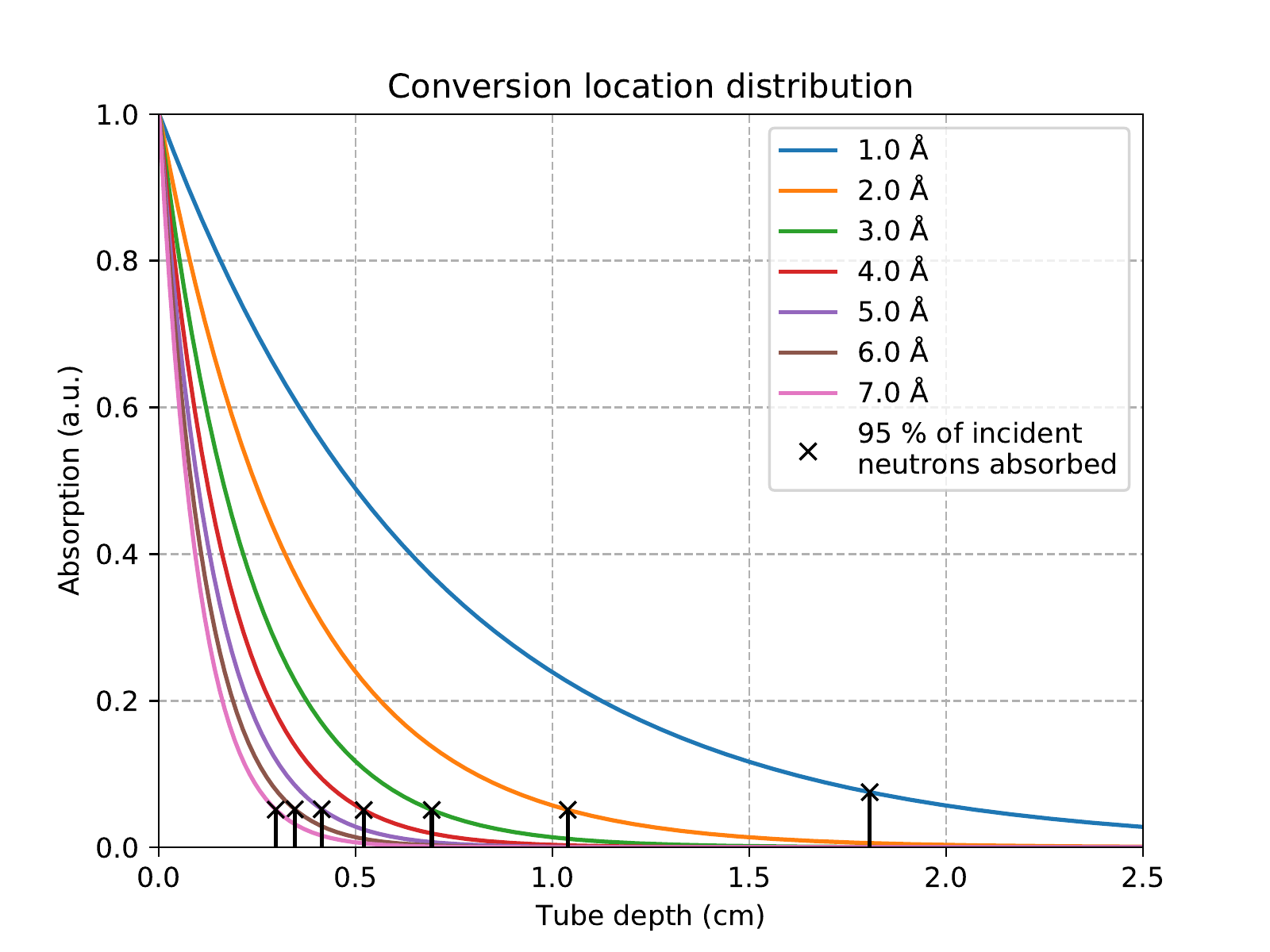}
        \caption{}
        \label{fig:he3_depth}
    \end{subfigure}%
    \begin{subfigure}{0.5\textwidth}
        \centering
        \includegraphics[height=2.5in]{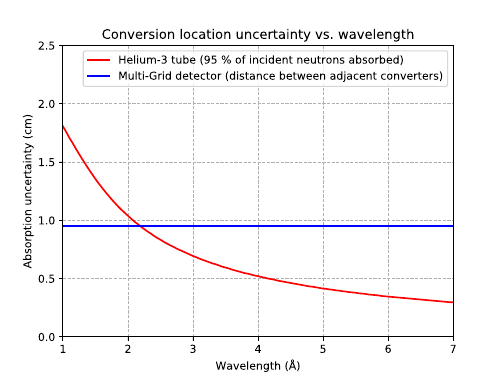}
        \caption{}
        \label{fig:he3_depth_vs_angstrom}
    \end{subfigure}
    \caption{Conversion location uncertainty in the helium-3 tube and Multi-Grid detector. In (a), the distributions of the neutron converter locations in the helium-3 tube are presented for a selection of wavelengths between 1 and 7~\AA{} (blue to pink lines). The data have been scaled such that all curves start at 1 on the y-axis, for added clarity of the distributions along the depth of the tube. The black crosses indicate the depth where 95\% of the detected neutrons have been absorbed for each distribution. This depth is treated as the conversion location uncertainty of the tube. In (b), the conversion location uncertainty of the helium-3 tube (red) is compared to the Multi-Grid detector (blue).}
    \label{fig:he3_unc_abs}
\end{figure}

In figure~\ref{fig:he3_unc_abs}, the conversion location uncertainties for the helium-3 tube and Multi-Grid detector are compared. The uncertainty in the helium-3 tube was acquired using a one-dimensional calculation of the neutron absorption as a function of depth, which was done for a series of different incident wavelengths. As an estimate of the detection location uncertainty, the depth where 95\% of the detected neutrons are converted is used. Only the range is considered, as this is estimated to be the dominant factor of the neutron detection distribution. However, factors such as tube shape and conversion products ranges also contribute to the distribution. Using a high percentage, 95\%, is an attempt to account for these missing factors.

For the four shortest wavelengths measured, the Multi-Grid detector and the helium-3 tube have equivalent energy resolutions. However, for longer wavelengths, the helium-3 tube performs better, as the majority of neutrons are absorbed within less than 0.95~cm from the surface. This is in contrast with the Multi-Grid detector, which still has the fixed 0.95~cm timing difference.

\begin{figure}[h!]
    \centering
    \begin{subfigure}{0.5\textwidth}
        \centering
        \includegraphics[height=2.5in]{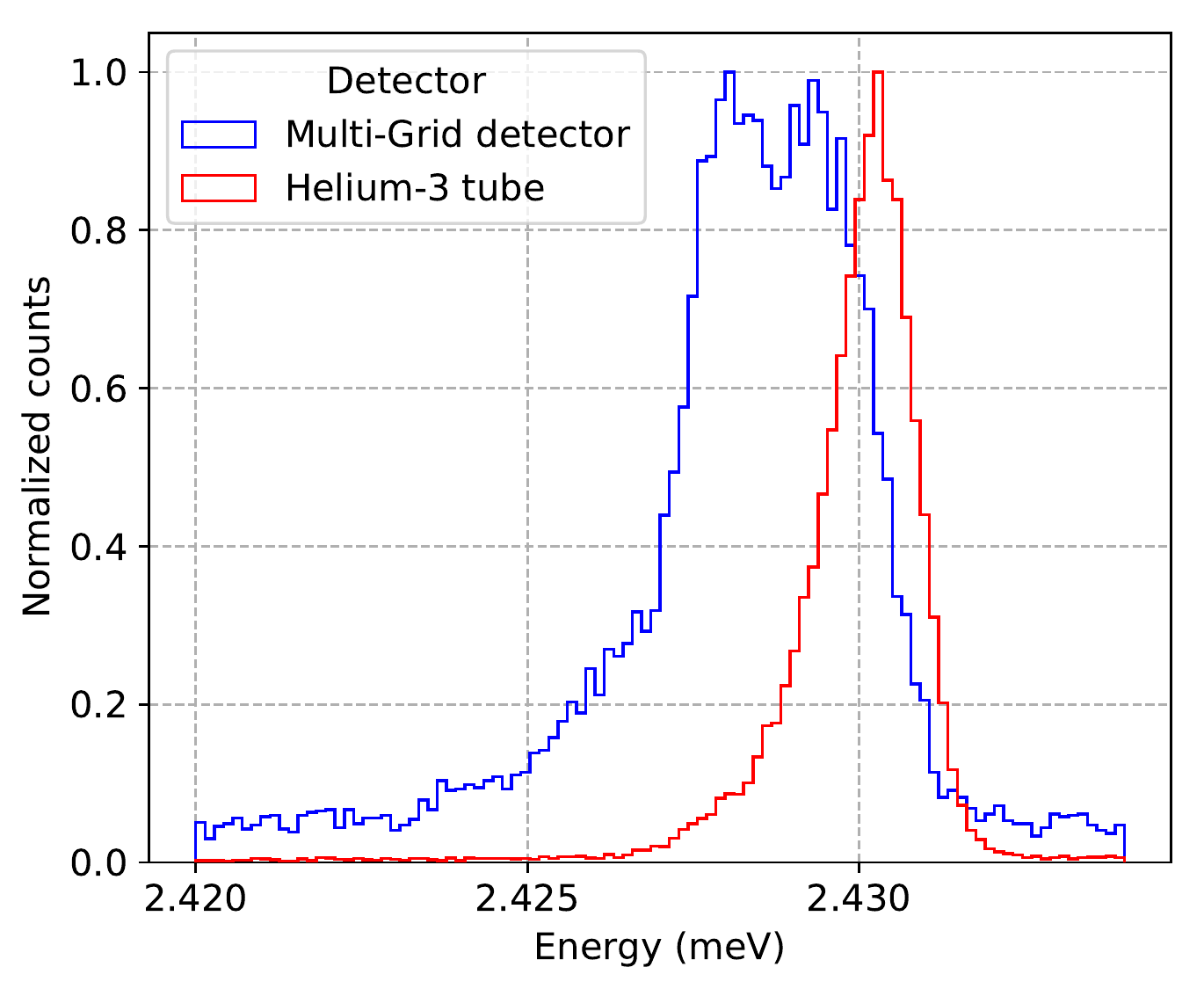}
        \caption{}
        \label{fig:energy_double_peak}
    \end{subfigure}%
    \begin{subfigure}{0.5\textwidth}
        \centering
        \includegraphics[height=2.5in]{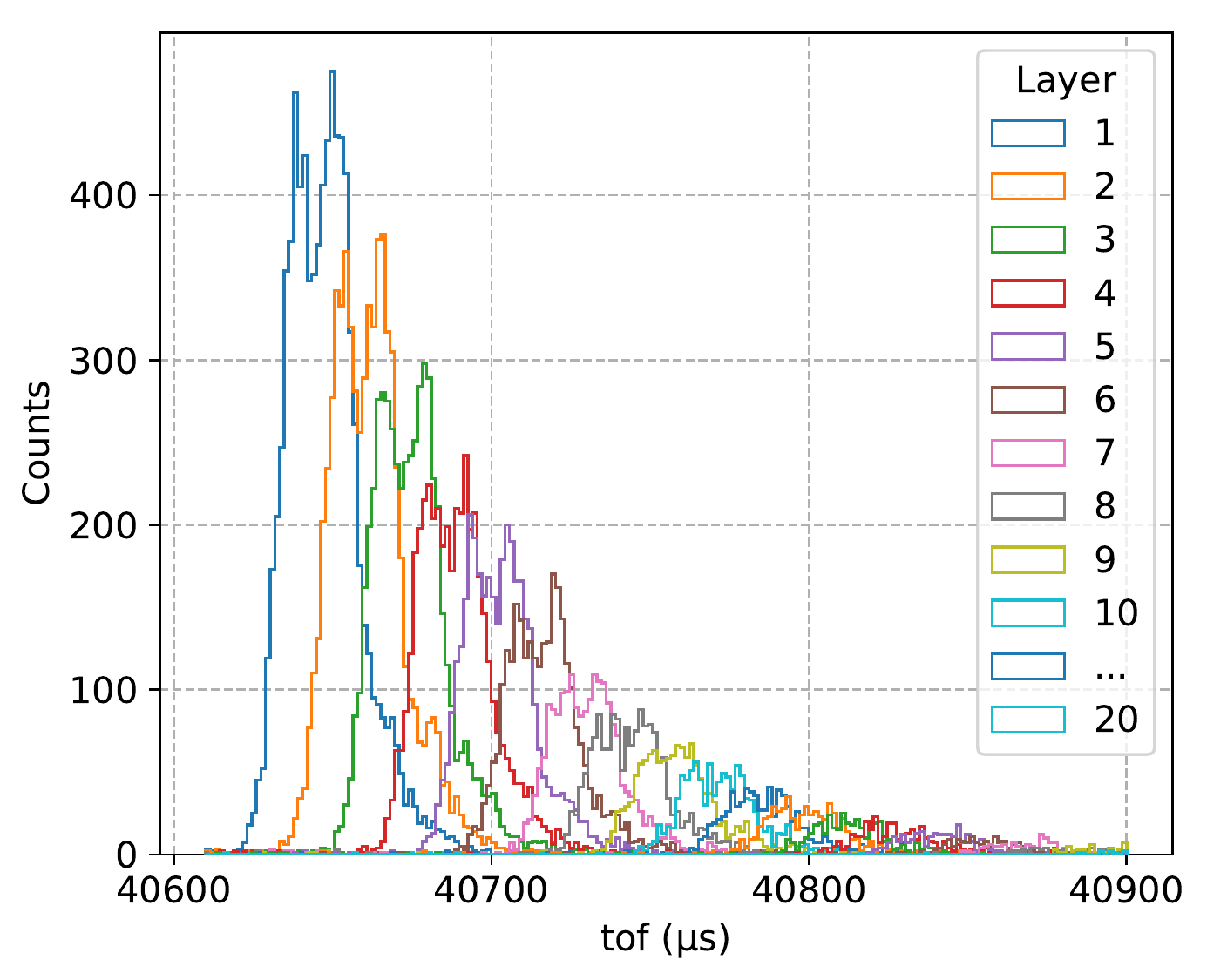}
        \caption{}
        \label{fig:tof_double_peak}
    \end{subfigure}
    \caption{Demonstration of the neutron absorption timing difference between the Multi-Grid detector and the helium-3 tube. In (a), the energy distribution for the Multi-Grid detector (blue) is plotted together with the helium-3 tube (red), for a peak at approximately 4.3~\AA{}. A peak at a longer wavelength has intentionally been chosen, as the double peak is only visible at long wavelengths. The histograms are normalized to the max values, to facilitate shape comparison. In (b), $tof$ from each of the twenty layers in the Multi-Grid detector are histogrammed individually. Note the double peak from each layer.}
    \label{fig:double_peaks}
\end{figure}

This is further illustrated in figure~\ref{fig:double_peaks}, where the energy-distribution for a peak at 4.3~\AA{} is presented together with the corresponding $tof$-distribution. In figure~\ref{fig:energy_double_peak}, a double peak is seen for the Multi-Grid detector (blue), while the helium-3 tube (red) has a single peak. The double peak is a consequence of the discrete 0.95~cm distance between consecutive conversion films. This is more clearly visualized in figure~\ref{fig:tof_double_peak}, where $tof$-data from each wire layer are plotted separately. It is seen that each wire layer has a double peak. This is because each wire layer is adjacent to two converter films, where neutrons can be absorbed in either one. The distance between the peaks is approximately 15 $\mu$s, which, for a 4.3 \AA{} neutron, translates to 0.91 cm. This is close to the expected 0.95 cm timing difference between adjacent converter films. 

It is worth noting, that the better the peaks are separated, the more information is gained concerning which coating surface the absorption took place. This could potentially be used to increase the spatial resolution of the Multi-Grid detector down to the specific surface the conversion took place, providing the peaks are sufficiently well separated. Although this investigation is beyond the scope of this work, it might be worth exploring in the future.

From the analysis a clear double peak is seen from the Multi-Grid detector for the longest wavelengths measured. However, in practical applications of the Multi-Grid detector at an instrument, such as at CNCS and SEQUOIA, this effect has not been observable. This is due to the approximately 1 order of magnitude longer Fermi-chopper-to-detector distance in a typical instrument, which leads to additional pulse broadening. A comparison can be made between the current setup, which has a few tens of centimeters Fermi-chopper-to-detector distance, with the CNCS and SEQUOIA instruments, where the equivalent distance is a few meters. As the pulse broadening is proportional to the distance, a corresponding one order of magnitude more broadening is seen at CNCS and SEQUOIA for the same wavelength range measured here \cite{CNCS, sequoia_paper}. The effect of the double-sided coating is below the instrument resolution, and no difference in energy resolution is seen between the helium-3 tube and Multi-Grid detector.

\FloatBarrier

\section{Conclusions}

Using a series of $\sim\mu$s short neutron pulses, chopped by a table-top prototype Fermi-chopper, it is possible to investigate subtle time- and energy resolved effects in the boron-10 based and helium-3 based thermal neutron detectors, reaching the resolution of a few $\mu$eV. Properties which can be studied include internal neutron scattering in the Multi-Grid detector, as well as a comparison in energy resolution with the helium-3 tube. In addition to this, the neutron detection efficiency of the Multi-Grid detector can be derived using a complementary measurement with the helium-3 tube.

The derived efficiency of the Multi-Grid detector is observed to match the theoretical prediction for wavelengths above 4~\AA{}, and be close at 1.3~\AA{}. For the remaining wavelengths, a strong discrepancy is seen, caused by saturation in the helium-3 detector system, as the detector and read-out system was unable to cope with the peak flux at these wavelengths. The Multi-Grid detector shows no sign of saturation. This re-emphasizes the need for neutron detector systems with high count rate capabilities.

Concerning internal neutron scattering in the Multi-Grid detector, it is seen to be primarily caused by neutron scattering in the aluminum structure within the Multi-Grid detector. Above 1~\AA{}, coherent scattering is the dominant component until approximately 4.7~\AA{}, where it drops to zero. This is reflected by the results, which show that above the aluminum cut-off wavelength at 4.7~\AA{}, the scattering is below the time-independent background level.

A detailed study on how the internally scattered neutrons affect the line shape was done using a geometrical cut, discarding the region where the beam hits the detector. This removes all events from directly absorbed neutrons, while keeping events from neutrons that are only absorbed after first being internally scattered. This allows an investigation of which part of the peak is due to scattering.

By studying data from Detector 1 (non-coated radial blades) and Detector 2 (coated radial blades), the net effect on the line shape is compared. Between 1 and 4 \AA{}, the average improvement induced by coating the radial blades is seen to increase with distance from the peak center, starting at 7~$\pm$~6~\% between [-5$\sigma$, 0$\sigma$], and reaching 31~$\pm$~8~\% at [-15$\sigma$, -10$\sigma$]. Note that, although the shoulder is a only small fraction of the main peak, it potentially affects the ability to extract the small quasi elastic signals which may occur in this region. Because of this, a suppression of 7-31\% in the shoulder region might be important. 

The energy resolution of the Multi-Grid detector and helium-3 tube is observed to broadly match, within 5 $\mu$eV, over the measured wavelengths. For the shortest wavelengths, around 1.5~\AA{}, the Multi-Grid detector and the helium-3 tube have equivalent resolutions. However, the trend implies that for shorter wavelengths than those measured here the Multi-Grid detector might have the advantage. Above 1.5~\AA{}, the helium-3 tube performs better in terms of energy resolution. This results from the fact that, for the helium-3 tube, neutrons can be converted anywhere in the gas along the 2.5 cm depth of the tube, while the absorption in the Multi-Grid detector only occurs at discrete intervals of 0.95~cm. This results in a difference in timing resolution. Due to this, at longer wavelengths, the obtained energy spectra from the Multi-Grid detector have a double peak, where each peak originates from one of the two conversion films adjacent to the same wire layer. This effect becomes more evident for longer wavelengths, as the additional path difference is more important for slow neutrons than for fast ones. However, in practical applications of the Multi-Grid detector at an instrument, such as at CNCS and SEQUOIA, this effect has not been observable. This is due to the longer flight path and additional pulse broadening at these facilities, which causes the 0.95~cm timing difference to be below instrument resolution.

Finally, note that the measured resolutions are a function of the detector specifications. The energy resolution of the helium-3 tube can be varied by choosing a different gas pressure in the tube, and, to a lesser extent, by changing the geometry. Conversely, the resolution of the Multi-Grid detector can be adjusted by varying the distance between layers. It can also be changed by choosing a single-sided coating approach, whereupon the uncertainty concerning which coating the conversion took place would vanish. That is, the double peak observed for long wavelengths would disappear, as it is unambiguous in which conversion layer the reaction occurred. However, a consequence of this would be either to accept a reduced efficiency, or a doubling of the detector depth. In both scenarios, the internal scattering would have a larger impact on the line shape, as the ratio between converter material and aluminum within the detector would be smaller. Thus, there would be less converter material to absorb the scattered neutrons. This results in a design trade-off between energy resolution, background reduction, and efficiency.

\section*{Acknowledgments}
This work was supported by the BrightnESS project (EU Horizon 2020, INFRADEV-3-2015, grant number 676548) and carried out as a part of the collaboration between ESS, Lund, Sweden, and ILL, Grenoble, France. The work originally started as a collaboration between ILL, the Link\"oping University, Link\"oping, Sweden, and ESS within the context of the International Collaboration on the development of Neutron Detectors (www.icnd.org). Support from the Swedish research council VR-RFI (\#2017-00646\_9) for the Accelerator based ion-technological center, and from the Swedish Foundation for Strategic Research (contract RIF14-0053) for the Tandem Accelerator Laboratory in Uppsala, is gratefully acknowledged. Furthermore, the authors would like to thank HZB, Berlin, Germany, for the beam time at the ESS test beamline V20. The data used for the analysis in this work are stored at http://researchdata.gla.ac.uk/id/eprint/1085.

\FloatBarrier
\newpage
\section*{References}
\bibliography{sources}

\end{document}